%% file: tfp.tex
\newif \iffull     \fullfalse
\newif \ifdraft    \draftfalse

\makeatletter \@input{texdirectives} \makeatother

\documentclass[runningheads,envcountsame]{llncs}

\usepackage{floatrow}
\usepackage{amsmath}

\usepackage{amssymb,amsmath}

\usepackage{enumitem}
\setlist{wide=0pt,align=left,leftmargin=1em,noitemsep}
\setlist[itemize,1]{label=--}

\usepackage{listings}
\usepackage{hyperref}
\hypersetup{
    pdftitle={Space-Efficient Latent Contracts},
    pdfauthor={Michael Greenberg},
    bookmarksnumbered=true,
    bookmarksopen=true,
    bookmarksopenlevel=1,
    hidelinks,
    naturalnames=true,
    pdfstartview=Fit,
    pdfpagemode=UseOutlines,
    pdfpagelayout=SinglePage
}

\usepackage{local}

\usepackage{tikz}
\usetikzlibrary{matrix,arrows}

\tikzset{align at top/.style={baseline=(current bounding box.north)}}

\tikzset{
    E/.style={
     shorten >=.25em,#1-to,
     to path={-- node[inner sep=0pt,at end,sloped] {${}_{ \mathsf{E} }^{\phantom{*}}$}
              (\tikztotarget) \tikztonodes}
    },
    E/.default=
}
\tikzset{
    E*/.style={
     shorten >=.25em,#1-to,
     to path={-- node[inner sep=0pt,at end,sloped] {${}_{ \mathsf{E} }^*$}
              (\tikztotarget) \tikztonodes}
    },
    E*/.default=
}

\input{cpcf}

\begin{document}

\mainmatter

\title{Space-Efficient Latent Contracts}
\ifdraft\subtitle{DRAFT --- do not distribute}\fi

\author{Michael Greenberg}
\authorrunning{Greenberg} 
%
\tocauthor{Michael Greenberg}
\institute{Pomona College \\
\email{michael@cs.pomona.edu}}

\maketitle

\begin{abstract}

Standard higher-order contract monitoring breaks tail recursion and
leads to space leaks that can change a program's asymptotic
complexity; space-efficiency restores tail recursion and bounds the
amount of space used by contracts.  Space-efficient contract
monitoring for contracts enforcing simple type disciplines (a/k/a
gradual typing) is well studied. Prior work establishes a
space-efficient semantics for manifest contracts without
dependency~\cite{Greenberg15space}; we adapt that work to a latent
calculus with dependency. We guarantee space efficiency when no
dependency is used; we cannot \textit{generally} guarantee space
efficiency when dependency is used, but instead offer a framework for
making such programs space efficient on a case-by-case basis.
\end{abstract}

\section{Introduction}
\label{sec:intro}

Findler and Felleisen~\cite{Findler02contracts} brought
design-by-contract~\cite{Meyer92Eiffel} into the higher-order world,
allowing programmers to write pre- and post-conditions on functions to be
checked at runtime.
Pre- and post-conditions are easy in first-order languages, where it's
very clear who is to blame when a contract is violated: if the
pre-condition fails, blame the caller; if the post-condition fails,
blame the callee.
In higher-order languages, however, it's harder to tell who calls
whom! Who should be to blame when a pre-condition on a higher-order
function fails? For example, consider the following contract:
\[   (   \mathsf{pred}(  \lambda \mathit{x} \mathord{:}  \mathsf{Int}  .~  \mathit{x}  \, \ottsym{>} \, \ottsym{0} )  \mapsto  \mathsf{pred}(  \lambda \mathit{y} \mathord{:}  \mathsf{Int}  .~  \mathit{y}  \, \ge \, \ottsym{0} )   )  \mapsto  \mathsf{pred}(  \lambda \mathit{z} \mathord{:}  \mathsf{Int}  .~  \mathit{z}  \,  \mathsf{mod}  \, \ottsym{2} \,  =  \, \ottsym{0} )   \]
This contract applies to a function (call it $\mathit{f}$, with type
$  (   \mathsf{Int}  \mathord{ \rightarrow }  \mathsf{Int}   )  \mathord{ \rightarrow }  \mathsf{Int}  $) that takes another function (call it $\mathit{g}$,
with type $  \mathsf{Int}  \mathord{ \rightarrow }  \mathsf{Int}  $) as input. The contract says that $\mathit{g}$
will only be called with positives and only return naturals; $\mathit{f}$
must return an even number.
If $\mathit{f}$ returns an odd number, $\mathit{f}$ is to blame; if $\mathit{g}$
returns a negative number, then it, too is to blame. But what if
$\mathit{g}$ is \textit{called} with a non-positive number, say, $ {-1} $?
Who is to blame then?
Findler and Felleisen's insight was that even in a higher-order
setting, there are only two parties to blame. Here, $\mathit{g}$ was given
to $\mathit{f}$, so any bad values given to $\mathit{g}$ here are due to some
nefarious action on $\mathit{f}$'s part---blame $\mathit{f}$!
That is, the higher-order case generalizes pre- and post-conditions so
that the negative positions of a contract all blame the caller while
the positive positions all blame the callee.

Dependent contracts---where the codomain contract can refer to the
function's argument---are particularly useful. For example, the square
root function, $ \mathsf{sqrt} $, satisfies the contract:
$ \mathit{x} \mathord{:}  \mathsf{pred}(  \lambda \mathit{y} \mathord{:}  \mathsf{Real}  .~  \mathit{y}  \, \ge \, \ottsym{0} )  \mapsto  \mathsf{pred}(   \lambda \mathit{z} \mathord{:}  \mathsf{Real}  .~   \mathsf{abs}    ~   ( \mathit{x} \,  -  \, \mathit{z} \,  *  \, \mathit{z} )   \, \ottsym{<} \,  \epsilon  )  $
That is, $ \mathsf{sqrt} $ takes a non-negative
real, $\mathit{x}$, and returns a non-negative real $\mathit{z}$ that's within
$ \epsilon $ of the square root of $\mathit{x}$. (The dependent variable
$\mathit{x}$ is bound in the codomain; the variable $\mathit{y}$ is local to the domain
predicate.)

\subsection{Contracts leak space}

While implementations of contracts have proven quite successful
(particularly so in Racket~\cite{Racket,RacketContracts}), there is a
problem: contracts leak space. Why?

The default implementation of contracts works by wrapping a function
in a \textit{function proxy}. For example, to check that $\mathit{f} =
 \lambda \mathit{x} \mathord{:}  \mathsf{Int}  .~  \mathit{x}  \,  +  \, \ottsym{1}$ satisfies the contract $\ottnt{C} =   \mathsf{pred}(  \lambda \mathit{z} \mathord{:}  \mathsf{Int}  .~  \mathit{z}  \,  \mathsf{mod}  \, \ottsym{2} \,  =  \, \ottsym{0} )  \mapsto  \mathsf{pred}(  \lambda \mathit{z} \mathord{:}  \mathsf{Int}  .~  \mathit{z}  \,  \mathsf{mod}  \, \ottsym{2} \,  =  \, \ottsym{0} )  $, we monitor the
function by wrapping it in a function proxy $ \mathsf{mon}^{ \ottnt{l} }( \ottnt{C} ,  \mathit{f} ) $. When
this proxy is called with an input $\ottnt{v}$, we first check that
$\ottnt{v}$ satisfies $\ottnt{C}$'s domain contract (i.e., that $\ottnt{v}$ is
even), then we run $\mathit{f}$ on $\ottnt{v}$ to get some result $\ottnt{v'}$, and
then check that $\ottnt{v'}$ satisfies $\ottnt{C}$'s codomain contract (that
the result is even). Here the contract will always fail blaming
$\ottnt{l}$: one of $\ottnt{v}$ and $\ottnt{v'}$ will always be odd.

Contracts leak space in two ways. First, there is no bound on the
number of function proxies that can appear on a given function.
More grievously, contracts break tail recursion.
To demonstrate the issue with with tail calls, we'll use the simplest
example of mutual recursion: detecting parity.
\[\begin{array}{lcl}
   \mathsf{let}~  \mathsf{odd}  &=&  \lambda \mathit{x} \mathord{:}  \mathsf{Int}  .~   \mathsf{if} ~   ( \mathit{x} \,  =  \, \ottsym{0} )   ~   \mathsf{false}   ~~   (   \mathsf{even}   ~   ( \mathit{x} \,  -  \, \ottsym{1} )   )    \\
   \mathsf{and}~  \mathsf{even}  &=&  \lambda \mathit{x} \mathord{:}  \mathsf{Int}  .~   \mathsf{if} ~   ( \mathit{x} \,  =  \, \ottsym{0} )   ~   \mathsf{true}   ~~   (   \mathsf{odd}   ~   ( \mathit{x} \,  -  \, \ottsym{1} )   )   
\end{array}\]
Functional programmers will expect this program to run in constant
space, because it is \textit{tail recursive}.
Adding a contract breaks the tail recursion.
If we add a contract to $ \mathsf{odd} $ and call $  \mathsf{odd}   ~  \ottsym{5} $, what contract
checks accumulate (Fig.~\ref{fig:spaceleak})?\footnote{Readers may
  observe that the contract betrays a deeper knowledge of numbers than
  the functions themselves. We offer this example as minimal, not naturally occurring.}
\begin{figure}[t]
\[\begin{array}{lcl}
   \mathsf{let}~  \mathsf{odd}  &=&
    \mathsf{mon}^{  l_{ \mathsf{odd} }  }(  \mathit{x} \mathord{:}  \mathsf{pred}(  \lambda \mathit{x} \mathord{:}  \mathsf{Int}  .~  \mathit{x}  \, \ge \, \ottsym{0} )  \mapsto  \mathsf{pred}(  \lambda \mathit{b} \mathord{:}  \mathsf{Bool}  .~  \mathit{b}  \,  \mathsf{or}  \,  ( \mathit{x} \,  \mathsf{mod}  \, \ottsym{2} \,  =  \, \ottsym{0} )  )   ,  {} \\  &  &  \quad   \lambda \mathit{x} \mathord{:}  \mathsf{Int}  .~   \mathsf{if} ~   ( \mathit{x} \,  =  \, \ottsym{0} )   ~   \mathsf{false}   ~~   (   \mathsf{even}   ~   ( \mathit{x} \,  -  \, \ottsym{1} )   )    )  \\

   \mathsf{and}~  \mathsf{even}  &=&  \lambda \mathit{x} \mathord{:}  \mathsf{Int}  .~   \mathsf{if} ~   ( \mathit{x} \,  =  \, \ottsym{0} )   ~   \mathsf{true}   ~~   (   \mathsf{odd}   ~   ( \mathit{x} \,  -  \, \ottsym{1} )   )   
\end{array}\]

\[ \begin{array}{rl}
  &   \mathsf{odd}   ~  \ottsym{5}  \\
   \longrightarrow ^{*}_{  \mathsf{C}  } &  \mathsf{mon}^{  l_{ \mathsf{odd} }  }(  \mathsf{pred}_{  [  \mathit{x}  \mapsto  \ottsym{5}  ]  }( \dots )  ,    \mathsf{even}   ~  \ottsym{4}  )  \\
   \longrightarrow ^{*}_{  \mathsf{C}  } &  \mathsf{mon}^{  l_{ \mathsf{odd} }  }(  \mathsf{pred}_{  [  \mathit{x}  \mapsto  \ottsym{5}  ]  }( \dots )  ,   \mathsf{mon}^{  l_{ \mathsf{odd} }  }(  \mathsf{pred}_{  [  \mathit{x}  \mapsto  \ottsym{3}  ]  }( \dots )  ,  {} \\   &  ~~   \mathsf{odd}   ~   \mathsf{mon}^{  l_{ \mathsf{odd} }  }(  \mathsf{pred}(  \lambda \mathit{x} \mathord{:}  \mathsf{Int}  .~  \mathit{x}  \, \ge \, \ottsym{0} )  ,  \ottsym{3} )   )  )  \\
   \longrightarrow ^{*}_{  \mathsf{C}  } &  \mathsf{mon}^{  l_{ \mathsf{odd} }  }(  \mathsf{pred}_{  [  \mathit{x}  \mapsto  \ottsym{5}  ]  }( \dots )  ,   \mathsf{mon}^{  l_{ \mathsf{odd} }  }(  \mathsf{pred}_{  [  \mathit{x}  \mapsto  \ottsym{3}  ]  }( \dots )  ,    \mathsf{even}   ~  \ottsym{2}  )  )  \\
   \longrightarrow ^{*}_{  \mathsf{C}  } &  \mathsf{mon}^{  l_{ \mathsf{odd} }  }(  \mathsf{pred}_{  [  \mathit{x}  \mapsto  \ottsym{5}  ]  }( \dots )  ,   \mathsf{mon}^{  l_{ \mathsf{odd} }  }(  \mathsf{pred}_{  [  \mathit{x}  \mapsto  \ottsym{3}  ]  }( \dots )  ,   \mathsf{mon}^{  l_{ \mathsf{odd} }  }(  \mathsf{pred}_{  [  \mathit{x}  \mapsto  \ottsym{1}  ]  }( \dots )  ,  {} \\   &  ~~   \mathsf{odd}   ~   \mathsf{mon}^{  l_{ \mathsf{odd} }  }(  \mathsf{pred}(  \lambda \mathit{x} \mathord{:}  \mathsf{Int}  .~  \mathit{x}  \, \ge \, \ottsym{0} )  ,  \ottsym{1} )   )  )  )  \\
   \longrightarrow ^{*}_{  \mathsf{C}  } &  \mathsf{mon}^{  l_{ \mathsf{odd} }  }(  \mathsf{pred}_{  [  \mathit{x}  \mapsto  \ottsym{5}  ]  }( \dots )  ,   \mathsf{mon}^{  l_{ \mathsf{odd} }  }(  \mathsf{pred}_{  [  \mathit{x}  \mapsto  \ottsym{3}  ]  }( \dots )  ,   \mathsf{mon}^{  l_{ \mathsf{odd} }  }(  \mathsf{pred}_{  [  \mathit{x}  \mapsto  \ottsym{1}  ]  }( \dots )  ,    \mathsf{even}   ~  \ottsym{0}  )  )  ) 
\end{array} \]
\caption{Contracts break tail recursion}
\label{fig:spaceleak}
\end{figure}
Notice how the checks accumulate in the codomain? Even though the
mutually recursive calls to $ \mathsf{even} $ and $ \mathsf{odd} $ are syntactically
tail calls, we can't bound the number of codomain checks that
occur. That is, we can't bound the size of the stack, and tail
recursion is broken!
Even though there's only one function proxy on $ \mathsf{odd} $, our
contracts create a space leak.

\subsection{Overview and contributions}

Space efficiency for gradual types~\cite{Siek06gradual} (a/k/a
contracts constrained to type tests) is well
studied~\cite{Herman07space,Herman10space,Siek10threesomes,Garcia13threesomes,Siek15coercions};
Greenberg~\cite{Greenberg15space} developed a space-efficient
semantics for general, non-dependent contracts.
He used a manifest calculus, conflating contracts and types; however,
contracts are typically implemented in latent calculi, where contracts
are distinct from whatever types may exist. Greenberg ``believe[s] it
would be easy to design a latent version of eidetic \lambdah,
following the translations in Greenberg, Pierce, and Weirich
(GPW)''~\cite{Greenberg10contracts}; in this paper, we show that
belief to be well founded by giving a space-efficient semantics for a
(dependent!) variant of contract PCF
(CPCF)~\cite{Dimoulas11cpcf,Dimoulas11indy}.

The rest of this paper discusses a formulation of contracts that
enjoys sound space efficiency; that is, where we slightly change the
implementation of contracts so that (a) programs are observationally
equivalent to the standard semantics, but (b) contracts consume a
bounded amount of space. In this paper, we've omitted some of the more
detailed examples and motivation---we refer curious readers to
Greenberg~\cite{Greenberg15space}, though we intend the paper to be
self-contained.

We follow Greenberg's general structure, defining two forms of
dependent CPCF: \CPCFc is the \textit{classic} semantics; \CPCFe
follows the space-efficient eidetic semantics. We are able to prove
space efficiency without dependency, bounding the amount of space
consumed by contracts; we are unable to prove space efficiency in
general with dependency, but instead offer a framework that allows for
dependent contracts to be \textit{made} space efficient.

We offer two primary contributions: adapting Greenberg's work to a
latent calculus and extending the possibility of space efficiency to
dependent contracts.

There are some other, smaller, contributions as well.
First, adding in nontermination moves beyond Greenberg's strongly normalizing
calculi, showing that the POPL 2015 paper's result isn't an artifact of
strong normalization (where we can, in theory, bound the size of the
any term's evaluation in advance, not just contracts).
Second, the simpler type system here makes it clear which type system
invariants are necessary for space-efficiency and which are
bookkeeping for proving that the more complicated manifest type system
is sound.
Third, by separating contracts and types, we can give tighter space
bounds\iffull---the $\mathsf{types}$ function from Greenberg collects types
that are never used in a contract, while we collect exactly contracts\fi.
Finally, we explore how space efficiency can be attained in dependent
contracts. While we can't give a guarantee for dependent contracts, we
show that it's possible to achieve and discuss different ways to do
so.

\section{Classic and space-efficient Contract PCF}
\label{sec:classic}

\TODO{change these colors}

\definecolor{salmon}{RGB}{253,237,216}
\definecolor{periwinkle}{RGB}{215,211,232}

\newcommand{\C}[1]{\highlight[salmon]{#1}}
\newcommand{\SE}[1]{\highlight[periwinkle]{#1}}

We present classic and space-efficient CPCF as separate calculi
sharing syntax and some typing rules (Fig.~\ref{fig:syntax} and
Fig.~\ref{fig:typing}), and a single, parameterized operational
semantics with some rules held completely in common \iffull
(Fig.~\ref{fig:sharedopsem}) \else (omitted to save space) \fi and
others specialized to each system (Fig.~\ref{fig:specificopsem}).
The formal presentation is modal, with two modes: $\mathsf{C}$ for
\underline{c}lassic and $\mathsf{E}$ for
space-\underline{e}fficient. While much is shared between the two
modes---types, $\ottnt{T}$; the core syntax of expressions, $\ottnt{e}$; most of the
typing rules---we use colors to highlight parts that belong to only
one system. Classic CPCF is typeset in $\C{\text{salmon}}$ while
space-efficient CPCF is in $\SE{\text{periwinkle}}$.

\begin{figure}[t]
  \[\begin{array}{l@{\quad}r@{~~~}c@{~~~}l}
    \textbf{Types} &
      \ottnt{B} &::=&  \mathsf{Bool}  \BNFALT  \mathsf{Int}  \BNFALT  \dots  \\
    & \ottnt{T} &::=& \ottnt{B} \BNFALT  \ottnt{T_{{\mathrm{1}}}} \mathord{ \rightarrow } \ottnt{T_{{\mathrm{2}}}}  \\

    \textbf{Terms} &
    \ottnt{e} &::=& \mathit{x} \BNFALT \ottnt{k} \BNFALT \ottnt{e_{{\mathrm{1}}}} \, \ottnt{op} \, \ottnt{e_{{\mathrm{2}}}} \BNFALT  \ottnt{e_{{\mathrm{1}}}}  ~  \ottnt{e_{{\mathrm{2}}}}  \BNFALT
                 \lambda \mathit{x} \mathord{:} \ottnt{T} .~  \ottnt{e}  \BNFALT  \mu ( \mathit{x} \mathord{:} \ottnt{T} ) . ~  \ottnt{e}  \BNFALT  \mathsf{if} ~  \ottnt{e_{{\mathrm{1}}}}  ~  \ottnt{e_{{\mathrm{2}}}}  ~~  \ottnt{e_{{\mathrm{3}}}}  \BNFALT \\
      &&&  \mathsf{err}^ \ottnt{l}  \BNFALT  \mathsf{mon}^{ \ottnt{l} }( \ottnt{C} ,  \ottnt{e} )  \BNFALT \SE{ \mathsf{mon}( \ottnt{c} ,  \ottnt{e} ) } \\
   & \ottnt{op} &::=&  \mathsf{add1}  \BNFALT  \mathsf{sub1}  \BNFALT  \dots  \\
   &  \ottnt{k} &::=&  \mathsf{true}  \BNFALT  \mathsf{false}  \BNFALT \ottsym{0} \BNFALT \ottsym{1} \BNFALT  \dots  \\
   &  \ottnt{w} &::=& \ottnt{v} \BNFALT  \mathsf{err}^ \ottnt{l}  \\
   &  \ottnt{v} &::=& \ottnt{k} \BNFALT  \lambda \mathit{x} \mathord{:} \ottnt{T} .~  \ottnt{e}  \BNFALT \C{ \mathsf{mon}^{ \ottnt{l} }(  \mathit{x} \mathord{:} \ottnt{C_{{\mathrm{1}}}} \mapsto \ottnt{C_{{\mathrm{2}}}}  ,  \ottnt{v} ) } \BNFALT \SE{ \mathsf{mon}(  \mathit{x} \mathord{:} \ottnt{c_{{\mathrm{1}}}} \mapsto \ottnt{c_{{\mathrm{2}}}}  ,   \lambda \mathit{x} \mathord{:} \ottnt{T} .~  \ottnt{e}  ) } \\
   &  \ottnt{C} &::=&  \mathsf{pred}_{ \sigma }( \ottnt{e} )  \BNFALT  \mathit{x} \mathord{:} \ottnt{C_{{\mathrm{1}}}} \mapsto \ottnt{C_{{\mathrm{2}}}}  \\
   &  \SE{\ottnt{c}} &::=& \SE{\vphantom{(}\ottnt{r}} \BNFALT \SE{\vphantom{(} \mathit{x} \mathord{:} \ottnt{c_{{\mathrm{1}}}} \mapsto \ottnt{c_{{\mathrm{2}}}} } \\
   &  \SE{\ottnt{r}} &::=& \SE{\vphantom{(} \mathsf{nil} } \BNFALT \SE{  \mathsf{pred}^{ \ottnt{l} }_{ \sigma }( \ottnt{e} )  ; \ottnt{r} }
%
%
  \end{array}\]
  \caption{Syntax of $\C{\text{classic}}$ and $\SE{\text{space-efficient}}$ CPCF}
  \label{fig:syntax}
\end{figure}

\subsection{Contract PCF (CPCF)}
\label{sec:cpcf}

Plain CPCF is an extension of Plotkin's 1977 PCF~\cite{Plotkin77pcf},
developed first by Dimoulas and
Felleisen~\cite{Dimoulas11cpcf,Dimoulas11indy} (our syntax is in
Fig.~\ref{fig:syntax}). It is a simply typed language with
recursion.  The typing rules are straightforward
(Fig.~\ref{fig:typing}). The operational semantics for the generic
fragment also uses conventional rules
({\iffull{}Fig.~\ref{fig:sharedopsem}\else{}omitted to save
  space\fi}). Dimoulas and Felleisen use evaluation contexts to offer
a concise description of their system; we write out our relation in
full, giving congruence rules (\E{*L}, \E{*R}, \E{If}) and error
propagating rules (\E{*Raise}) explicitly---we will need to restrict
congruence for casts, and our methods are more transparent written with explicit congruence rules
than using the subtly nested evaluation contexts of Herman et
al.~\cite{Herman07space,Herman10space}\iffull, which are error
prone~\cite{Greenberg13thesis}\fi.

\textit{Contracts} are CPCF's distinguishing feature.
Contracts, $\ottnt{C}$, are installed via monitors, written
$ \mathsf{mon}^{ \ottnt{l} }( \ottnt{C} ,  \ottnt{e} ) $; such a monitor says ``ensure
that $\ottnt{e}$ satisfies the contract $\ottnt{C}$; if not, the blame lies
with label $\ottnt{l}$''.
Monitors only apply to appropriate types (\T{Mon}).
There are two kinds of contracts in CPCF: \textit{predicate contracts}
over base type, written $ \mathsf{pred}_{ \sigma }( \ottnt{e} ) $, and \textit{function
  contracts}, written $ \mathit{x} \mathord{:} \ottnt{C_{{\mathrm{1}}}} \mapsto \ottnt{C_{{\mathrm{2}}}} $.

Predicate contracts $ \mathsf{pred}_{ \sigma }( \ottnt{e} ) $ have two parts: a predicate on
base types, $\ottnt{e}$, which identifies which values satisfy the contract;
and a closing substitution $\sigma$ which keeps track of values
substituted into the contract. For example, if $ \iota $ is the identity substitution mapping variables to themselves:
\begin{itemize}
\item $ \mathsf{pred}_{  \iota  }(  \lambda \mathit{x} \mathord{:}  \mathsf{Int}  .~  \mathit{x}  \, \ottsym{>} \, \ottsym{0} ) $ identifies the positives;
\item $ \mathsf{pred}_{  \iota  }(  \lambda \mathit{x} \mathord{:}  \mathsf{Int}  .~  \mathit{x}  \, \ottsym{>} \, \mathit{y} ) $ identifies numbers greater than an unspecified number
$\mathit{y}$; and,
\item $ \mathsf{pred}_{  [  \mathit{y}  \mapsto  \ottsym{47}  ]  }(  \lambda \mathit{x} \mathord{:}  \mathsf{Int}  .~  \mathit{x}  \, \ottsym{>} \, \mathit{y} ) $ identifies numbers greater than 47.
\end{itemize}
When the closing substitution $\sigma$ is the identity mapping $ \iota $, we write
$ \mathsf{pred}( \ottnt{e} ) $ instead of $ \mathsf{pred}_{  \iota  }( \ottnt{e} ) $.
In \CPCFc, closing substitutions will map each variable to either (a)
itself or (b) a value.
Substitution into contracts is a non-issue without dependency: each
contract is just closed. Having introduced dependency, we use explicit
closing substitutions rather than direct substitution for three
reasons: first, it simplifies our space efficiency proof for simple
contracts (Sec.~\ref{sec:simplebounds}); second, explicitness lets
us distinguish the contract $ \mathsf{pred}_{  [  \mathit{x}  \mapsto  \ottsym{0}  ]  }(  \lambda \mathit{x} \mathord{:}  \mathsf{Int}  .~  \mathit{x}  \,  =  \, \ottsym{0} ) $ from
$ \mathsf{pred}_{  [  \mathit{x}  \mapsto  \ottsym{0}  ]  }(  \lambda \mathit{x} \mathord{:}  \mathsf{Int}  .~  \ottsym{0}  \,  =  \, \ottsym{0} ) $; third, it emphasizes that
contracts are just another form of closure.
Predicates are solely over \textit{base types}, not functions.

Function contracts $ \mathit{x} \mathord{:} \ottnt{C_{{\mathrm{1}}}} \mapsto \ottnt{C_{{\mathrm{2}}}} $ are satisfied by functions
satisfying their parts: functions whose inputs all satisfy $\ottnt{C_{{\mathrm{1}}}}$
and whose outputs all satisfy $\ottnt{C_{{\mathrm{2}}}}$. Function contracts are
dependent: the codomain contract $\ottnt{C_{{\mathrm{2}}}}$ can refer back to the input
to the function. For example, the contract $ \mathit{x} \mathord{:}  \mathsf{pred}(  \lambda \mathit{z} \mathord{:}  \mathsf{Int}  .~  \mathit{z}  \, \ottsym{>} \, \ottsym{0} )  \mapsto  \mathsf{pred}(  \lambda \mathit{y} \mathord{:}  \mathsf{Int}  .~  \mathit{y}  \, \ottsym{>} \, \mathit{x} )  $ is satisfied by increasing functions on
the positives. Note that $\mathit{x}$ is bound in the codomain, but $\mathit{z}$
is not.\footnote{Concrete syntax for such predicates can be written
  much more nicely, but we ignore such concerns here.}
When function contracts aren't dependent, we omit the binder at the front,
e.g., $  \mathsf{pred}(  \lambda \mathit{x} \mathord{:}  \mathsf{Int}  .~  \mathit{x}  \, \ottsym{>} \, \ottsym{0} )  \mapsto  \mathsf{pred}(  \lambda \mathit{x} \mathord{:}  \mathsf{Int}  .~  \mathit{x}  \, \ottsym{>} \, \ottsym{0} )  $ means
operators on positives.
We check that contracts are satisfied at runtime.

\begin{figure}[tp]
  \hdr{Typing rules}{\qquad \fbox{$ \mathrm{\Gamma}   \vdash   \ottnt{e}  :  \ottnt{T} $}}
  \threesidebyside
    {\ottusedrule{\ottdruleTVar{}}}
    {\ottusedrule{\ottdruleTConst{}}}
    {\ottusedrule{\ottdruleTBlame{}}}
  \sidebyside
    {\ottusedrule{\ottdruleTAbs{}}}
    {\ottusedrule{\ottdruleTRec{}}}
  \sidebyside
    {\ottusedrule{\ottdruleTOp{}}}
    {\ottusedrule{\ottdruleTApp{}}}
  {\ottusedrule{\ottdruleTIf{}}}
  \sidebyside
    {\ottusedrule{\ottdruleTMon{}}}
    {\ottusedrule{\SE{\ottdruleTMonC{}}}}

  \hdr{Contract typing}{
    \qquad \fbox{$\mathrm{\Gamma}  \vdash  \ottnt{C}  \ottsym{:}  \ottnt{T}$}
    \qquad \fbox{$\mathrm{\Gamma}  \vdash  \ottnt{c}  \ottsym{:}  \ottnt{T}$}}
  \sidebysidesqueeze[.45][.47][0.03]
    {\ottusedrule{\ottdruleTPred{}}}
    {\ottusedrule{\ottdruleTFun{}}}
  \sidebyside[.29][.65][b]
    {\ottusedrule{\SE{\ottdruleTCNil{}}}}
    {\ottusedrule{\SE{\ottdruleTCPred{}}}}
  {\ottusedrule{\SE{\ottdruleTCFun{}}}}

  \hdr{Closing substitutions}{\qquad \fbox{$\mathrm{\Gamma}  \vdash  \sigma$}}
  \sidebyside
    {\ottusedrule{\ottdruleTId{}}}
    {\ottusedrule{\ottdruleTMap{}}}

  \vspace*{-1em}
  \caption{Typing rules of $\C{\text{classic}}$ and
    $\SE{\text{space-efficient}}$ CPCF}
  \label{fig:typing}
\end{figure}

We use explicit, delayed substitutions to keep track of which values
are substituted into predicate contracts. To help with our proof of space efficiency, we don't track variables that don't appear in the predicate:
\[ \begin{array}{rcl}
    \mathsf{pred}_{ \sigma }( \ottnt{e} )   [  \ottnt{v}  /  \mathit{x}  ]  &=& \begin{cases}
     \mathsf{pred}_{  \sigma  [  \mathit{x}  \mapsto  \ottnt{v}  ]  }( \ottnt{e} )  &  \mathit{x}  \in   \operatorname{fv} ( \sigma  \ottsym{(}  \ottnt{e}  \ottsym{)} )   \\
     \mathsf{pred}_{ \sigma }( \ottnt{e} )  & \text{otherwise}
  \end{cases}
\end{array} \]
Alpha equivalence allows us to give fresh names to variables in the
domain of $\sigma$ by consistently renaming those
variables inside of the predicate $\ottnt{e}$.
Only holding on to substitutions that close up free variables in
$\ottnt{e}$ is a way of modeling closures. A dependent predicate closes
over some finite number of variables; a compiled representation would
generate a closure with a corresponding number of slots in the closing
environment.
Restricting substitutions to exactly those variables appearing free
in the predicate serves another purpose: we can easily recover
space-efficiency bounds for programs without dependent contracts
(Sec.~\ref{sec:simplebounds}).

{\iffull
\begin{figure}[t]
  \hdr{Operational semantics}{\qquad \fbox{$\ottnt{e_{{\mathrm{1}}}} \,  \longrightarrow _{ \mathsf{m} }  \, \ottnt{e_{{\mathrm{2}}}}$}}

  \sidebyside
    {\ottusedrule{\ottdruleEBeta{}}}
    {\ottusedrule{\ottdruleEDelta{}}}
  {\ottusedrule{\ottdruleEFix{}}}
  \sidebyside
    {\ottusedrule{\ottdruleEIfTrue{}}}
    {\ottusedrule{\ottdruleEIfFalse{}}}
  \sidebyside
    {\ottusedrule{\ottdruleEAppL{}}}
    {\ottusedrule{\ottdruleEAppR{}}}
  \sidebyside
    {\ottusedrule{\ottdruleEOpL{}}}
    {\ottusedrule{\ottdruleEOpR{}}}
  \sidebyside
    {\ottusedrule{\ottdruleEIf{}}}
    {\ottusedrule{\ottdruleEIfRaise{}}}
  \sidebyside
    {\ottusedrule{\ottdruleEAppLRaise{}}}
    {\ottusedrule{\ottdruleEAppRRaise{}}}
  \sidebyside
    {\ottusedrule{\ottdruleEOpLRaise{}}}
    {\ottusedrule{\ottdruleEOpRRaise{}}}

  \vspace*{-1em}
  \caption{Shared operational semantics of CPCF}
  \label{fig:sharedopsem}
\end{figure}
\fi}

\subsection{Classic Contract PCF (\CPCFc)}
\label{sec:cpcfc}

\begin{figure}[t!]
  {\ottusedrule{\C{\ottdruleEMonPred{}}}}
  {\ottusedrule{\C{\ottdruleEMonApp{}}}}
  \sidebyside[.47][.48]
    {\ottusedrule{\C{\ottdruleEMon{}}}}
    {\ottusedrule{\C{\ottdruleEMonRaise{}}}}
  \vspace*{-.1in}
  {\ottusedrule{\SE{\ottdruleEMonLabel{}}}}
  {\ottusedrule{\SE{\ottdruleEMonCNil{}}}}
  {\ottusedrule{\SE{\ottdruleEMonCPred{}}}}
  {\ottusedrule{\SE{\ottdruleEMonCApp{}}}}
  \sidebyside[.48][.47]
    {\ottusedrule{\SE{\ottdruleEMonC{}}}}
    {\ottusedrule{\SE{\ottdruleEMonCRaise{}}}}
  \vspace*{-.1in}
  {\ottusedrule{\SE{\ottdruleEMonCJoin{}}}}

  \caption{Operational semantics of $\C{\text{classic}}$ and
    $\SE{\text{space-efficient}}$ CPCF}
  \label{fig:specificopsem}
\end{figure}

Classic CPCF gives a straightforward semantics to contracts
(\iffull{}Figs.~\ref{fig:sharedopsem} and~\ref{fig:specificopsem}\else{}Fig.~\ref{fig:specificopsem}\fi), largely
following the seminal work by Findler and
Felleisen~\cite{Findler02contracts}. To check a predicate contract, we
simply test it (\E{MonPred}), returning either the value or an
appropriately labeled error. Function contracts are deferred:
$ \mathsf{mon}^{ \ottnt{l} }(  \mathit{x} \mathord{:} \ottnt{C_{{\mathrm{1}}}} \mapsto \ottnt{C_{{\mathrm{2}}}}  ,  \ottnt{v} ) $ is a \textit{value}, called a \textit{function
  proxy}. When a function proxy is applied, it unwraps the
proxy, monitoring the argument with the domain contract, running the
function, and then monitoring the return value with the codomain contract
(\E{MonApp}).

Our semantics may seem to be \textit{lax}, where no monitor is applied
to dependent uses of the argument in the codomain
monitor~\cite{Greenberg10contracts}.
In fact, it is agnostic: we could be \textit{picky} by requiring that
function contract monitors $ \mathsf{mon}^{ \ottnt{l} }(  \mathit{x} \mathord{:} \ottnt{C_{{\mathrm{1}}}} \mapsto \ottnt{C_{{\mathrm{2}}}}  ,  \ottnt{e} ) $ have the
substitution $ [  \mathit{x}  \mapsto   \mathsf{mon}^{ \ottnt{l} }( \ottnt{C_{{\mathrm{1}}}} ,  \mathit{x} )   ] $ throughout $\ottnt{C_{{\mathrm{2}}}}$; we could be
\textit{indy} by having $ [  \mathit{x}  \mapsto   \mathsf{mon}^{ \ottnt{l'} }( \ottnt{C_{{\mathrm{1}}}} ,  \mathit{x} )   ] $ throughout
$\ottnt{C_{{\mathrm{2}}}}$~\cite{Dimoulas11indy}.
We default to a lax rule to make our proof of soundness easier, but
we'll have as a corollary that classic and space-efficient semantics
yield the same result regardless of what the closing substitutions do
in the codomain (Sec.~\ref{sec:soundness}).

Standard congruence rules allow for evaluation inside of monitors
(\E{Mon}) and the propagation of errors (\E{MonRaise}).

{\iffull
\subsubsection{Metatheory}
\label{sec:cpcfcmetatheory}

We prove \CPCFc's type system sound with a minimum fuss, using the
usual syntactic methods. The only subtlety is that we must be careful
when proving the substitution property, since we have slightly changed
the definition.

In this section, we only consider the typing rules for \CPCFc, i.e.,
those typeset on white and $\C{\text{salmon}}$.

{\iffull
\begin{lemma}[Weakening]
  \label{lem:cpcfcweakening}
  \begin{itemize}
  \item If $ \mathrm{\Gamma}   \vdash   \ottnt{e}  :  \ottnt{T} $ and $ \mathit{x}  \not\in   \operatorname{fv} ( \ottnt{e} )  $ then $  \mathrm{\Gamma} , \mathit{x} \mathord{:} \ottnt{T'}    \vdash   \ottnt{e}  :  \ottnt{T} $.
  \item If $\mathrm{\Gamma}  \vdash  \ottnt{C}  \ottsym{:}  \ottnt{T}$ and $ \mathit{x}  \not\in   \operatorname{fv} ( \ottnt{C} )  $ then $ \mathrm{\Gamma} , \mathit{x} \mathord{:} \ottnt{T'}   \vdash  \ottnt{C}  \ottsym{:}  \ottnt{T}$.
  \end{itemize}
  \begin{proof}
    By mutual induction on the terms and contracts.
  \end{proof}
\end{lemma}

\begin{lemma}[Substitution]
  \label{lem:cpcfcsubstitution}
  If $ \emptyset   \vdash   \ottnt{v}  :  \ottnt{T_{{\mathrm{1}}}} $, then
  \begin{itemize}
  \item $  \mathrm{\Gamma}_{{\mathrm{1}}} , \mathit{x} \mathord{:} \ottnt{T_{{\mathrm{1}}}}   \ottsym{,}  \mathrm{\Gamma}_{{\mathrm{2}}}   \vdash   \ottnt{e}  :  \ottnt{T_{{\mathrm{2}}}} $ implies $ \mathrm{\Gamma}_{{\mathrm{1}}}  \ottsym{,}  \mathrm{\Gamma}_{{\mathrm{2}}}   \vdash    \ottnt{e}  [  \ottnt{v} / \mathit{x}  ]   :  \ottnt{T_{{\mathrm{2}}}} $, and
  \item $ \mathrm{\Gamma}_{{\mathrm{1}}} , \mathit{x} \mathord{:} \ottnt{T_{{\mathrm{1}}}}   \ottsym{,}  \mathrm{\Gamma}_{{\mathrm{2}}}  \vdash  \ottnt{C}  \ottsym{:}  \ottnt{T_{{\mathrm{2}}}}$ implies $\mathrm{\Gamma}_{{\mathrm{1}}}  \ottsym{,}  \mathrm{\Gamma}_{{\mathrm{2}}}  \vdash   \ottnt{C}  [  \ottnt{v}  /  \mathit{x}  ]   \ottsym{:}  \ottnt{T_{{\mathrm{2}}}}$.
  \end{itemize}
  \begin{proof}
    By mutual induction the terms and contracts. \iffull
    \begin{itemize}
    \item[($\ottnt{e}  \ottsym{=}  \mathit{y}$)] By case analysis on $\mathit{x}  \ottsym{=}  \mathit{y}$, using
      weakening (Lemma~\ref{lem:cpcfcweakening}) as appropriate.
    \item[($\ottnt{e}  \ottsym{=}  \ottnt{k}$)] Immediate.
    \item[($\ottnt{e}  \ottsym{=}  \ottnt{e_{{\mathrm{1}}}} \, \ottnt{op} \, \ottnt{e_{{\mathrm{2}}}}$)] By the IHs.
    \item[($\ottnt{e}  \ottsym{=}   \ottnt{e_{{\mathrm{1}}}}  ~  \ottnt{e_{{\mathrm{2}}}} $)] By the IHs.
    \item[($\ottnt{e}  \ottsym{=}   \lambda \mathit{y} \mathord{:} \ottnt{T} .~  \ottnt{e_{{\mathrm{1}}}} $)] By case analysis on $\mathit{x}  \ottsym{=}  \mathit{y}$,
      narrowing when the two are equal.
    \item[($\ottnt{e}  \ottsym{=}   \mu ( \mathit{y} \mathord{:} \ottnt{T} ) . ~  \ottnt{e_{{\mathrm{1}}}} $)] By case analysis on $\mathit{x}  \ottsym{=}  \mathit{y}$,
      narrowing when the two are equal.
    \item[($\ottnt{e}  \ottsym{=}   \mathsf{if} ~  \ottnt{e_{{\mathrm{1}}}}  ~  \ottnt{e_{{\mathrm{2}}}}  ~~  \ottnt{e_{{\mathrm{3}}}} $)] By the IHs.
    \item[($\ottnt{e}  \ottsym{=}   \mathsf{err}^ \ottnt{l} $)] Immediate.
    \item[($\ottnt{e}  \ottsym{=}   \mathsf{mon}^{ \ottnt{l} }( \ottnt{C} ,  \ottnt{e} ) $)] By the IHs.
    \item[($\ottnt{C}  \ottsym{=}   \mathsf{pred}_{ \sigma }( \ottnt{e} ) $)] \else The interesting case is when
      $\ottnt{C}  \ottsym{=}   \mathsf{pred}_{ \sigma }( \ottnt{e} ) $. \fi
      We have $ \mathrm{\Gamma}_{{\mathrm{1}}} , \mathit{x} \mathord{:} \ottnt{T_{{\mathrm{1}}}}   \ottsym{,}  \mathrm{\Gamma}_{{\mathrm{2}}}  \vdash   \mathsf{pred}_{ \sigma }( \ottnt{e} )   \ottsym{:}  \ottnt{T_{{\mathrm{2}}}}$ and we must show that $\mathrm{\Gamma}_{{\mathrm{1}}}  \ottsym{,}  \mathrm{\Gamma}_{{\mathrm{2}}}  \vdash    \mathsf{pred}_{ \sigma }( \ottnt{e} )   [  \ottnt{v}  /  \mathit{x}  ]   \ottsym{:}  \ottnt{T_{{\mathrm{2}}}}$.
      By inversion, we know that:
      $\ottnt{T_{{\mathrm{2}}}}  \ottsym{=}  \ottnt{B}$ for some base type $\ottnt{B}$;
      $  \mathrm{\Gamma}_{{\mathrm{1}}} , \mathit{x} \mathord{:} \ottnt{T_{{\mathrm{1}}}}   \ottsym{,}  \mathrm{\Gamma}_{{\mathrm{2}}}  \ottsym{,}  \mathrm{\Gamma}'   \vdash   \ottnt{e}  :   \ottnt{B} \mathord{ \rightarrow }  \mathsf{Bool}   $;
      and $\mathrm{\Gamma}'  \vdash  \sigma$.

      There are two cases. In both cases, we use \T{Pred}.
      \begin{itemize}
      \item[($ \mathit{x}  \in   \operatorname{fv} ( \ottnt{e} )  $)] The substitution is actually stored in
        $\sigma$. We find $  \mathrm{\Gamma}_{{\mathrm{1}}}  \ottsym{,}  \mathrm{\Gamma}_{{\mathrm{2}}}  \ottsym{,}  \mathrm{\Gamma}' , \mathit{x} \mathord{:} \ottnt{T_{{\mathrm{1}}}}    \vdash   \ottnt{e}  :  \ottnt{B} $ by the IH
        on $\ottnt{e}$ and $ \mathrm{\Gamma}' , \mathit{x} \mathord{:} \ottnt{T_{{\mathrm{1}}}}   \vdash   \sigma  [  \mathit{x}  \mapsto  \ottnt{v}  ] $ by \T{Map} and
        our assumption.
      \item[($ \mathit{x}  \not\in   \operatorname{fv} ( \ottnt{e} )  $)] The substitution is ignored, and
        $ \ottnt{e}  [  \ottnt{v} / \mathit{x}  ]   \ottsym{=}  \ottnt{e}$. We find $ \mathrm{\Gamma}_{{\mathrm{1}}}  \ottsym{,}  \mathrm{\Gamma}_{{\mathrm{2}}}  \ottsym{,}  \mathrm{\Gamma}'   \vdash   \ottnt{e}  :  \ottnt{B} $ by the IH
        and $\mathrm{\Gamma}'  \vdash  \sigma$ by assumption.
      \end{itemize}
      \iffull
    \item[($\ottnt{C}  \ottsym{=}   \mathit{x} \mathord{:} \ottnt{C_{{\mathrm{1}}}} \mapsto \ottnt{C_{{\mathrm{2}}}} $)] By the IHs.
    \end{itemize}
    \fi
  \end{proof}
\end{lemma}

As a corollary, closing substitutions $\sigma$ close up exactly
their context $\mathrm{\Gamma}'$.

\begin{lemma}[Closing substitutions]
  \label{lem:cpcfcclosure}
  If $\mathrm{\Gamma}'  \vdash  \sigma$ and $ \mathrm{\Gamma}  \ottsym{,}  \mathrm{\Gamma}'   \vdash   \ottnt{e}  :  \ottnt{T} $ then $ \mathrm{\Gamma}   \vdash   \sigma  \ottsym{(}  \ottnt{e}  \ottsym{)}  :  \ottnt{T} $.
  \begin{proof}
    By induction on $\mathrm{\Gamma}'$, using substitution (Lemma~\ref{lem:cpcfcsubstitution}).
  \end{proof}
\end{lemma}

\fi}

\begin{lemma}[Progress]
  \label{lem:cpcfcprogress}
  If $ \emptyset   \vdash   \ottnt{e}  :  \ottnt{T} $ then either (a) $\ottnt{e}$ is some value,
  $\ottnt{v}$; (b) $\ottnt{e}$ is some error, $ \mathsf{err}^ \ottnt{l} $; or, (c) $\ottnt{e} \,  \longrightarrow _{  \mathsf{C}  }  \, \ottnt{e'}$.
  \begin{proof}
    By induction on the typing derivation.
    {\iffull
    \begin{itemize}
    \item[(\T{Var})] Contradictory---variables aren't well typed in
      the empty context.
    \item[(\T{Const})] $\ottnt{k}$ is a value.
    \item[(\T{Blame})] $ \mathsf{err}^ \ottnt{l} $ is an error.
    \item[(\T{Abs})] Function abstractions are values.
    \item[(\T{Rec})] Takes a step by \E{Fix}.
    \item[(\T{Op})] By the IH, $\ottnt{e_{{\mathrm{1}}}}$ is a value, error, or
      steps. In the latter two cases, we step by \E{OpLRaise} or
      \E{OpL}, respectively.

      By the IH, $\ottnt{e_{{\mathrm{2}}}}$ is a value, error, or steps. In the latter
      two cases we step by \E{OpRRaise} or \E{OpR}.

      If both are values, then we step by \E{Delta}.
    \item[(\T{App})] By the IH, $\ottnt{e_{{\mathrm{1}}}}$ is a value, error, or
      steps. In the latter two cases, we step by \E{AppLRaise} or
      \E{AppL}, respectively.

      By the IH, $\ottnt{e_{{\mathrm{2}}}}$ is a value, error, or steps. In the latter
      two cases we step by \E{AppRRaise} or \E{AppR}.

      If both are values, we know that $\ottnt{e_{{\mathrm{1}}}}$ is either a function
      or a function proxy; we step by either \E{Beta} or \E{MonApp},
      depending on the shape of $\ottnt{e_{{\mathrm{1}}}}$.
    \item[(\T{If})] By the IH, $\ottnt{e_{{\mathrm{1}}}}$ is a value, error, or
      steps. In the latter two cases, we step by \E{IfRaise} or
      \E{If}, respectively./

      If $\ottnt{e_{{\mathrm{1}}}}$ is a value, it must be either $ \mathsf{true} $ or
      $ \mathsf{false} $; we step by \E{IfTrue} or \E{IfFalse}, depending on
      the shape of $\ottnt{e_{{\mathrm{1}}}}$.
    \item[(\T{Mon})] By the IH, $\ottnt{e}$ is either a value, error, or
      steps. In the latter two cases, we step by \E{MonRaise} or
      \E{Mon}, respectively.

      If $\ottnt{e}$ is a value, we step by \E{MonPred} or \E{MonApp},
      depending on the shape of the contract $\ottnt{C}$.
    \end{itemize}
    \fi}
  \end{proof}
\end{lemma}

\begin{lemma}[Preservation]
  \label{lem:cpcfcpreservation}
  If $ \emptyset   \vdash   \ottnt{e}  :  \ottnt{T} $ and $\ottnt{e} \,  \longrightarrow _{  \mathsf{C}  }  \, \ottnt{e'}$ then $ \emptyset   \vdash   \ottnt{e'}  :  \ottnt{T} $.
  \begin{proof}
    By induction on the typing derivation, with cases on the step
    taken.
    {\iffull
    \begin{itemize}
    \item[(\T{Var})] Contradictory---variables aren't well typed in
      the empty context.
    \item[(\T{Const})] Contradictory---constants are normal forms.
    \item[(\T{Blame})] Contradictory---errors are normal forms.
    \item[(\T{Abs})] Contradictory---functions are normal forms.
    \item[(\T{Rec})] We must have stepped by \E{Fix}; by substitution
      (Lemma~\ref{lem:cpcfcsubstitution}) and \T{Fix}.
    \item[(\T{Op})] If we stepped by an \E{OpRaise*} rule, we are done
      by \T{Blame}. If we stepped by a congruence rule, \E{Op*}, we
      are done by the IH and \T{Op}.
      If we stepped by \E{Delta}, by the assumption that built-in
      operations have sound denotations.
    \item[(\T{App})] If we stepped by an \E{AppRaise*} rule, we are
      done by \T{Blame}.If we stepped by a congruence rule, \E{App*},
      we are done by the IH and \T{App}.
      If we stepped by \E{Beta}, by
      substitution. (Lemma~\ref{lem:cpcfcsubstitution})
      If we stepped by \E{MonApp}, then: the domain is well typed by
      \T{Mon}; we can apply the proxied function by \T{App}; the
      codomain check is well formed by substitution
      (Lemma~\ref{lem:cpcfcsubstitution}); and the entire resulting
      term is well typed by \T{Mon} again.
    \item[(\T{If})] If we stepped by \E{IfRaise}, by \T{Blame}. If we
      stepped by \E{If}, by the IH and \T{If}. If we stepped by
      \E{IfTrue} or \E{IfFalse}, by assumption.
    \item[(\T{Mon})] If we stepped by \E{MonRaise}, the resulting
      error is well typed by \T{Blame}. If we stepped by \E{Mon}, by
      the IH and \T{Mon}. If we stepped by \E{MonPred}, then: we have
      $ \emptyset   \vdash   \sigma  \ottsym{(}  \ottnt{e}  \ottsym{)}  :   \ottnt{B} \mathord{ \rightarrow }  \mathsf{Bool}   $ by closing substitution
      (Lemma~\ref{lem:cpcfcclosure}); we can type the condition by
      \T{App}; the then branch is well typed by assumption; the false
      branch is well typed by \T{Blame}; and the whole lot is well
      typed by \T{If}.
    \end{itemize}
    \fi}
  \end{proof}
\end{lemma}

\fi}

\subsection{Space-efficient Contract PCF (\CPCFe)}
\label{sec:cpcfe}

\begin{figure}[t]
  {\[ \begin{array}{rcl}
       \mathsf{label} ^{ \ottnt{l} }(  \mathsf{pred}_{ \sigma }( \ottnt{e_{{\mathrm{1}}}} )  )  &=&  \mathsf{pred}^{ \ottnt{l} }_{ \sigma }( \ottnt{e_{{\mathrm{1}}}} )  \\
       \mathsf{label} ^{ \ottnt{l} }(  \mathit{x} \mathord{:} \ottnt{C_{{\mathrm{1}}}} \mapsto \ottnt{C_{{\mathrm{2}}}}  )  &=&  \mathit{x} \mathord{:}  \mathsf{label} ^{ \ottnt{l} }( \ottnt{C_{{\mathrm{1}}}} )  \mapsto  \mathsf{label} ^{ \ottnt{l} }( \ottnt{C_{{\mathrm{2}}}} )   \\
      && \\
       \mathsf{join} ( \mathsf{nil} , \ottnt{r_{{\mathrm{2}}}} )  &=& \ottnt{r_{{\mathrm{2}}}} \\
       \mathsf{join} (   \mathsf{pred}^{ \ottnt{l} }_{ \sigma }( \ottnt{e} )  ; \ottnt{r_{{\mathrm{1}}}}  , \ottnt{r_{{\mathrm{2}}}} )  &=&   \mathsf{pred}^{ \ottnt{l} }_{ \sigma }( \ottnt{e} )  ;  \mathsf{drop} (  \mathsf{join} ( \ottnt{r_{{\mathrm{1}}}} , \ottnt{r_{{\mathrm{2}}}} )  ,  \mathsf{pred}_{ \sigma }( \ottnt{e} )  )   \\
       \mathsf{join} (  \mathit{x} \mathord{:} \ottnt{c_{{\mathrm{11}}}} \mapsto \ottnt{c_{{\mathrm{12}}}}  ,  \mathit{x} \mathord{:} \ottnt{c_{{\mathrm{21}}}} \mapsto \ottnt{c_{{\mathrm{22}}}}  )  &=&  \mathit{x} \mathord{:}  \mathsf{join} ( \ottnt{c_{{\mathrm{21}}}} , \ottnt{c_{{\mathrm{11}}}} )  \mapsto  \mathsf{join} (  \mathsf{wrap} ( \ottnt{c_{{\mathrm{12}}}} , \mathit{x} , \ottnt{c_{{\mathrm{22}}}} )  , \ottnt{c_{{\mathrm{22}}}} )   \\
      && \\
       \mathsf{drop} ( \mathsf{nil} ,  \mathsf{pred}_{ \sigma }( \ottnt{e} )  )  &=&  \mathsf{nil}  \\
       \mathsf{drop} (   \mathsf{pred}^{ \ottnt{l} }_{ \sigma_{{\mathrm{2}}} }( \ottnt{e_{{\mathrm{2}}}} )  ; \ottnt{r}  ,  \mathsf{pred}_{ \sigma_{{\mathrm{1}}} }( \ottnt{e_{{\mathrm{1}}}} )  )  &=& \\[.5em]
      \multicolumn{3}{r}{\begin{cases}
           \mathsf{drop} ( \ottnt{r} ,  \mathsf{pred}_{ \sigma_{{\mathrm{2}}} }( \ottnt{e_{{\mathrm{2}}}} )  )  &  \mathsf{pred}_{ \sigma_{{\mathrm{1}}} }( \ottnt{e_{{\mathrm{1}}}} )  \, \supset \,  \mathsf{pred}_{ \sigma_{{\mathrm{2}}} }( \ottnt{e_{{\mathrm{2}}}} )  \\
            \mathsf{pred}^{ \ottnt{l} }_{ \sigma_{{\mathrm{2}}} }( \ottnt{e_{{\mathrm{2}}}} )  ;  \mathsf{drop} ( \ottnt{r} ,  \mathsf{pred}_{ \sigma_{{\mathrm{1}}} }( \ottnt{e_{{\mathrm{1}}}} )  )   &  \mathsf{pred}_{ \sigma_{{\mathrm{1}}} }( \ottnt{e_{{\mathrm{1}}}} )  \, \not \supset \,  \mathsf{pred}_{ \sigma_{{\mathrm{2}}} }( \ottnt{e_{{\mathrm{2}}}} ) 
      \end{cases}} \\
      && \\
       \mathsf{wrap} (  \mathsf{pred}^{ \ottnt{l} }_{ \sigma }( \ottnt{e} )  , \mathit{x} , \ottnt{c} )  &=&
      \begin{cases}
         \mathsf{pred}^{ \ottnt{l} }_{ \sigma }( \ottnt{e} )  &  \mathit{x}  \not\in   \operatorname{fv} ( \sigma  \ottsym{(}  \ottnt{e}  \ottsym{)} )   \\
         \mathsf{pred}^{ \ottnt{l} }_{  \sigma  [  \mathit{x}  \mapsto   \mathsf{mon}(  \mathsf{join} ( \ottnt{c'} , \ottnt{c} )  ,  \ottnt{e} )   ]  }( \ottnt{e} )  & \sigma  \ottsym{(}  \mathit{x}  \ottsym{)}  \ottsym{=}   \mathsf{mon}( \ottnt{c'} ,  \ottnt{e} )  \\
         \mathsf{pred}^{ \ottnt{l} }_{  \sigma  [  \mathit{x}  \mapsto   \mathsf{mon}( \ottnt{c} ,  \sigma  \ottsym{(}  \mathit{x}  \ottsym{)} )   ]  }( \ottnt{e} )  & \text{otherwise}
      \end{cases} \\
       \mathsf{wrap} ( \mathsf{nil} , \mathit{x} , \ottnt{c} )  &=&  \mathsf{nil}  \\
       \mathsf{wrap} (   \mathsf{pred}^{ \ottnt{l} }_{ \sigma }( \ottnt{e} )  ; \ottnt{r}  , \mathit{x} , \ottnt{c} )  &=&   \mathsf{wrap} (  \mathsf{pred}^{ \ottnt{l} }_{ \sigma }( \ottnt{e} )  , \mathit{x} , \ottnt{c} )  ;  \mathsf{wrap} ( \ottnt{r} , \mathit{x} , \ottnt{c} )   \\
       \mathsf{wrap} (  \mathit{y} \mathord{:} \ottnt{c_{{\mathrm{1}}}} \mapsto \ottnt{c_{{\mathrm{2}}}}  , \mathit{x} , \ottnt{c} )  &=&  \mathit{y} \mathord{:}  \mathsf{wrap} ( \ottnt{c_{{\mathrm{1}}}} , \mathit{x} , \ottnt{c} )  \mapsto  \mathsf{wrap} ( \ottnt{c_{{\mathrm{2}}}} , \mathit{x} , \ottnt{c} )  
  \end{array} \]}

  \caption{Contract labeling and predicate stack management}
  \label{fig:stackops}
\end{figure}

How can we recover tail calls in CPCF? \CPCFc will happily wrap
arbitrarily many function proxies around a value, and there's no bound
on the number of codomain contract checks that can accumulate.
The key idea is \textit{joining} contracts. We'll make two changes to
the language: we'll bound function proxies so each function has at most one, and we'll bound stacks to avoid redundant checking.
We ultimately show that contracts without dependency use constant
space, but that the story for dependent functions is more complex
(Sec.~\ref{sec:bounds}).

Fortuitously, our notion of join solves both of our problems, working identically for both simple and dependent contracts.
To ensure a function value can have only one proxy, we change the
semantics of monitoring: when monitoring a proxied value, we join the new monitor and the old
one.
To bound the size of stacks contract checks, we join pending contracts to avoid
redundancy.

The $ \mathsf{join} $ operation works on \textit{labeled contracts}, which (a)
move the label from the monitor into the contract and (b) allow us to
keep track of many predicates at once (Fig.~\ref{fig:stackops}).
Concretely, labeled contracts use the metavariable $\ottnt{c}$ (as opposed
to $\ottnt{C}$), comprising function contracts as usual ($ \mathit{x} \mathord{:} \ottnt{c_{{\mathrm{1}}}} \mapsto \ottnt{c_{{\mathrm{2}}}} $)
and \textit{predicate stacks}, $\ottnt{r}$ (Fig.~\ref{fig:syntax}). A predicate
  stack $\ottnt{r}$ is a list of \textit{labeled predicates}
$ \mathsf{pred}^{ \ottnt{l} }( \ottnt{e} ) $, where $ \mathsf{nil} $ is the empty stack.

The $ \mathsf{join} $ operation takes two labeled contracts and combines
them, eliminating redundant contracts as it goes. To join a new and an
old predicate stack, we keep new contracts and eliminate redundant old
ones; only more ``recent'' contracts are kept. Joining functions works
contravariantly, being careful to maintain correct substitution
behavior using $ \mathsf{wrap} $.

Finally, we establish what we mean by ``redundant'' using
\textit{predicate implication}: when does one contract imply another?
\begin{definition}[Predicate implication]
  \label{def:implaxioms}
  Let $ \mathsf{pred}_{ \sigma_{{\mathrm{1}}} }( \ottnt{e_{{\mathrm{1}}}} )  \, \supset \,  \mathsf{pred}_{ \sigma_{{\mathrm{2}}} }( \ottnt{e_{{\mathrm{2}}}} ) $ be a relation on
  predicates such that:
  \begin{enumerate}
  \item[(\textbf{Reflexivity})] \label{implrefl} If $\emptyset  \vdash   \mathsf{pred}_{ \sigma }( \ottnt{e} )   \ottsym{:}  \ottnt{B}$ then $ \mathsf{pred}_{ \sigma }( \ottnt{e} )  \, \supset \,  \mathsf{pred}_{ \sigma }( \ottnt{e} ) $.
  \item[(\textbf{Transitivity})] \label{impltrans} If
    $ \mathsf{pred}_{ \sigma_{{\mathrm{1}}} }( \ottnt{e_{{\mathrm{1}}}} )  \, \supset \,  \mathsf{pred}_{ \sigma_{{\mathrm{2}}} }( \ottnt{e_{{\mathrm{2}}}} ) $ and $ \mathsf{pred}_{ \sigma_{{\mathrm{2}}} }( \ottnt{e_{{\mathrm{2}}}} )  \, \supset \,  \mathsf{pred}_{ \sigma_{{\mathrm{3}}} }( \ottnt{e_{{\mathrm{3}}}} ) $, then $ \mathsf{pred}_{ \sigma_{{\mathrm{1}}} }( \ottnt{e_{{\mathrm{1}}}} )  \, \supset \,  \mathsf{pred}_{ \sigma_{{\mathrm{3}}} }( \ottnt{e_{{\mathrm{3}}}} ) $.
  \item[(\textbf{Substitutivity})] \label{implsubst} When
    $ \mathrm{\Gamma}_{\ottmv{i}\,{\mathrm{1}}} , \mathit{x} \mathord{:} \ottnt{T'}   \ottsym{,}  \mathrm{\Gamma}_{\ottmv{i}\,{\mathrm{2}}}  \vdash   \mathsf{pred}_{ \sigma_{\ottmv{i}} }( \ottnt{e_{\ottmv{i}}} )   \ottsym{:}  \ottnt{T}$ and $ \emptyset   \vdash   \ottnt{v}  :  \ottnt{T'} $, if $ \mathsf{pred}_{ \sigma_{{\mathrm{1}}} }( \ottnt{e_{{\mathrm{1}}}} )  \, \supset \,  \mathsf{pred}_{ \sigma_{{\mathrm{2}}} }( \ottnt{e_{{\mathrm{2}}}} ) $ then
    $  \mathsf{pred}_{ \sigma_{{\mathrm{1}}} }( \ottnt{e_{{\mathrm{1}}}} )   [  \ottnt{v}  /  \mathit{x}  ]  \, \supset \,   \mathsf{pred}_{ \sigma_{{\mathrm{2}}} }( \ottnt{e_{{\mathrm{2}}}} )   [  \ottnt{v}  /  \mathit{x}  ] $.
  \item[(\textbf{Adequacy})] \label{impladeq} If $\emptyset  \vdash   \mathsf{pred}_{ \sigma_{\ottmv{i}} }( \ottnt{e_{\ottmv{i}}} )   \ottsym{:}  \ottnt{T}$ and $ \mathsf{pred}_{ \sigma_{{\mathrm{1}}} }( \ottnt{e_{{\mathrm{1}}}} )  \, \supset \,  \mathsf{pred}_{ \sigma_{{\mathrm{2}}} }( \ottnt{e_{{\mathrm{2}}}} ) $ then $ \forall   \ottnt{k}  \in \mathcal{K}_{ \ottnt{B} }   . ~   \sigma_{{\mathrm{1}}}  \ottsym{(}  \ottnt{e_{{\mathrm{1}}}}  \ottsym{)}  ~  \ottnt{k}  \,  \longrightarrow _{ \mathsf{m} }  \,  \mathsf{true}  $ implies $ \sigma_{{\mathrm{2}}}  \ottsym{(}  \ottnt{e_{{\mathrm{2}}}}  \ottsym{)}  ~  \ottnt{k}  \,  \longrightarrow _{ \mathsf{m} }  \,  \mathsf{true} $.
  \item[(\textbf{Decidability})] \label{impldec} For all $\emptyset  \vdash   \mathsf{pred}_{ \sigma_{{\mathrm{1}}} }( \ottnt{e_{{\mathrm{1}}}} )   \ottsym{:}  \ottnt{B}$ and $\emptyset  \vdash   \mathsf{pred}_{ \sigma_{{\mathrm{2}}} }( \ottnt{e_{{\mathrm{2}}}} )   \ottsym{:}  \ottnt{B}$, it is decidable whether $ \mathsf{pred}_{ \sigma_{{\mathrm{1}}} }( \ottnt{e_{{\mathrm{1}}}} )  \, \supset \,  \mathsf{pred}_{ \sigma_{{\mathrm{2}}} }( \ottnt{e_{{\mathrm{2}}}} ) $ or $ \mathsf{pred}_{ \sigma_{{\mathrm{1}}} }( \ottnt{e_{{\mathrm{1}}}} )  \, \not \supset \,  \mathsf{pred}_{ \sigma_{{\mathrm{2}}} }( \ottnt{e_{{\mathrm{2}}}} ) $.
  \end{enumerate}
\end{definition}
The entire development of space-efficiency is parameterized over this
implication relation, $ \supset $, characterizing when one
first-order contract subsumes another.
We write $ \not \supset $ for the negation of $ \supset $.
The $ \supset $ relation is a \textit{total pre-order} (a/k/a a
\textit{preference relation})---it would be a total order, but it may
not necessarily enjoy anti-symmetry. For example, we could have
$ \mathsf{pred}_{  \iota  }(  \lambda \mathit{x} \mathord{:}  \mathsf{Int}  .~  \mathit{x}  \, \ge \, \ottsym{0} )  \, \supset \,  \mathsf{pred}_{  \iota  }(  \lambda \mathit{x} \mathord{:}  \mathsf{Int}  .~  \mathit{x}  \,  +  \, \ottsym{1} \, \ottsym{>} \, \ottsym{0} ) $ and vice versa,
even though the two predicates aren't equal. You can also
view $ \supset $ as a total order \textit{up-to contextual
  equivalence}.

There is at least one workable implication relation: syntactic
equality. We say $ \mathsf{pred}_{ \sigma_{{\mathrm{1}}} }( \ottnt{e_{{\mathrm{1}}}} )  \, \supset \,  \mathsf{pred}_{ \sigma_{{\mathrm{2}}} }( \ottnt{e_{{\mathrm{2}}}} ) $ iff $\ottnt{e_{{\mathrm{1}}}}  \ottsym{=}  \ottnt{e_{{\mathrm{2}}}}$ and $\sigma_{{\mathrm{1}}} = \sigma_{{\mathrm{2}}}$. Since we've been careful to
store only those values that are actually referenced in the closure of
the predicate, the steps to determine these equalities are finite and
computable at runtime.
For example, suppose we wish to show that $ \mathsf{pred}_{  [  \mathit{y}  \mapsto  \ottsym{47}  ]  }(  \lambda \mathit{x} \mathord{:}  \mathsf{Int}  .~  \mathit{x}  \, \ottsym{>} \, \mathit{y} )  \, \supset \,  \mathsf{pred}_{  [  \mathit{y}  \mapsto  \ottsym{47}  ]  }(  \lambda \mathit{x} \mathord{:}  \mathsf{Int}  .~  \mathit{x}  \, \ottsym{>} \, \mathit{y} ) $. The code part---the
predicate $ \lambda \mathit{x} \mathord{:}  \mathsf{Int}  .~  \mathit{x}  \, \ottsym{>} \, \mathit{y}$---is the same; an implementation might
observe that the function pointers are equal. The environment has only
one slot, for $\mathit{y}$, with the value $\ottsym{47}$ in it; an
implementation might compare the two environments slot-by-slot.
{\iffull
We \textit{could} have given an operational semantics which behaves
more like an implementation, explicitly generating conditionals and
merge operations in the terms themselves, but we believe our slightly
more abstract presentation is more digestible.
\fi}

\subsubsection{Substitution in the codomain: lax, picky, and indy}

We extend Greenberg's notion of join to account for dependency with a
new function, $ \mathsf{wrap} $.
Greenberg, Pierce, and Weirich identified two variants of latent
contracts in the literature, differing in their treatment of the
dependent substitution of arguments in the codomain: \textit{picky},
where we monitor the value substituted in the codomain with the domain
contract; and \textit{lax}, where the actual parameter value
substituted into the codomain is
unmonitored~\cite{Greenberg10contracts}. There is a third variant,
\textit{indy}, which applies a monitor to the argument value but uses
a different blame label~\cite{Dimoulas11indy}.
These different models of substitution all exhibit different behavior
for \textit{abusive} contracts, where the codomain contract violates
the domain contract.

There is another source of substitutions in the codomain: multiple function proxies.
How do the monitors unfold when we have two function proxies? In the
classic lax semantics, we find (leaving the domain check unevaluated):
\[ \begin{array}{rl}
           &   \mathsf{mon}(  \mathit{x} \mathord{:} \ottnt{c_{{\mathrm{11}}}} \mapsto \ottnt{c_{{\mathrm{12}}}}  ,   \mathsf{mon}(  \mathit{x} \mathord{:} \ottnt{c_{{\mathrm{21}}}} \mapsto \ottnt{c_{{\mathrm{22}}}}  ,  \mathit{f} )  )   ~  \ottnt{v}  \\
   \longrightarrow _{  \mathsf{C}  }  &  \mathsf{mon}(  \ottnt{c_{{\mathrm{12}}}}  [  \ottnt{v}  /  \mathit{x}  ]  ,    \mathsf{mon}(  \mathit{x} \mathord{:} \ottnt{c_{{\mathrm{21}}}} \mapsto \ottnt{c_{{\mathrm{22}}}}  ,  \mathit{f} )   ~   \mathsf{mon}( \ottnt{c_{{\mathrm{11}}}} ,  \ottnt{v} )   )  \\
   \longrightarrow _{  \mathsf{C}  }  &  \mathsf{mon}(  \ottnt{c_{{\mathrm{12}}}}  [  \ottnt{v}  /  \mathit{x}  ]  ,   \mathsf{mon}(  \ottnt{c_{{\mathrm{22}}}}  [   \mathsf{mon}( \ottnt{c_{{\mathrm{11}}}} ,  \ottnt{v} )   /  \mathit{x}  ]  ,   \mathit{f}  ~   \mathsf{mon}( \ottnt{c_{{\mathrm{21}}}} ,   \mathsf{mon}( \ottnt{c_{{\mathrm{11}}}} ,  \ottnt{v} )  )   )  ) 
\end{array} \]
Even though we're using the lax semantics, we substitute contracts
into the codomain.
\iffull
\begin{figure}[p]
\else
\begin{figure}[t]
\fi
\[ \begin{array}{rcl}
  \ottnt{C_{{\mathrm{1}}}} &=&  \mathit{f} \mathord{:}  (   \mathsf{pred}(  \lambda \mathit{x} \mathord{:}  \mathsf{Int}  .~  \mathit{x}  \, \ottsym{>} \, \ottsym{0} )  \mapsto  \mathsf{pred}(  \lambda \mathit{x} \mathord{:}  \mathsf{Int}  .~  \mathit{x}  \, \ottsym{>} \, \ottsym{0} )   )  \mapsto  \mathsf{pred}(  \lambda \mathit{x} \mathord{:}  \mathsf{Int}  .~  \mathit{x}  \, \ottsym{>} \, \ottsym{0} )   \\
  \ottnt{C_{{\mathrm{2}}}} &=&  \mathit{f} \mathord{:}  (   \mathsf{pred}(  \lambda \mathit{x} \mathord{:}  \mathsf{Int}  .~   \mathsf{true}   )  \mapsto  \mathsf{pred}(  \lambda \mathit{x} \mathord{:}  \mathsf{Int}  .~   \mathsf{true}   )   )  \mapsto  \mathsf{pred}(   \lambda \mathit{x} \mathord{:}  \mathsf{Int}  .~  \mathit{f}   ~  \ottsym{0}  \,  =  \, \ottsym{0} )  
\end{array}\]
\iffull
Referring to the domains as $\ottnt{C_{\ottmv{i}\,{\mathrm{1}}}}$ and codomains as $\ottnt{C_{\ottmv{i}\,{\mathrm{2}}}}$
\fi
\[ \begin{array}{rl}
           &   \mathsf{mon}^{ \ottnt{l_{{\mathrm{1}}}} }( \ottnt{C_{{\mathrm{1}}}} ,   \mathsf{mon}^{ \ottnt{l_{{\mathrm{2}}}} }( \ottnt{C_{{\mathrm{2}}}} ,   \lambda \mathit{f} \mathord{:}  (   \mathsf{Int}  \mathord{ \rightarrow }  \mathsf{Int}   )  .~   \mathit{f}  ~  \ottsym{5}   )  )   ~   (  \lambda \mathit{x} \mathord{:}  \mathsf{Int}  .~  \mathit{x}  )   \\
   \longrightarrow _{  \mathsf{C}  }  &  \mathsf{mon}^{ \ottnt{l_{{\mathrm{1}}}} }(  \ottnt{C_{{\mathrm{12}}}}  [   (  \lambda \mathit{x} \mathord{:}  \mathsf{Int}  .~  \mathit{x}  )   /  \mathit{f}  ]  ,  {} \\   &  ~~   \mathsf{mon}^{ \ottnt{l_{{\mathrm{2}}}} }( \ottnt{C_{{\mathrm{2}}}} ,   \lambda \mathit{f} \mathord{:}  (   \mathsf{Int}  \mathord{ \rightarrow }  \mathsf{Int}   )  .~   \mathit{f}  ~  \ottsym{5}   )   ~   \mathsf{mon}^{ \ottnt{l_{{\mathrm{1}}}} }( \ottnt{C_{{\mathrm{11}}}} ,   (  \lambda \mathit{x} \mathord{:}  \mathsf{Int}  .~  \mathit{x}  )  )   )  \\
\iffull
   \longrightarrow _{  \mathsf{C}  }  &  \mathsf{mon}^{ \ottnt{l_{{\mathrm{1}}}} }(  \ottnt{C_{{\mathrm{12}}}}  [   (  \lambda \mathit{x} \mathord{:}  \mathsf{Int}  .~  \mathit{x}  )   /  \mathit{f}  ]  ,   \mathsf{mon}^{ \ottnt{l_{{\mathrm{2}}}} }(  \ottnt{C_{{\mathrm{22}}}}  [   \mathsf{mon}^{ \ottnt{l_{{\mathrm{1}}}} }( \ottnt{C_{{\mathrm{11}}}} ,   \lambda \mathit{x} \mathord{:}  \mathsf{Int}  .~  \mathit{x}  )   /  \mathit{f}  ]  ,  {} \\   &  ~~   (  \lambda \mathit{f} \mathord{:}  (   \mathsf{Int}  \mathord{ \rightarrow }  \mathsf{Int}   )  .~   \mathit{f}  ~  \ottsym{5}   )   ~   \mathsf{mon}^{ \ottnt{l_{{\mathrm{2}}}} }( \ottnt{C_{{\mathrm{21}}}} ,   \mathsf{mon}^{ \ottnt{l_{{\mathrm{1}}}} }( \ottnt{C_{{\mathrm{11}}}} ,   (  \lambda \mathit{x} \mathord{:}  \mathsf{Int}  .~  \mathit{x}  )  )  )   )  )  \\
   \longrightarrow _{  \mathsf{C}  }  &  \mathsf{mon}^{ \ottnt{l_{{\mathrm{1}}}} }(  \ottnt{C_{{\mathrm{12}}}}  [   (  \lambda \mathit{x} \mathord{:}  \mathsf{Int}  .~  \mathit{x}  )   /  \mathit{f}  ]  ,   \mathsf{mon}^{ \ottnt{l_{{\mathrm{2}}}} }(  \ottnt{C_{{\mathrm{22}}}}  [   \mathsf{mon}^{ \ottnt{l_{{\mathrm{1}}}} }( \ottnt{C_{{\mathrm{11}}}} ,   \lambda \mathit{x} \mathord{:}  \mathsf{Int}  .~  \mathit{x}  )   /  \mathit{f}  ]  ,  {} \\   &  ~~   \mathsf{mon}^{ \ottnt{l_{{\mathrm{2}}}} }( \ottnt{C_{{\mathrm{21}}}} ,   \mathsf{mon}^{ \ottnt{l_{{\mathrm{1}}}} }( \ottnt{C_{{\mathrm{11}}}} ,   (  \lambda \mathit{x} \mathord{:}  \mathsf{Int}  .~  \mathit{x}  )  )  )   ~  \ottsym{5}  )  )  \\
   \longrightarrow _{  \mathsf{C}  }  &  \mathsf{mon}^{ \ottnt{l_{{\mathrm{1}}}} }(  \ottnt{C_{{\mathrm{12}}}}  [   (  \lambda \mathit{x} \mathord{:}  \mathsf{Int}  .~  \mathit{x}  )   /  \mathit{f}  ]  ,   \mathsf{mon}^{ \ottnt{l_{{\mathrm{2}}}} }(  \ottnt{C_{{\mathrm{22}}}}  [   \mathsf{mon}^{ \ottnt{l_{{\mathrm{1}}}} }( \ottnt{C_{{\mathrm{11}}}} ,   \lambda \mathit{x} \mathord{:}  \mathsf{Int}  .~  \mathit{x}  )   /  \mathit{f}  ]  ,  {} \\  &  ~~   \mathsf{mon}^{ \ottnt{l_{{\mathrm{2}}}} }(  \mathsf{pred}(  \lambda \mathit{x} \mathord{:}  \mathsf{Int}  .~   \mathsf{true}   )  ,  {} \\   &  \quad   \mathsf{mon}^{ \ottnt{l_{{\mathrm{1}}}} }( \ottnt{C_{{\mathrm{11}}}} ,   (  \lambda \mathit{x} \mathord{:}  \mathsf{Int}  .~  \mathit{x}  )  )   ~   \mathsf{mon}^{ \ottnt{l_{{\mathrm{2}}}} }(  \mathsf{pred}(  \lambda \mathit{x} \mathord{:}  \mathsf{Int}  .~   \mathsf{true}   )  ,  \ottsym{5} )   )  )  )  \\
  \longrightarrow ^{*}_{  \mathsf{C}  }  &  \mathsf{mon}^{ \ottnt{l_{{\mathrm{1}}}} }(  \ottnt{C_{{\mathrm{12}}}}  [   (  \lambda \mathit{x} \mathord{:}  \mathsf{Int}  .~  \mathit{x}  )   /  \mathit{f}  ]  ,   \mathsf{mon}^{ \ottnt{l_{{\mathrm{2}}}} }(  \ottnt{C_{{\mathrm{22}}}}  [   \mathsf{mon}^{ \ottnt{l_{{\mathrm{1}}}} }( \ottnt{C_{{\mathrm{11}}}} ,   \lambda \mathit{x} \mathord{:}  \mathsf{Int}  .~  \mathit{x}  )   /  \mathit{f}  ]  ,  {} \\  &  ~~   \mathsf{mon}^{ \ottnt{l_{{\mathrm{2}}}} }(  \mathsf{pred}(  \lambda \mathit{x} \mathord{:}  \mathsf{Int}  .~   \mathsf{true}   )  ,    \mathsf{mon}^{ \ottnt{l_{{\mathrm{1}}}} }( \ottnt{C_{{\mathrm{11}}}} ,   (  \lambda \mathit{x} \mathord{:}  \mathsf{Int}  .~  \mathit{x}  )  )   ~  \ottsym{5}  )  )  )  \\
   \longrightarrow _{  \mathsf{C}  }  &  \mathsf{mon}^{ \ottnt{l_{{\mathrm{1}}}} }(  \ottnt{C_{{\mathrm{12}}}}  [   (  \lambda \mathit{x} \mathord{:}  \mathsf{Int}  .~  \mathit{x}  )   /  \mathit{f}  ]  ,   \mathsf{mon}^{ \ottnt{l_{{\mathrm{2}}}} }(  \ottnt{C_{{\mathrm{22}}}}  [   \mathsf{mon}^{ \ottnt{l_{{\mathrm{1}}}} }( \ottnt{C_{{\mathrm{11}}}} ,   \lambda \mathit{x} \mathord{:}  \mathsf{Int}  .~  \mathit{x}  )   /  \mathit{f}  ]  ,  {} \\  &  ~~   \mathsf{mon}^{ \ottnt{l_{{\mathrm{2}}}} }(  \mathsf{pred}(  \lambda \mathit{x} \mathord{:}  \mathsf{Int}  .~   \mathsf{true}   )  ,   \mathsf{mon}^{ \ottnt{l_{{\mathrm{1}}}} }(  \mathsf{pred}(  \lambda \mathit{x} \mathord{:}  \mathsf{Int}  .~  \mathit{x}  \, \ottsym{>} \, \ottsym{0} )  ,  {} \\   &  \quad   (  \lambda \mathit{x} \mathord{:}  \mathsf{Int}  .~  \mathit{x}  )   ~   \mathsf{mon}^{ \ottnt{l_{{\mathrm{1}}}} }(  \mathsf{pred}(  \lambda \mathit{x} \mathord{:}  \mathsf{Int}  .~  \mathit{x}  \, \ottsym{>} \, \ottsym{0} )  ,  \ottsym{5} )   )  )  )  )  \\
  \longrightarrow ^{*}_{  \mathsf{C}  }  &  \mathsf{mon}^{ \ottnt{l_{{\mathrm{1}}}} }(  \ottnt{C_{{\mathrm{12}}}}  [   (  \lambda \mathit{x} \mathord{:}  \mathsf{Int}  .~  \mathit{x}  )   /  \mathit{f}  ]  ,   \mathsf{mon}^{ \ottnt{l_{{\mathrm{2}}}} }(  \ottnt{C_{{\mathrm{22}}}}  [   \mathsf{mon}^{ \ottnt{l_{{\mathrm{1}}}} }( \ottnt{C_{{\mathrm{11}}}} ,   \lambda \mathit{x} \mathord{:}  \mathsf{Int}  .~  \mathit{x}  )   /  \mathit{f}  ]  ,  {} \\  &  ~~   \mathsf{mon}^{ \ottnt{l_{{\mathrm{2}}}} }(  \mathsf{pred}(  \lambda \mathit{x} \mathord{:}  \mathsf{Int}  .~   \mathsf{true}   )  ,   \mathsf{mon}^{ \ottnt{l_{{\mathrm{1}}}} }(  \mathsf{pred}(  \lambda \mathit{x} \mathord{:}  \mathsf{Int}  .~  \mathit{x}  \, \ottsym{>} \, \ottsym{0} )  ,    (  \lambda \mathit{x} \mathord{:}  \mathsf{Int}  .~  \mathit{x}  )   ~  \ottsym{5}  )  )  )  )  \\
   \longrightarrow _{  \mathsf{C}  }  &  \mathsf{mon}^{ \ottnt{l_{{\mathrm{1}}}} }(  \ottnt{C_{{\mathrm{12}}}}  [   (  \lambda \mathit{x} \mathord{:}  \mathsf{Int}  .~  \mathit{x}  )   /  \mathit{f}  ]  ,   \mathsf{mon}^{ \ottnt{l_{{\mathrm{2}}}} }(  \ottnt{C_{{\mathrm{22}}}}  [   \mathsf{mon}^{ \ottnt{l_{{\mathrm{1}}}} }( \ottnt{C_{{\mathrm{11}}}} ,   \lambda \mathit{x} \mathord{:}  \mathsf{Int}  .~  \mathit{x}  )   /  \mathit{f}  ]  ,  {} \\  &  ~~   \mathsf{mon}^{ \ottnt{l_{{\mathrm{2}}}} }(  \mathsf{pred}(  \lambda \mathit{x} \mathord{:}  \mathsf{Int}  .~   \mathsf{true}   )  ,   \mathsf{mon}^{ \ottnt{l_{{\mathrm{1}}}} }(  \mathsf{pred}(  \lambda \mathit{x} \mathord{:}  \mathsf{Int}  .~  \mathit{x}  \, \ottsym{>} \, \ottsym{0} )  ,  \ottsym{5} )  )  )  )  \\
  \longrightarrow ^{*}_{  \mathsf{C}  }  &  \mathsf{mon}^{ \ottnt{l_{{\mathrm{1}}}} }(  \ottnt{C_{{\mathrm{12}}}}  [   (  \lambda \mathit{x} \mathord{:}  \mathsf{Int}  .~  \mathit{x}  )   /  \mathit{f}  ]  ,   \mathsf{mon}^{ \ottnt{l_{{\mathrm{2}}}} }(  \ottnt{C_{{\mathrm{22}}}}  [   \mathsf{mon}^{ \ottnt{l_{{\mathrm{1}}}} }( \ottnt{C_{{\mathrm{11}}}} ,   \lambda \mathit{x} \mathord{:}  \mathsf{Int}  .~  \mathit{x}  )   /  \mathit{f}  ]  ,  {} \\  &  ~~   \mathsf{mon}^{ \ottnt{l_{{\mathrm{2}}}} }(  \mathsf{pred}(  \lambda \mathit{x} \mathord{:}  \mathsf{Int}  .~   \mathsf{true}   )  ,  \ottsym{5} )  )  )  \\
\fi
  \longrightarrow ^{*}_{  \mathsf{C}  }  &  \mathsf{mon}^{ \ottnt{l_{{\mathrm{1}}}} }(  \ottnt{C_{{\mathrm{12}}}}  [   (  \lambda \mathit{x} \mathord{:}  \mathsf{Int}  .~  \mathit{x}  )   /  \mathit{f}  ]  ,   \mathsf{mon}^{ \ottnt{l_{{\mathrm{2}}}} }(  \ottnt{C_{{\mathrm{22}}}}  [   \mathsf{mon}^{ \ottnt{l_{{\mathrm{1}}}} }( \ottnt{C_{{\mathrm{11}}}} ,   \lambda \mathit{x} \mathord{:}  \mathsf{Int}  .~  \mathit{x}  )   /  \mathit{f}  ]  ,  \ottsym{5} )  )  \\
   \longrightarrow _{  \mathsf{C}  }  &  \mathsf{mon}^{ \ottnt{l_{{\mathrm{1}}}} }(  \ottnt{C_{{\mathrm{12}}}}  [   (  \lambda \mathit{x} \mathord{:}  \mathsf{Int}  .~  \mathit{x}  )   /  \mathit{f}  ]  ,  {} \\  &  ~~   \mathsf{if} ~   (   (  \lambda \mathit{x} \mathord{:}  \mathsf{Int}  .~    \mathsf{mon}^{ \ottnt{l_{{\mathrm{1}}}} }( \ottnt{C_{{\mathrm{11}}}} ,   \lambda \mathit{x} \mathord{:}  \mathsf{Int}  .~  \mathit{x}  )   ~  \ottsym{0}  \,  =  \, \ottsym{0}  )   ~  \ottsym{5}  )   ~  \ottsym{5}  ~~   \mathsf{err}^ \ottnt{l_{{\mathrm{2}}}}   )  \\
   \longrightarrow _{  \mathsf{C}  }  &  \mathsf{mon}^{ \ottnt{l_{{\mathrm{1}}}} }(  \ottnt{C_{{\mathrm{12}}}}  [   (  \lambda \mathit{x} \mathord{:}  \mathsf{Int}  .~  \mathit{x}  )   /  \mathit{f}  ]  ,   \mathsf{if} ~   (   \mathsf{mon}^{ \ottnt{l_{{\mathrm{1}}}} }( \ottnt{C_{{\mathrm{11}}}} ,   \lambda \mathit{x} \mathord{:}  \mathsf{Int}  .~  \mathit{x}  )   ~  \ottsym{0} \,  =  \, \ottsym{0}  )   ~  \ottsym{5}  ~~   \mathsf{err}^ \ottnt{l_{{\mathrm{2}}}}   )  \\
\iffull
   \longrightarrow _{  \mathsf{C}  }  &  \mathsf{mon}^{ \ottnt{l_{{\mathrm{1}}}} }(  \ottnt{C_{{\mathrm{12}}}}  [   (  \lambda \mathit{x} \mathord{:}  \mathsf{Int}  .~  \mathit{x}  )   /  \mathit{f}  ]  ,  {} \\  &  ~~   \mathsf{if} ~   (  \mathsf{mon}^{ \ottnt{l_{{\mathrm{1}}}} }(  \mathsf{pred}(  \lambda \mathit{x} \mathord{:}  \mathsf{Int}  .~  \mathit{x}  \, \ottsym{>} \, \ottsym{0} )  ,  {} \\   &  \qquad  \qquad  \quad   (  \lambda \mathit{x} \mathord{:}  \mathsf{Int}  .~  \mathit{x}  )   ~   \mathsf{mon}^{ \ottnt{l_{{\mathrm{1}}}} }(  \mathsf{pred}(  \lambda \mathit{x} \mathord{:}  \mathsf{Int}  .~  \mathit{x}  \, \ottsym{>} \, \ottsym{0} )  ,  \ottsym{0} )   )  \,  =  \, \ottsym{0} )   ~  {} \\  &  ~~  \ottsym{5}  ~~   \mathsf{err}^ \ottnt{l_{{\mathrm{2}}}}   )  \\
   \longrightarrow _{  \mathsf{C}  }  &  \mathsf{mon}^{ \ottnt{l_{{\mathrm{1}}}} }(  \ottnt{C_{{\mathrm{12}}}}  [   (  \lambda \mathit{x} \mathord{:}  \mathsf{Int}  .~  \mathit{x}  )   /  \mathit{f}  ]  ,  {} \\  &  ~~   \mathsf{if} ~   (  \mathsf{mon}^{ \ottnt{l_{{\mathrm{1}}}} }(  \mathsf{pred}(  \lambda \mathit{x} \mathord{:}  \mathsf{Int}  .~  \mathit{x}  \, \ottsym{>} \, \ottsym{0} )  ,  {} \\   &  \qquad  \qquad  \quad   (  \lambda \mathit{x} \mathord{:}  \mathsf{Int}  .~  \mathit{x}  )   ~   (  \mathsf{if} ~   (   (  \lambda \mathit{x} \mathord{:}  \mathsf{Int}  .~  \mathit{x} \, \ottsym{>} \, \ottsym{0}  )   ~  \ottsym{0}  )   ~  \ottsym{0}  ~~   \mathsf{err}^ \ottnt{l_{{\mathrm{2}}}}   )   )  \,  =  \, \ottsym{0} )   ~  {} \\  &  ~~  \ottsym{5}  ~~   \mathsf{err}^ \ottnt{l_{{\mathrm{2}}}}   )  \\
   \longrightarrow _{  \mathsf{C}  }  &  \mathsf{mon}^{ \ottnt{l_{{\mathrm{1}}}} }(  \ottnt{C_{{\mathrm{12}}}}  [   (  \lambda \mathit{x} \mathord{:}  \mathsf{Int}  .~  \mathit{x}  )   /  \mathit{f}  ]  ,  {} \\  &  ~~   \mathsf{if} ~   (  \mathsf{mon}^{ \ottnt{l_{{\mathrm{1}}}} }(  \mathsf{pred}(  \lambda \mathit{x} \mathord{:}  \mathsf{Int}  .~  \mathit{x}  \, \ottsym{>} \, \ottsym{0} )  ,  {} \\   &  \qquad  \qquad  \quad   (  \lambda \mathit{x} \mathord{:}  \mathsf{Int}  .~  \mathit{x}  )   ~   (  \mathsf{if} ~   ( \ottsym{0} \, \ottsym{>} \, \ottsym{0} )   ~  \ottsym{0}  ~~   \mathsf{err}^ \ottnt{l_{{\mathrm{2}}}}   )   )  \,  =  \, \ottsym{0} )   ~  {} \\  &  ~~  \ottsym{5}  ~~   \mathsf{err}^ \ottnt{l_{{\mathrm{2}}}}   )  \\
   \longrightarrow _{  \mathsf{C}  }  &  \mathsf{mon}^{ \ottnt{l_{{\mathrm{1}}}} }(  \ottnt{C_{{\mathrm{12}}}}  [   (  \lambda \mathit{x} \mathord{:}  \mathsf{Int}  .~  \mathit{x}  )   /  \mathit{f}  ]  ,  {} \\  &  ~~   \mathsf{if} ~   (  \mathsf{mon}^{ \ottnt{l_{{\mathrm{1}}}} }(  \mathsf{pred}(  \lambda \mathit{x} \mathord{:}  \mathsf{Int}  .~  \mathit{x}  \, \ottsym{>} \, \ottsym{0} )  ,  {} \\   &  \qquad  \qquad  \quad   (  \lambda \mathit{x} \mathord{:}  \mathsf{Int}  .~  \mathit{x}  )   ~   (  \mathsf{if} ~   \mathsf{false}   ~  \ottsym{0}  ~~   \mathsf{err}^ \ottnt{l_{{\mathrm{2}}}}   )   )  \,  =  \, \ottsym{0} )   ~  {} \\  &  ~~  \ottsym{5}  ~~   \mathsf{err}^ \ottnt{l_{{\mathrm{2}}}}   )  \\
   \longrightarrow _{  \mathsf{C}  }  &  \mathsf{mon}^{ \ottnt{l_{{\mathrm{1}}}} }(  \ottnt{C_{{\mathrm{12}}}}  [   (  \lambda \mathit{x} \mathord{:}  \mathsf{Int}  .~  \mathit{x}  )   /  \mathit{f}  ]  ,  {} \\  &  ~~   \mathsf{if} ~   (  \mathsf{mon}^{ \ottnt{l_{{\mathrm{1}}}} }(  \mathsf{pred}(  \lambda \mathit{x} \mathord{:}  \mathsf{Int}  .~  \mathit{x}  \, \ottsym{>} \, \ottsym{0} )  ,    (  \lambda \mathit{x} \mathord{:}  \mathsf{Int}  .~  \mathit{x}  )   ~   \mathsf{err}^ \ottnt{l_{{\mathrm{2}}}}   )  \,  =  \, \ottsym{0} )   ~  \ottsym{5}  ~~   \mathsf{err}^ \ottnt{l_{{\mathrm{2}}}}   )  \\
\fi
  \longrightarrow ^{*}_{  \mathsf{C}  }  &  \mathsf{err}^ \ottnt{l_{{\mathrm{2}}}} 
\end{array} \]
\caption{Abusive function proxies in \CPCFc}
\label{fig:abusive}
\end{figure}
For the space-efficient semantics to be sound, it must behave
\textit{exactly} like the classic semantics: no matter
what joins happen, \CPCFe must replicate the contract substitutions
done in \CPCFc.
We can construct an abusive contract in \CPCFc---even though it has
lax semantics---by having the inner function proxy abuse the outer
one (Fig.~\ref{fig:abusive}).
Why was blame raised? Because $\ottnt{c_{{\mathrm{2}}}}$'s codomain contract
\textit{abused} $\ottnt{c_{{\mathrm{1}}}}$'s domain contract. Even though \CPCFc has a
lax semantics, wrapping multiple function proxies leads to monitoring
domains from one contract in the codomain of another---a situation
ripe for abuse.

Space-efficiency means joining contracts, so how can we emulate this classic-semantics substitution behavior?
We use the $ \mathsf{wrap} $ function, forcing a substitution when two
function contracts are joined. By keeping track of these substitutions
at every join, any joins that happen in the future will be working on
contracts which already have appropriate substitutions.

\CPCFe uses \textit{labeled contracts} (Fig.~\ref{fig:syntax});
substitution for labeled predicate contracts is explicit and delayed,
as for ordinary contracts:
\[ \begin{array}{rcl}
    \mathsf{pred}^{ \ottnt{l} }_{ \sigma }( \ottnt{e} )   [  \ottnt{v}  /  \mathit{x}  ]  &=& \begin{cases}
     \mathsf{pred}^{ \ottnt{l} }_{  \sigma  [  \mathit{x}  \mapsto  \ottnt{v}  ]  }( \ottnt{e} )  &  \mathit{x}  \in   \operatorname{fv} ( \sigma  \ottsym{(}  \ottnt{e}  \ottsym{)} )   \\
     \mathsf{pred}^{ \ottnt{l} }_{ \sigma }( \ottnt{e} )  & \text{otherwise}
  \end{cases} \\
   \mathsf{nil}  [  \ottnt{v}  /  \mathit{x}  ]  &=&  \mathsf{nil}  \\
    (   \mathsf{pred}^{ \ottnt{l} }_{ \sigma }( \ottnt{e} )  ; \ottnt{r}  )   [  \ottnt{v}  /  \mathit{x}  ]  &=&     \mathsf{pred}^{ \ottnt{l} }_{ \sigma }( \ottnt{e} )   [  \ottnt{v}  /  \mathit{x}  ]  ; \ottnt{r}   [  \ottnt{v}  /  \mathit{x}  ] 
\end{array} \]
We do \textit{not} do any joining when a substitution
occurs (but see Sec.~\ref{sec:extensions}).
In \CPCFe, closing substitutions map each variable to (a)
itself ($ [  \mathit{x}  \mapsto  \mathit{x}  ] $), (b) a monitor on itself
($ [  \mathit{x}  \mapsto   \mathsf{mon}( \ottnt{c} ,  \mathit{x} )   ] $), or (c) a value.
We add an evaluation rule taking ordinary contract monitors
$ \mathsf{mon}^{ \ottnt{l} }( \ottnt{C} ,  \ottnt{e} ) $ to labeled-contract monitors $ \mathsf{mon}( \ottnt{c} ,  \ottnt{e} ) $ by means
of the labeling function $ \mathsf{label} $ (\E{MonLabel}).

Space-efficiency comes by restricting congruence to only apply when
there are abutting monitors (cf. \E{MonC} here in \CPCFe to \E{Mon} in
\CPCFc).
When two monitors collide, we \textit{join} them (\E{MonCJoin}).
Checking function contracts is as usual (\E{MonCApp} is the same as
\E{MonApp}, only the latter works over labeled contracts); checking
predicate stacks proceeds straightforwardly predicate-by-predicate
(\E{MonCNil} and \E{MonCPred}).

{\iffull
\subsubsection{Metatheory}
\label{sec:cpcfemetatheory}

We prove \CPCFe's type system sound; the proof follows that for
\CPCFc, though there are more evaluation rules to consider here.
We only consider the typing rules for \CPCFe, i.e.,
those typeset on white and $\SE{\text{periwinkle}}$.

{\iffull
\begin{lemma}[Weakening]
  \label{lem:cpcfeweakening}
  \begin{itemize}
  \item If $ \mathrm{\Gamma}   \vdash   \ottnt{e}  :  \ottnt{T} $ and $ \mathit{x}  \not\in   \operatorname{fv} ( \ottnt{e} )  $ then $  \mathrm{\Gamma} , \mathit{x} \mathord{:} \ottnt{T'}    \vdash   \ottnt{e}  :  \ottnt{T} $.
  \item If $\mathrm{\Gamma}  \vdash  \ottnt{C}  \ottsym{:}  \ottnt{T}$ and $ \mathit{x}  \not\in   \operatorname{fv} ( \ottnt{C} )  $ then $ \mathrm{\Gamma} , \mathit{x} \mathord{:} \ottnt{T'}   \vdash  \ottnt{C}  \ottsym{:}  \ottnt{T}$.
  \item If $\mathrm{\Gamma}  \vdash  \ottnt{c}  \ottsym{:}  \ottnt{T}$ and $ \mathit{x}  \not\in   \operatorname{fv} ( \ottnt{c} )  $ then $ \mathrm{\Gamma} , \mathit{x} \mathord{:} \ottnt{T'}   \vdash  \ottnt{c}  \ottsym{:}  \ottnt{T}$.
  \end{itemize}
  \begin{proof}
    By mutual induction on the terms and contracts.
  \end{proof}
\end{lemma}

\begin{lemma}[Substitution]
  \label{lem:cpcfesubstitution}
  If $ \emptyset   \vdash   \ottnt{v}  :  \ottnt{T_{{\mathrm{1}}}} $, then
  \begin{itemize}
  \item $  \mathrm{\Gamma}_{{\mathrm{1}}} , \mathit{x} \mathord{:} \ottnt{T_{{\mathrm{1}}}}   \ottsym{,}  \mathrm{\Gamma}_{{\mathrm{2}}}   \vdash   \ottnt{e}  :  \ottnt{T_{{\mathrm{2}}}} $ implies $ \mathrm{\Gamma}_{{\mathrm{1}}}  \ottsym{,}  \mathrm{\Gamma}_{{\mathrm{2}}}   \vdash    \ottnt{e}  [  \ottnt{v} / \mathit{x}  ]   :  \ottnt{T_{{\mathrm{2}}}} $, and
  \item $ \mathrm{\Gamma}_{{\mathrm{1}}} , \mathit{x} \mathord{:} \ottnt{T_{{\mathrm{1}}}}   \ottsym{,}  \mathrm{\Gamma}_{{\mathrm{2}}}  \vdash  \ottnt{C}  \ottsym{:}  \ottnt{T_{{\mathrm{2}}}}$ implies $\mathrm{\Gamma}_{{\mathrm{1}}}  \ottsym{,}  \mathrm{\Gamma}_{{\mathrm{2}}}  \vdash   \ottnt{C}  [  \ottnt{v}  /  \mathit{x}  ]   \ottsym{:}  \ottnt{T_{{\mathrm{2}}}}$.
  \item $ \mathrm{\Gamma}_{{\mathrm{1}}} , \mathit{x} \mathord{:} \ottnt{T_{{\mathrm{1}}}}   \ottsym{,}  \mathrm{\Gamma}_{{\mathrm{2}}}  \vdash  \ottnt{c}  \ottsym{:}  \ottnt{T_{{\mathrm{2}}}}$ implies $\mathrm{\Gamma}_{{\mathrm{1}}}  \ottsym{,}  \mathrm{\Gamma}_{{\mathrm{2}}}  \vdash   \ottnt{c}  [  \ottnt{v}  /  \mathit{x}  ]   \ottsym{:}  \ottnt{T_{{\mathrm{2}}}}$.
  \end{itemize}
  \begin{proof}
    By mutual induction the terms, contracts, and labeled contracts. \iffull
    \begin{itemize}
    \item[($\ottnt{e}  \ottsym{=}  \mathit{y}$)] By case analysis on $\mathit{x}  \ottsym{=}  \mathit{y}$, using
      weakening (Lemma~\ref{lem:cpcfeweakening}) as appropriate.
    \item[($\ottnt{e}  \ottsym{=}  \ottnt{k}$)] Immediate.
    \item[($\ottnt{e}  \ottsym{=}  \ottnt{e_{{\mathrm{1}}}} \, \ottnt{op} \, \ottnt{e_{{\mathrm{2}}}}$)] By the IHs.
    \item[($\ottnt{e}  \ottsym{=}   \ottnt{e_{{\mathrm{1}}}}  ~  \ottnt{e_{{\mathrm{2}}}} $)] By the IHs.
    \item[($\ottnt{e}  \ottsym{=}   \lambda \mathit{y} \mathord{:} \ottnt{T} .~  \ottnt{e_{{\mathrm{1}}}} $)] By case analysis on $\mathit{x}  \ottsym{=}  \mathit{y}$,
      narrowing when the two are equal.
    \item[($\ottnt{e}  \ottsym{=}   \mu ( \mathit{y} \mathord{:} \ottnt{T} ) . ~  \ottnt{e_{{\mathrm{1}}}} $)] By case analysis on $\mathit{x}  \ottsym{=}  \mathit{y}$,
      narrowing when the two are equal.
    \item[($\ottnt{e}  \ottsym{=}   \mathsf{if} ~  \ottnt{e_{{\mathrm{1}}}}  ~  \ottnt{e_{{\mathrm{2}}}}  ~~  \ottnt{e_{{\mathrm{3}}}} $)] By the IHs.
    \item[($\ottnt{e}  \ottsym{=}   \mathsf{err}^ \ottnt{l} $)] Immediate.
    \item[($\ottnt{e}  \ottsym{=}   \mathsf{mon}^{ \ottnt{l} }( \ottnt{C} ,  \ottnt{e} ) $)] By the IHs.
    \item[($\ottnt{C}  \ottsym{=}   \mathsf{pred}_{ \sigma }( \ottnt{e} ) $)] \else The interesting case is when
      $\ottnt{C}  \ottsym{=}   \mathsf{pred}_{ \sigma }( \ottnt{e} ) $. \fi
      We have $ \mathrm{\Gamma}_{{\mathrm{1}}} , \mathit{x} \mathord{:} \ottnt{T_{{\mathrm{1}}}}   \ottsym{,}  \mathrm{\Gamma}_{{\mathrm{2}}}  \vdash   \mathsf{pred}_{ \sigma }( \ottnt{e} )   \ottsym{:}  \ottnt{T_{{\mathrm{2}}}}$ and we must show that $\mathrm{\Gamma}_{{\mathrm{1}}}  \ottsym{,}  \mathrm{\Gamma}_{{\mathrm{2}}}  \vdash    \mathsf{pred}_{ \sigma }( \ottnt{e} )   [  \ottnt{v}  /  \mathit{x}  ]   \ottsym{:}  \ottnt{T_{{\mathrm{2}}}}$.
      By inversion, we know that:
      $\ottnt{T_{{\mathrm{2}}}}  \ottsym{=}  \ottnt{B}$ for some base type $\ottnt{B}$;
      $  \mathrm{\Gamma}_{{\mathrm{1}}} , \mathit{x} \mathord{:} \ottnt{T_{{\mathrm{1}}}}   \ottsym{,}  \mathrm{\Gamma}_{{\mathrm{2}}}  \ottsym{,}  \mathrm{\Gamma}'   \vdash   \ottnt{e}  :   \ottnt{B} \mathord{ \rightarrow }  \mathsf{Bool}   $;
      and $\mathrm{\Gamma}'  \vdash  \sigma$.

      There are two cases. In both cases, we use \T{Pred}.
      \begin{itemize}
      \item[($ \mathit{x}  \in   \operatorname{fv} ( \ottnt{e} )  $)] The substitution is actually stored in
        $\sigma$. We find $  \mathrm{\Gamma}_{{\mathrm{1}}}  \ottsym{,}  \mathrm{\Gamma}_{{\mathrm{2}}}  \ottsym{,}  \mathrm{\Gamma}' , \mathit{x} \mathord{:} \ottnt{T_{{\mathrm{1}}}}    \vdash   \ottnt{e}  :  \ottnt{B} $ by the IH
        on $\ottnt{e}$ and $ \mathrm{\Gamma}' , \mathit{x} \mathord{:} \ottnt{T_{{\mathrm{1}}}}   \vdash   \sigma  [  \mathit{x}  \mapsto  \ottnt{v}  ] $ by \T{Map} and
        our assumption.
      \item[($ \mathit{x}  \not\in   \operatorname{fv} ( \ottnt{e} )  $)] The substitution is ignored, and
        $ \ottnt{e}  [  \ottnt{v} / \mathit{x}  ]   \ottsym{=}  \ottnt{e}$. We find $ \mathrm{\Gamma}_{{\mathrm{1}}}  \ottsym{,}  \mathrm{\Gamma}_{{\mathrm{2}}}  \ottsym{,}  \mathrm{\Gamma}'   \vdash   \ottnt{e}  :  \ottnt{B} $ by the IH
        and $\mathrm{\Gamma}'  \vdash  \sigma$ by assumption.
      \end{itemize}
      \iffull
    \item[($\ottnt{C}  \ottsym{=}   \mathit{x} \mathord{:} \ottnt{C_{{\mathrm{1}}}} \mapsto \ottnt{C_{{\mathrm{2}}}} $)] By the IHs.
    \item[($\ottnt{c}  \ottsym{=}  \mathsf{nil}$)] Immediate.
    \item[($\ottnt{c}  \ottsym{=}    \mathsf{pred}^{ \ottnt{l} }_{ \sigma }( \ottnt{e} )  ; \ottnt{r} $)] As for the predicate contracts
      above, using the IH to show that $\ottnt{r}$ is still well typed.
    \item[($\ottnt{c}  \ottsym{=}   \mathit{x} \mathord{:} \ottnt{c_{{\mathrm{1}}}} \mapsto \ottnt{c_{{\mathrm{2}}}} $)] By the IHS.
    \end{itemize}
    \fi
  \end{proof}
\end{lemma}

As a corollary, closing substitutions $\sigma$ close up exactly
their context $\mathrm{\Gamma}'$.

\begin{lemma}[Closing substitutions]
  \label{lem:cpcfeclosure}
  If $\mathrm{\Gamma}'  \vdash  \sigma$ and $ \mathrm{\Gamma}  \ottsym{,}  \mathrm{\Gamma}'   \vdash   \ottnt{e}  :  \ottnt{T} $ then $ \mathrm{\Gamma}   \vdash   \sigma  \ottsym{(}  \ottnt{e}  \ottsym{)}  :  \ottnt{T} $.
  \begin{proof}
    By induction on $\mathrm{\Gamma}'$, using substitution (Lemma~\ref{lem:cpcfesubstitution}).
  \end{proof}
\end{lemma}
\fi}

\begin{lemma}[Progress]
  \label{lem:cpcfeprogress}
  If $ \emptyset   \vdash   \ottnt{e}  :  \ottnt{T} $ then either (a) $\ottnt{e}$ is some value,
  $\ottnt{v}$; (b) $\ottnt{e}$ is some error, $ \mathsf{err}^ \ottnt{l} $; or, (c) $\ottnt{e} \,  \longrightarrow _{  \mathsf{E}  }  \, \ottnt{e'}$.
  \begin{proof}
    By induction on the typing derivation.
    {\iffull
    \begin{itemize}
    \item[(\T{Var})] Contradictory---variables aren't well typed in
      the empty context.
    \item[(\T{Const})] $\ottnt{k}$ is a value.
    \item[(\T{Blame})] $ \mathsf{err}^ \ottnt{l} $ is an error.
    \item[(\T{Abs})] Function abstractions are values.
    \item[(\T{Rec})] Takes a step by \E{Fix}.
    \item[(\T{Op})] By the IH, $\ottnt{e_{{\mathrm{1}}}}$ is a value, error, or
      steps. In the latter two cases, we step by \E{OpLRaise} or
      \E{OpL}, respectively.

      By the IH, $\ottnt{e_{{\mathrm{2}}}}$ is a value, error, or steps. In the latter
      two cases we step by \E{OpRRaise} or \E{OpR}.

      If both are values, then we step by \E{Delta}.
    \item[(\T{App})] By the IH, $\ottnt{e_{{\mathrm{1}}}}$ is a value, error, or
      steps. In the latter two cases, we step by \E{AppLRaise} or
      \E{AppL}, respectively.

      By the IH, $\ottnt{e_{{\mathrm{2}}}}$ is a value, error, or steps. In the latter
      two cases we step by \E{AppRRaise} or \E{AppR}.

      If both are values, we know that $\ottnt{e_{{\mathrm{1}}}}$ is either a function
      or a function proxy; we step by either \E{Beta} or \E{MonCApp},
      depending on the shape of $\ottnt{e_{{\mathrm{1}}}}$.
    \item[(\T{If})] By the IH, $\ottnt{e_{{\mathrm{1}}}}$ is a value, error, or
      steps. In the latter two cases, we step by \E{IfRaise} or
      \E{If}, respectively./

      If $\ottnt{e_{{\mathrm{1}}}}$ is a value, it must be either $ \mathsf{true} $ or
      $ \mathsf{false} $; we step by \E{IfTrue} or \E{IfFalse}, depending on
      the shape of $\ottnt{e_{{\mathrm{1}}}}$.
    \item[(\T{Mon})] We step by \E{MonLabel}.
    \item[(\T{MonC})] If the inner term $\ottnt{e}$ is a monitor, we step by
      \E{MonCJoin}.
      Otherwise, by the IH, the inner term is a value, an error, or
      steps. If it's an error, we step by \E{MonCRaise}. If it steps,
      we step by \E{MonC} (knowing already that that $\ottnt{e}$ isn't a
      monitor).
      If $\ottnt{e}$ is a value, we either step by \E{MonCNil}, step by
      \E{MonCPred}, or have a value already (when $\ottnt{e}$ is a
      function).
    \end{itemize}
    \fi}
  \end{proof}
\end{lemma}

{\iffull
\begin{lemma}
  \label{lem:cpcfelabel}
  If $\mathrm{\Gamma}  \vdash  \ottnt{C}  \ottsym{:}  \ottnt{T}$ then $\mathrm{\Gamma}  \vdash   \mathsf{label} ^{ \ottnt{l} }( \ottnt{C} )   \ottsym{:}  \ottnt{T}$.
  \begin{proof}
    By induction on the typing derivation, using \T{CPred} and
    \T{CNil} in the predicate case and \T{CFun} in the function case.
  \end{proof}
\end{lemma}

\begin{lemma}
  \label{lem:cpcfepredstack}
  If $\mathrm{\Gamma}  \vdash  \ottnt{r_{{\mathrm{1}}}}  \ottsym{:}  \ottnt{T}$ and $\mathrm{\Gamma}  \vdash  \ottnt{r_{{\mathrm{2}}}}  \ottsym{:}  \ottnt{T}$ then $\mathrm{\Gamma}  \vdash   \mathsf{join} ( \ottnt{r_{{\mathrm{1}}}} , \ottnt{r_{{\mathrm{2}}}} )   \ottsym{:}  \ottnt{T}$.
  \begin{proof}
    By induction on the typing derivation of $\ottnt{r_{{\mathrm{1}}}}$.
  \end{proof}
\end{lemma}

\begin{lemma}
  \label{lem:cpcfejoin}
  If $\mathrm{\Gamma}  \vdash  \ottnt{c_{{\mathrm{1}}}}  \ottsym{:}  \ottnt{T}$ and $\mathrm{\Gamma}  \vdash  \ottnt{c_{{\mathrm{2}}}}  \ottsym{:}  \ottnt{T}$ then $\mathrm{\Gamma}  \vdash   \mathsf{join} ( \ottnt{c_{{\mathrm{1}}}} , \ottnt{c_{{\mathrm{2}}}} )   \ottsym{:}  \ottnt{T}$.
  \begin{proof}
    By induction on the typing derivation of $\ottnt{c_{{\mathrm{1}}}}$, using
    Lemma~\ref{lem:cpcfepredstack} in the base case and substitution
    (Lemma~\ref{lem:cpcfesubstitution}).
  \end{proof}
\end{lemma}
\fi}

\begin{lemma}[Preservation]
  \label{lem:cpcfepreservation}
  If $ \emptyset   \vdash   \ottnt{e}  :  \ottnt{T} $ and $\ottnt{e} \,  \longrightarrow _{  \mathsf{E}  }  \, \ottnt{e'}$ then $ \emptyset   \vdash   \ottnt{e'}  :  \ottnt{T} $.
  \begin{proof}
    By induction on the typing derivation, with cases on the step
    taken.
    {\iffull
    \begin{itemize}
    \item[(\T{Var})] Contradictory---variables aren't well typed in
      the empty context.
    \item[(\T{Const})] Contradictory---constants are normal forms.
    \item[(\T{Blame})] Contradictory---errors are normal forms.
    \item[(\T{Abs})] Contradictory---functions are normal forms.
    \item[(\T{Rec})] We must have stepped by \E{Fix}; by substitution
      (Lemma~\ref{lem:cpcfesubstitution}) and \T{Fix}.
    \item[(\T{Op})] If we stepped by an \E{OpRaise*} rule, we are done
      by \T{Blame}. If we stepped by a congruence rule, \E{Op*}, we
      are done by the IH and \T{Op}.
      If we stepped by \E{Delta}, by the assumption that built-in
      operations have sound denotations.
    \item[(\T{App})] If we stepped by an \E{AppRaise*} rule, we are
      done by \T{Blame}.If we stepped by a congruence rule, \E{App*},
      we are done by the IH and \T{App}.
      If we stepped by \E{Beta}, by
      substitution. (Lemma~\ref{lem:cpcfesubstitution})
      If we stepped by \E{MonCApp}, then: the domain is well typed by
      \T{Mon}; we can apply the proxied function by \T{App}; the
      \textit{unmonitored} codomain well formed by substitution
      (Lemma~\ref{lem:cpcfesubstitution}); and the entire resulting
      term is well typed by \T{Mon} again.
    \item[(\T{If})] If we stepped by \E{IfRaise}, by \T{Blame}. If we
      stepped by \E{If}, by the IH and \T{If}. If we stepped by
      \E{IfTrue} or \E{IfFalse}, by assumption.
    \item[(\T{Mon})] Immediate from the assumptions, since labeling
      contracts preserves typing (Lemma~\ref{lem:cpcfelabel}).
    \item[(\T{MonC})] If we stepped by \E{MonCRaise}, the resulting
      error is well typed by \T{Blame}. If we stepped by \E{MonC}, by
      the IH and \T{Mon}. If we stepped by \E{MonCJoin}, then by
      Lemma~\ref{lem:cpcfejoin}; if by \E{MonCNil}, then by the
      assumptions . If we stepped by \E{MonCPred}, then: we have
      $ \emptyset   \vdash   \sigma  \ottsym{(}  \ottnt{e}  \ottsym{)}  :   \ottnt{B} \mathord{ \rightarrow }  \mathsf{Bool}   $ by closing substitution
      (Lemma~\ref{lem:cpcfeclosure}); we can type the condition by
      \T{App}; the then branch is well typed by assumption; the false
      branch is well typed by \T{Blame}; and the whole lot is well
      typed by \T{If}.
    \end{itemize}
    \fi}
  \end{proof}
\end{lemma}

{\iffull
\begin{lemma}[Determinism]
  \label{lem:secpcfdeterministic}
  If $\ottnt{e_{{\mathrm{1}}}} \,  \longrightarrow _{  \mathsf{E}  }  \, \ottnt{e_{{\mathrm{2}}}}$ and $\ottnt{e_{{\mathrm{1}}}} \,  \longrightarrow _{  \mathsf{E}  }  \, \ottnt{e'_{{\mathrm{2}}}}$, then $\ottnt{e_{{\mathrm{2}}}}  \ottsym{=}  \ottnt{e'_{{\mathrm{2}}}}$.
  \begin{proof}
    By induction on the first derivation. Recall that we exclude the
    \CPCFc rules, so only \E{MonLabel} fires on monitors. \E{MonC} and
    \E{MonCJoin} carefully avoid overlapping.
  \end{proof}
\end{lemma}
\fi}

\fi}

\section{Soundness for space efficiency}
\label{sec:soundness}

\CPCFc and \CPCFe are operationally equivalent, even though their cast
semantics differ. We can make this connection formal by proving that
every CPCF term either: (a) diverges in both \CPCFc and \CPCFe or
(b) reduces to equivalent terms in both \CPCFc and \CPCFe.

One minor technicality: some of the forms in our language are necessary only for runtime or
only appear in one of the two calculi. We characterize \textit{source programs}
as those which omit runtime terms.
\begin{definition}[Source program]
  \label{def:sourceprogram}
A well typed \textit{source program} does not use \T{Blame} or
\T{MonC} (and so \T{CNil}, \T{CPred}, and \T{CFun} cannot be
used).
\end{definition}

{\iffull
\begin{lemma}[Associativity of $ \mathsf{join} $ on predicate stacks]
  \label{lem:joinpsassociative}

  $ \mathsf{join} ( \ottnt{r_{{\mathrm{1}}}} ,  \mathsf{join} ( \ottnt{r_{{\mathrm{2}}}} , \ottnt{r_{{\mathrm{3}}}} )  )   \ottsym{=}   \mathsf{join} (  \mathsf{join} ( \ottnt{r_{{\mathrm{1}}}} , \ottnt{r_{{\mathrm{2}}}} )  , \ottnt{r_{{\mathrm{3}}}} ) $.
  \begin{proof}
    By induction on $\ottnt{r_{{\mathrm{1}}}}$.
    \begin{itemize}
    \item[($\ottnt{r_{{\mathrm{1}}}}  \ottsym{=}  \mathsf{nil}$)] Immediate.
    \item[($\ottnt{r_{{\mathrm{1}}}}  \ottsym{=}    \mathsf{pred}^{ \ottnt{l} }( \ottnt{e} )  ; \ottnt{r'_{{\mathrm{1}}}} $)] The IH is
      $ \mathsf{join} ( \ottnt{r'_{{\mathrm{1}}}} ,  \mathsf{join} ( \ottnt{r_{{\mathrm{2}}}} , \ottnt{r_{{\mathrm{3}}}} )  )   \ottsym{=}   \mathsf{join} (  \mathsf{join} ( \ottnt{r'_{{\mathrm{1}}}} , \ottnt{r_{{\mathrm{2}}}} )  , \ottnt{r_{{\mathrm{3}}}} ) $. We
      calculate:
      \[ \begin{array}{rcl}
              \mathsf{join} ( \ottnt{r_{{\mathrm{1}}}} ,  \mathsf{join} ( \ottnt{r_{{\mathrm{2}}}} , \ottnt{r_{{\mathrm{3}}}} )  ) 
         &=&  \mathsf{join} ( \ottsym{(}    \mathsf{pred}^{ \ottnt{l} }_{ \sigma }( \ottnt{e} )  ; \ottnt{r'_{{\mathrm{1}}}}   \ottsym{)} ,  \mathsf{join} ( \ottnt{r_{{\mathrm{2}}}} , \ottnt{r_{{\mathrm{3}}}} )  )  \\
         &=&   \mathsf{pred}^{ \ottnt{l} }_{ \sigma }( \ottnt{e} )  ;  \mathsf{drop} (  \mathsf{join} ( \ottnt{r'_{{\mathrm{1}}}} ,  \mathsf{join} ( \ottnt{r_{{\mathrm{2}}}} , \ottnt{r_{{\mathrm{3}}}} )  )  ,  \mathsf{pred}_{ \sigma }( \ottnt{e} )  )   \\
         &=&   \mathsf{pred}^{ \ottnt{l} }_{ \sigma }( \ottnt{e} )  ;  \mathsf{drop} (  \mathsf{join} (  \mathsf{join} ( \ottnt{r'_{{\mathrm{1}}}} , \ottnt{r_{{\mathrm{2}}}} )  , \ottnt{r_{{\mathrm{3}}}} )  ,  \mathsf{pred}_{ \sigma }( \ottnt{e} )  )   \text{\quad (IH)} \\
         &=&  \mathsf{join} (   \mathsf{pred}^{ \ottnt{l} }_{ \sigma }( \ottnt{e} )  ;  \mathsf{drop} (  \mathsf{join} ( \ottnt{r'_{{\mathrm{1}}}} , \ottnt{r_{{\mathrm{2}}}} )  ,  \mathsf{pred}_{ \sigma }( \ottnt{e} )  )   , \ottnt{r_{{\mathrm{3}}}} )  \\
         &=&  \mathsf{join} (  \mathsf{join} ( \ottsym{(}    \mathsf{pred}^{ \ottnt{l} }_{ \sigma }( \ottnt{e} )  ; \ottnt{r'_{{\mathrm{1}}}}   \ottsym{)} , \ottnt{r_{{\mathrm{2}}}} )  , \ottnt{r_{{\mathrm{3}}}} )  \\
         &=&  \mathsf{join} (  \mathsf{join} ( \ottnt{r_{{\mathrm{1}}}} , \ottnt{r_{{\mathrm{2}}}} )  , \ottnt{r_{{\mathrm{3}}}} ) 
      \end{array} \]
    \end{itemize}
  \end{proof}
\end{lemma}

\begin{lemma}[Associativity of join]
  \label{lem:joinassociative}
  ~ \\

  $ \mathsf{join} ( \ottnt{c_{{\mathrm{1}}}} ,  \mathsf{join} ( \ottnt{c_{{\mathrm{2}}}} , \ottnt{c_{{\mathrm{3}}}} )  )   \ottsym{=}   \mathsf{join} (  \mathsf{join} ( \ottnt{c_{{\mathrm{1}}}} , \ottnt{c_{{\mathrm{2}}}} )  , \ottnt{c_{{\mathrm{3}}}} ) $.
  \begin{proof}
    By induction on $\ottnt{c_{{\mathrm{1}}}}$.
    In the predicate stack case, by Lemma~\ref{lem:joinpsassociative}.
    In the function case, we calculate:
    \[ \begin{array}{rl}
        &  \mathsf{join} (  \mathit{x} \mathord{:} \ottnt{c_{{\mathrm{11}}}} \mapsto \ottnt{c_{{\mathrm{12}}}}  ,  \mathsf{join} (  \mathit{x} \mathord{:} \ottnt{c_{{\mathrm{21}}}} \mapsto \ottnt{c_{{\mathrm{22}}}}  ,  \mathit{x} \mathord{:} \ottnt{c_{{\mathrm{31}}}} \mapsto \ottnt{c_{{\mathrm{32}}}}  )  )  \\
      =&  \mathsf{join} (  \mathit{x} \mathord{:} \ottnt{c_{{\mathrm{11}}}} \mapsto \ottnt{c_{{\mathrm{12}}}}  ,  \mathit{x} \mathord{:}  \mathsf{join} ( \ottnt{c_{{\mathrm{31}}}} , \ottnt{c_{{\mathrm{21}}}} )  \mapsto  \mathsf{join} (  \mathsf{wrap} ( \ottnt{c_{{\mathrm{22}}}} , \mathit{x} , \ottnt{c_{{\mathrm{31}}}} )  , \ottnt{c_{{\mathrm{32}}}} )   )  \\
      =&  \mathit{x} \mathord{:}  \mathsf{join} (  \mathsf{join} ( \ottnt{c_{{\mathrm{31}}}} , \ottnt{c_{{\mathrm{21}}}} )  , \ottnt{c_{{\mathrm{11}}}} )  \mapsto {} \\  &  \qquad   \mathsf{join} (  \mathsf{wrap} ( \ottnt{c_{{\mathrm{12}}}} , \mathit{x} ,  \mathsf{join} ( \ottnt{c_{{\mathrm{31}}}} , \ottnt{c_{{\mathrm{21}}}} )  )  ,  \mathsf{join} (  \mathsf{wrap} ( \ottnt{c_{{\mathrm{22}}}} , \mathit{x} , \ottnt{c_{{\mathrm{31}}}} )  , \ottnt{c_{{\mathrm{32}}}} )  )   \\
      \multicolumn{2}{r}{\text{(IH)}} \\
      =&  \mathit{x} \mathord{:}  \mathsf{join} ( \ottnt{c_{{\mathrm{31}}}} ,  \mathsf{join} ( \ottnt{c_{{\mathrm{21}}}} , \ottnt{c_{{\mathrm{11}}}} )  )  \mapsto  \mathsf{join} (  \mathsf{wrap} (  \mathsf{join} (  \mathsf{wrap} ( \ottnt{c_{{\mathrm{12}}}} , \mathit{x} , \ottnt{c_{{\mathrm{21}}}} )  , \ottnt{c_{{\mathrm{22}}}} )  , \mathit{x} , \ottnt{c_{{\mathrm{31}}}} )  , \ottnt{c_{{\mathrm{32}}}} )   \\
      =&  \mathsf{join} (  \mathit{x} \mathord{:} \ottnt{c_{{\mathrm{31}}}} \mapsto \ottnt{c_{{\mathrm{32}}}}  ,  \mathit{x} \mathord{:}  \mathsf{join} ( \ottnt{c_{{\mathrm{21}}}} , \ottnt{c_{{\mathrm{11}}}} )  \mapsto  \mathsf{join} (  \mathsf{wrap} ( \ottnt{c_{{\mathrm{12}}}} , \mathit{x} , \ottnt{c_{{\mathrm{21}}}} )  , \ottnt{c_{{\mathrm{22}}}} )   )  \\
      =&  \mathsf{join} (  \mathit{x} \mathord{:} \ottnt{c_{{\mathrm{31}}}} \mapsto \ottnt{c_{{\mathrm{32}}}}  ,  \mathsf{join} (  \mathit{x} \mathord{:} \ottnt{c_{{\mathrm{11}}}} \mapsto \ottnt{c_{{\mathrm{12}}}}  ,  \mathit{x} \mathord{:} \ottnt{c_{{\mathrm{21}}}} \mapsto \ottnt{c_{{\mathrm{22}}}}  )  ) 
    \end{array}\]
  \end{proof}
\end{lemma}

\begin{lemma}[Idempotence of predicate stacks]
  \label{lem:contractidempotent}
  If:
  \begin{itemize}
  \item $ \emptyset   \vdash   \ottnt{k}  :  \ottnt{B} $,
  \item $ \ottnt{e}  ~  \ottnt{k}  \,  \longrightarrow ^{*}_{  \mathsf{E}  }  \,  \mathsf{true} $, and
  \item $\emptyset  \vdash   \mathsf{join} ( \ottnt{r_{{\mathrm{1}}}} , \ottnt{r_{{\mathrm{2}}}} )   \ottsym{:}  \ottnt{B}$,
  \end{itemize}
  then $ \mathsf{mon}(  \mathsf{join} ( \ottnt{r_{{\mathrm{1}}}} , \ottnt{r_{{\mathrm{2}}}} )  ,  \ottnt{k} )  \,  \longrightarrow ^{*}_{  \mathsf{E}  }  \, \ottnt{w}$ iff
  $ \mathsf{mon}(  \mathsf{join} ( \ottnt{r_{{\mathrm{1}}}} ,  \mathsf{drop} ( \ottnt{r_{{\mathrm{2}}}} ,  \mathsf{pred}_{ \sigma }( \ottnt{e} )  )  )  ,  \ottnt{k} )  \,  \longrightarrow ^{*}_{  \mathsf{E}  }  \, \ottnt{w}$.
  \begin{proof}
    By induction on length of $\ottnt{r_{{\mathrm{1}}}}$, observing that $ \sigma  \ottsym{(}  \ottnt{e}  \ottsym{)}  ~  \ottnt{k}  \,  \longrightarrow ^{*}_{  \mathsf{E}  }  \,  \mathsf{true} $ is redundantly (but successfully) checked on the
    left more than on the right.
  \end{proof}
\end{lemma}

\fi}

Greenberg identified the key property for proving soundness of a space
efficient semantics: to be sound, the space-efficient semantics must
recover a notion of congruence for checking. {\iffull In his manifest setting,
he calls it \textit{cast congruence}; since CPCF uses contract
monitors, we call it \textit{monitor congruence}. \fi}

\begin{lemma}[Monitor congruence (single step)]
  \label{lem:semonitorcongruencestep}
  If $ \emptyset   \vdash   \ottnt{e_{{\mathrm{1}}}}  :  \ottnt{T} $ and $\emptyset  \vdash  \ottnt{c}  \ottsym{:}  \ottnt{T}$ and $\ottnt{e_{{\mathrm{1}}}} \,  \longrightarrow _{  \mathsf{E}  }  \, \ottnt{e_{{\mathrm{2}}}}$,
  then $ \mathsf{mon}( \ottnt{c} ,  \ottnt{e_{{\mathrm{1}}}} )  \,  \longrightarrow ^{*}_{  \mathsf{E}  }  \, \ottnt{w}$ iff $ \mathsf{mon}( \ottnt{c} ,  \ottnt{e_{{\mathrm{2}}}} )  \,  \longrightarrow ^{*}_{  \mathsf{E}  }  \, \ottnt{w}$.
  \begin{proof}
    By cases on the step taken to find $\ottnt{e_{{\mathrm{1}}}} \,  \longrightarrow _{  \mathsf{E}  }  \, \ottnt{e_{{\mathrm{2}}}}$.
    In the easy case, there's no joining of coercions and the same
    rule can apply in both derivations. In the more interesting case,
    two contract monitors join.
    In either case, it suffices to show that the terms are ultimately
    confluent, since determinism \iffull(Lemma~\ref{lem:secpcfdeterministic})\fi
    will do the rest.
{\iffull
    \paragraph{Join-free reductions}
    \begin{itemize}
    \item[(\E{Fix})] We give this case explicitly, as the original proof
      in Greenberg~\cite{Greenberg15space} doesn't talk about
      fixpoints.

      We must show $ \mathsf{mon}( \ottnt{c} ,   \mu ( \mathit{x} \mathord{:} \ottnt{T} ) . ~  \ottnt{e}  )  \,  \longrightarrow ^{*}_{  \mathsf{E}  }  \, \ottnt{w}$ iff $ \mathsf{mon}( \ottnt{c} ,   \ottnt{e}  [   \mu ( \mathit{x} \mathord{:} \ottnt{T} ) . ~  \ottnt{e}  / \mathit{x}  ]  ) $.  The former steps to the latter
      immediately by \E{MonC} and \E{Fix}, so we are done by
      determinism (Lemma~\ref{lem:secpcfdeterministic}).
    \end{itemize}
    The following cases are all similar: \E{Delta}, \E{Beta},
    \E{IfTrue}, \E{IfFalse}, \E{AppL}, \E{AppR}, \E{OpL}, \E{OpR},
    \E{If}, \E{MonLabel}, \E{MonCApp}, and all of the \E{*Raise} rules.

    A word about \E{MonCApp}: one might think that this is a joining
    reduction, but in fact it is not: if we can unwrap in the inner
    step, then the function proxy isn't able to join with the outer
    monitor.

    \paragraph{Joining reductions}
    \begin{itemize}
    \item[(\E{MonCNil})] We must show that \[  \mathsf{mon}( \ottnt{c} ,   \mathsf{mon}( \mathsf{nil} ,  \ottnt{v_{{\mathrm{1}}}} )  )  \,  \longrightarrow ^{*}_{  \mathsf{E}  }  \, \ottnt{w} \Iff  \mathsf{mon}( \ottnt{c} ,  \ottnt{v_{{\mathrm{1}}}} )  \,  \longrightarrow ^{*}_{  \mathsf{E}  }  \, \ottnt{w}. \] The terms are confluent:
      \[  \mathsf{mon}( \ottnt{c} ,   \mathsf{mon}( \mathsf{nil} ,  \ottnt{v_{{\mathrm{1}}}} )  )  \,  \longrightarrow _{  \mathsf{E}  }  \,  \mathsf{mon}(  \mathsf{join} ( \mathsf{nil} , \ottnt{c} )  ,  \ottnt{v_{{\mathrm{1}}}} )  =
       \mathsf{mon}( \ottnt{c} ,  \ottnt{v_{{\mathrm{1}}}} ) . \]

    \item[(\E{MonCPred})] We must show that:
      \[ \begin{array}{c}
         \mathsf{mon}( \ottnt{r_{{\mathrm{2}}}} ,   \mathsf{mon}(  (   \mathsf{pred}^{ \ottnt{l} }_{ \sigma }( \ottnt{e} )  ; \ottnt{r_{{\mathrm{1}}}}  )  ,  \ottnt{v_{{\mathrm{1}}}} )  )  \,  \longrightarrow ^{*}_{  \mathsf{E}  }  \, \ottnt{w} \\ \Iff  \\
         \mathsf{mon}( \ottnt{r_{{\mathrm{2}}}} ,   \mathsf{if} ~   (  \sigma  \ottsym{(}  \ottnt{e}  \ottsym{)}  ~  \ottnt{v_{{\mathrm{1}}}}  )   ~   \mathsf{mon}( \ottnt{r_{{\mathrm{1}}}} ,  \ottnt{v_{{\mathrm{1}}}} )   ~~   \mathsf{err}^ \ottnt{l}   )  \,  \longrightarrow ^{*}_{  \mathsf{E}  }  \, \ottnt{w}
      \end{array}
      \]

      The left-hand side steps:
      \[ \begin{array}{l}
         \mathsf{mon}(  \mathsf{join} (  (   \mathsf{pred}^{ \ottnt{l} }_{ \sigma }( \ottnt{e} )  ; \ottnt{r_{{\mathrm{1}}}}  )  , \ottnt{r_{{\mathrm{2}}}} )  ,  \ottnt{v_{{\mathrm{1}}}} )  \,  \longrightarrow _{  \mathsf{C}  }  \, {} \\   \mathsf{mon}(   \mathsf{pred}^{ \ottnt{l} }_{ \sigma }( \ottnt{e} )  ;  \mathsf{drop} (  \mathsf{join} ( \ottnt{r_{{\mathrm{1}}}} , \ottnt{r_{{\mathrm{2}}}} )  ,  \mathsf{pred}_{ \sigma }( \ottnt{e} )  )   ,  \ottnt{v_{{\mathrm{1}}}} ) 
      \end{array} \]
      by \E{MonCJoin}, which in turn steps to \[  \mathsf{if} ~   (  \sigma  \ottsym{(}  \ottnt{e}  \ottsym{)}  ~  \ottnt{v_{{\mathrm{1}}}}  )   ~   \mathsf{mon}(  \mathsf{drop} (  \mathsf{join} ( \ottnt{r_{{\mathrm{1}}}} , \ottnt{r_{{\mathrm{2}}}} )  ,  \mathsf{pred}_{ \sigma }( \ottnt{e} )  )  ,  \ottnt{v_{{\mathrm{1}}}} )   ~~   \mathsf{err}^ \ottnt{l}  . \] By
      cases on the behavior of $ \sigma  \ottsym{(}  \ottnt{e}  \ottsym{)}  ~  \ottnt{v_{{\mathrm{1}}}} $.

      \begin{itemize}
      \item If $ \sigma  \ottsym{(}  \ottnt{e}  \ottsym{)}  ~  \ottnt{v_{{\mathrm{1}}}}  \,  \longrightarrow ^{*}_{  \mathsf{E}  }  \,  \mathsf{true} $, then both sides reduce by
        \E{IfTrue} (using \E{MonC} on the right). The left hand side
        reduces to $ \mathsf{mon}(  \mathsf{drop} (  \mathsf{join} ( \ottnt{r_{{\mathrm{1}}}} , \ottnt{r_{{\mathrm{2}}}} )  ,  \mathsf{pred}_{ \sigma }( \ottnt{e} )  )  ,  \ottnt{v_{{\mathrm{1}}}} ) $ while
        the right hand side is $ \mathsf{mon}(  \mathsf{join} ( \ottnt{r_{{\mathrm{1}}}} , \ottnt{r_{{\mathrm{2}}}} )  ,  \ottnt{v_{{\mathrm{1}}}} ) $. By
        idempotence of contracts (Lemma~\ref{lem:contractidempotent}).

      \item If $ \sigma  \ottsym{(}  \ottnt{e}  \ottsym{)}  ~  \ottnt{v_{{\mathrm{1}}}}  \,  \longrightarrow ^{*}_{  \mathsf{E}  }  \,  \mathsf{false} $, then both sides reduce
        by \E{IfFalse} (using \E{MonC} on the right); the right
        reduces by \E{MonRaise}, and both sides are $ \mathsf{err}^ \ottnt{l} $.

      \item Finally, if $ \sigma  \ottsym{(}  \ottnt{e}  \ottsym{)}  ~  \ottnt{v_{{\mathrm{1}}}} $ diverges, then both terms
        diverge.
      \end{itemize}
    \item[(\E{MonC})] We must show that:
      \[  \mathsf{mon}( \ottnt{c} ,   \mathsf{mon}( \ottnt{c'} ,  \ottnt{e} )  )  \,  \longrightarrow ^{*}_{  \mathsf{E}  }  \, \ottnt{w} \Iff  \mathsf{mon}( \ottnt{c} ,   \mathsf{mon}( \ottnt{c'} ,  \ottnt{e'} )  )  \,  \longrightarrow ^{*}_{  \mathsf{E}  }  \, \ottnt{w} \]
      given that $\ottnt{e} \,  \longrightarrow _{  \mathsf{E}  }  \, \ottnt{e'}$ and knowing that $\ottnt{e}$ isn't another
      monitor. Both sides step by \E{MonCJoin}, and then the left
      steps by \E{Mon}, to find $ \mathsf{mon}(  \mathsf{join} ( \ottnt{c'} , \ottnt{c} )  ,  \ottnt{e'} ) $.

    \item[(\E{MonCJoin})] We must show that:
      \[  \mathsf{mon}( \ottnt{c} ,   \mathsf{mon}( \ottnt{c_{{\mathrm{2}}}} ,   \mathsf{mon}( \ottnt{c_{{\mathrm{1}}}} ,  \ottnt{e} )  )  )  \,  \longrightarrow ^{*}_{  \mathsf{E}  }  \, \ottnt{w} \Iff  \mathsf{mon}( \ottnt{c} ,   \mathsf{mon}(  \mathsf{join} ( \ottnt{c_{{\mathrm{1}}}} , \ottnt{c_{{\mathrm{2}}}} )  ,  \ottnt{e} )  )  \,  \longrightarrow ^{*}_{  \mathsf{E}  }  \, \ottnt{w} \]

      The left-hand side joins to $ \mathsf{mon}(  \mathsf{join} ( \ottnt{c_{{\mathrm{1}}}} ,  \mathsf{join} ( \ottnt{c} , \ottnt{c_{{\mathrm{2}}}} )  )  ,  \ottnt{e} ) $
      while the right-hand side joins to
      $ \mathsf{mon}(  \mathsf{join} (  \mathsf{join} ( \ottnt{c_{{\mathrm{1}}}} , \ottnt{c_{{\mathrm{2}}}} )  , \ottnt{c} )  ,  \ottnt{e} ) $. We have confluence, by
      associativity of join (Lemma~\ref{lem:joinassociative}).

    \item[(\E{MonCRaise})] We must show that:
      \[  \mathsf{mon}( \ottnt{c} ,   \mathsf{mon}( \ottnt{c'} ,   \mathsf{err}^ \ottnt{l}  )  )  \,  \longrightarrow ^{*}_{  \mathsf{E}  }  \, \ottnt{w} \Iff  \mathsf{mon}( \ottnt{c} ,   \mathsf{err}^ \ottnt{l}  )  \,  \longrightarrow ^{*}_{  \mathsf{E}  }  \, \ottnt{w} \]
      The left-hand steps by \E{MonCJoin}, and then both sides can
      step by \E{MonCRaise} to $ \mathsf{err}^ \ottnt{l} $.
    \end{itemize}
\fi}
  \end{proof}
\end{lemma}

{\iffull
\begin{lemma}[Monitor congruence]
  \label{lem:semonitorcongruence}
  If
  \begin{itemize}
  \item $ \emptyset   \vdash   \ottnt{e_{{\mathrm{1}}}}  :  \ottnt{T} $,
  \item $\emptyset  \vdash  \ottnt{c}  \ottsym{:}  \ottnt{T}$, and
  \item $\ottnt{e_{{\mathrm{1}}}} \,  \longrightarrow ^{*}_{  \mathsf{E}  }  \, \ottnt{e_{{\mathrm{2}}}}$,
  \end{itemize}
  then $ \mathsf{mon}( \ottnt{c} ,  \ottnt{e_{{\mathrm{1}}}} )  \,  \longrightarrow ^{*}_{  \mathsf{E}  }  \, \ottnt{w}$ iff $ \mathsf{mon}( \ottnt{c} ,  \ottnt{e_{{\mathrm{2}}}} )  \,  \longrightarrow ^{*}_{  \mathsf{E}  }  \, \ottnt{e_{{\mathrm{2}}}}$. Diagrammatically:
  \begin{center}
  \begin{tikzpicture}[description/.style={fill=white,inner sep=2pt},align at top]
    \matrix (m) [matrix of math nodes, row sep=4pt, nodes in empty cells,
                 text height=1.5ex, text depth=0.25ex]
    {
      \ottnt{e_{{\mathrm{1}}}}        & & \ottnt{e_{{\mathrm{2}}}} \\
                    & \Downarrow & \\
       \mathsf{mon}( \ottnt{c} ,  \ottnt{e_{{\mathrm{1}}}} )  & &  \mathsf{mon}( \ottnt{c} ,  \ottnt{e_{{\mathrm{2}}}} )  \\[20pt]
      & \ottnt{w} & \\
    };

    \path[->] (m-1-1) edge[E*] (m-1-3);
    \path[->] (m-3-1) edge[E*] (m-4-2.north west);
    \path[->] (m-3-3) edge[E*] (m-4-2.north east);
  \end{tikzpicture}
  \end{center}
  \begin{proof}
    By induction on the derivation of $\ottnt{e_{{\mathrm{1}}}} \,  \longrightarrow ^{*}_{  \mathsf{E}  }  \, \ottnt{e_{{\mathrm{2}}}}$, using
    single-step monitor congruence
    (Lemma~\ref{lem:semonitorcongruencestep}).
  \end{proof}
\end{lemma}
\fi}

It is particularly satisfying that the key property for showing
soundness of space efficiency can be proved independently of the
inefficient semantics. Implementors can work entirely in the
context of the space-efficient semantics, knowing that congruence ensures soundness.
We show the observational equivalence of \CPCFc and \CPCFe by logical
relations (Fig.~\ref{fig:lr}), which gives us contextual
equivalence---the strongest equivalence we could ask for.

\begin{figure}[t]
  \hdr{Result rules}{\qquad \fbox{$\ottnt{e_{{\mathrm{1}}}}  \sim  \ottnt{e_{{\mathrm{2}}}}  \ottsym{:}  \ottnt{T}$}}
  \[ \begin{array}{rcl}
  \ottnt{k}  \sim  \ottnt{k}  \ottsym{:}  \ottnt{B} &\iff&  \mathsf{ty} ( \ottnt{k} )   \ottsym{=}  \ottnt{B} \\
  \ottnt{v_{{\mathrm{11}}}}  \sim  \ottnt{v_{{\mathrm{21}}}}  \ottsym{:}   \ottnt{T_{{\mathrm{1}}}} \mathord{ \rightarrow } \ottnt{T_{{\mathrm{2}}}}  &\iff&  \forall  \ottnt{e_{{\mathrm{12}}}}  \sim  \ottnt{e_{{\mathrm{22}}}}  \ottsym{:}  \ottnt{T_{{\mathrm{1}}}}  . ~   \ottnt{v_{{\mathrm{11}}}}  ~  \ottnt{e_{{\mathrm{12}}}}   \simeq   \ottnt{v_{{\mathrm{21}}}}  ~  \ottnt{e_{{\mathrm{22}}}}   \ottsym{:}  \ottnt{T_{{\mathrm{2}}}}  \\
   \mathsf{err}^ \ottnt{l}   \sim   \mathsf{err}^ \ottnt{l}   \ottsym{:}  \ottnt{T}
  \end{array} \]

  \hdr{Term rules}{\qquad \fbox{$\ottnt{e_{{\mathrm{1}}}}  \simeq  \ottnt{e_{{\mathrm{2}}}}  \ottsym{:}  \ottnt{T}$}}
  \[ \ottnt{e_{{\mathrm{1}}}}  \simeq  \ottnt{e_{{\mathrm{2}}}}  \ottsym{:}  \ottnt{T} \iff
      (\ottnt{e_{{\mathrm{1}}}}\text{ diverges} \mathrel{\wedge} \ottnt{e_{{\mathrm{2}}}}\text{ diverges}) \vee
      (\ottnt{e_{{\mathrm{1}}}} \,  \longrightarrow ^{*}_{  \mathsf{C}  }  \, \ottnt{w_{{\mathrm{1}}}} \wedge \ottnt{e_{{\mathrm{2}}}} \,  \longrightarrow ^{*}_{  \mathsf{E}  }  \, \ottnt{w_{{\mathrm{2}}}} \wedge \ottnt{w_{{\mathrm{1}}}}  \sim  \ottnt{w_{{\mathrm{2}}}}  \ottsym{:}  \ottnt{T})
  \]

  \hdr{Contract rules (invariant relation)}{\qquad \fbox{$\mathrm{\Gamma}  \vdash  \ottnt{C_{{\mathrm{1}}}}  \sim  \ottnt{C_{{\mathrm{2}}}}  \ottsym{:}  \ottnt{T}$}}

  \[ \begin{array}{rcl}
    \mathrm{\Gamma}  \vdash   \mathsf{pred}_{ \sigma_{{\mathrm{1}}} }( \ottnt{e_{{\mathrm{1}}}} )   \sim   \mathsf{pred}_{ \sigma_{{\mathrm{2}}} }( \ottnt{e_{{\mathrm{2}}}} )   \ottsym{:}  \ottnt{B} &\iff&  \mathrm{\Gamma}  \vdash  \sigma_{{\mathrm{1}}}  \quad   ( \ottnt{e_{{\mathrm{1}}}} )   \simeq  \sigma_{{\mathrm{2}}}  \ottsym{(}  \ottnt{e_{{\mathrm{2}}}}  \ottsym{)}  \ottsym{:}   \ottnt{B} \mathord{ \rightarrow }  \mathsf{Bool}    \\
    \mathrm{\Gamma}  \vdash   \mathit{x} \mathord{:} \ottnt{C_{{\mathrm{11}}}} \mapsto \ottnt{C_{{\mathrm{12}}}}   \sim   \mathit{x} \mathord{:} \ottnt{C_{{\mathrm{21}}}} \mapsto \ottnt{C_{{\mathrm{22}}}}   \ottsym{:}   \ottnt{T_{{\mathrm{1}}}} \mathord{ \rightarrow } \ottnt{T_{{\mathrm{2}}}}  &\iff& \\
    \multicolumn{3}{r}{\mathrm{\Gamma}  \vdash  \ottnt{C_{{\mathrm{11}}}}  \sim  \ottnt{C_{{\mathrm{21}}}}  \ottsym{:}  \ottnt{T_{{\mathrm{1}}}} \wedge  \mathrm{\Gamma} , \mathit{x} \mathord{:} \ottnt{T_{{\mathrm{1}}}}   \vdash  \ottnt{C_{{\mathrm{12}}}}  \sim  \ottnt{C_{{\mathrm{22}}}}  \ottsym{:}  \ottnt{T_{{\mathrm{2}}}}}
  \end{array} \]

  \hdr{Closing substitutions and open terms}{\qquad \fbox{$\mathrm{\Gamma}  \models  \delta$} \qquad \fbox{$\mathrm{\Gamma}  \vdash  \ottnt{e_{{\mathrm{1}}}}  \simeq  \ottnt{e_{{\mathrm{2}}}}  \ottsym{:}  \ottnt{T}$}}

  \[ \begin{array}{r@{~}c@{~}l}
    \mathrm{\Gamma}  \models  \delta &\iff&  \forall   \mathit{x}  \in   \operatorname{dom} ( \mathrm{\Gamma} )    . ~  \delta_{{\mathrm{1}}}  \ottsym{(}  \mathit{x}  \ottsym{)}  \sim  \delta_{{\mathrm{2}}}  \ottsym{(}  \mathit{x}  \ottsym{)}  \ottsym{:}   \mathrm{\Gamma} ( \mathit{x} )   \\
    \mathrm{\Gamma}  \vdash  \ottnt{e_{{\mathrm{1}}}}  \simeq  \ottnt{e_{{\mathrm{2}}}}  \ottsym{:}  \ottnt{T} &\iff&  \forall  \mathrm{\Gamma}  \models  \delta  . ~  \delta_{{\mathrm{1}}}  \ottsym{(}  \ottnt{e_{{\mathrm{1}}}}  \ottsym{)}  \simeq  \delta_{{\mathrm{2}}}  \ottsym{(}  \ottnt{e_{{\mathrm{2}}}}  \ottsym{)}  \ottsym{:}  \ottnt{T} 
  \end{array} \]

  \caption{Logical relation between classic and space-efficient CPCF}
  \label{fig:lr}
\end{figure}

\begin{lemma}[Similar contracts are logically related]
  \label{lem:lrcontract}
  If $\mathrm{\Gamma}  \vdash  \ottnt{C_{{\mathrm{1}}}}  \sim  \ottnt{C_{{\mathrm{2}}}}  \ottsym{:}  \ottnt{T}$ and $\mathrm{\Gamma}  \vdash  \ottnt{v_{{\mathrm{1}}}}  \simeq  \ottnt{v_{{\mathrm{2}}}}  \ottsym{:}  \ottnt{T}$ then $\mathrm{\Gamma}  \vdash   \mathsf{mon}^{ \ottnt{l} }( \ottnt{C_{{\mathrm{1}}}} ,  \ottnt{v_{{\mathrm{1}}}} )   \simeq   \mathsf{mon}^{ \ottnt{l} }( \ottnt{C_{{\mathrm{2}}}} ,  \ottnt{v_{{\mathrm{2}}}} )   \ottsym{:}  \ottnt{T}$.
  \begin{proof}
    By induction on the type index of the invariant relation $\mathrm{\Gamma}  \vdash  \ottnt{C_{{\mathrm{1}}}}  \sim  \ottnt{C_{{\mathrm{2}}}}  \ottsym{:}  \ottnt{T}$.
    {\iffull

    Let $\mathrm{\Gamma}  \models  \delta$ be given; we must show that
    $\delta_{{\mathrm{1}}}  \ottsym{(}   \mathsf{mon}^{ \ottnt{l} }( \ottnt{C_{{\mathrm{1}}}} ,  \ottnt{v_{{\mathrm{1}}}} )   \ottsym{)}  \simeq  \delta_{{\mathrm{2}}}  \ottsym{(}   \mathsf{mon}^{ \ottnt{l} }( \ottnt{C_{{\mathrm{2}}}} ,  \ottnt{v_{{\mathrm{2}}}} )   \ottsym{)}  \ottsym{:}  \ottnt{T_{{\mathrm{2}}}}$.

    The right-hand side steps by \E{MonLabel} to
    $ \mathsf{mon}(  \mathsf{label} ^{ \ottnt{l} }(  \delta_{{\mathrm{2}}} ( \ottnt{C_{{\mathrm{2}}}} )  )  ,  \delta_{{\mathrm{2}}}  \ottsym{(}  \ottnt{e_{{\mathrm{2}}}}  \ottsym{)} ) $.

    If $\ottnt{T}  \ottsym{=}  \ottnt{B}$, then $\ottnt{C_{\ottmv{i}}}  \ottsym{=}   \mathsf{pred}_{ \sigma_{\ottmv{i}} }( \ottnt{e_{\ottmv{i}}} ) $, and each side reduces to
    \[  \mathsf{if} ~   (  \delta_{\ottmv{i}}  \ottsym{(}  \sigma_{\ottmv{i}}  \ottsym{(}  \ottnt{e_{\ottmv{i}}}  \ottsym{)}  \ottsym{)}  ~  \delta_{\ottmv{i}}  \ottsym{(}  \ottnt{v_{\ottmv{i}}}  \ottsym{)}  )   ~  \delta_{\ottmv{i}}  \ottsym{(}  \ottnt{v_{\ottmv{i}}}  \ottsym{)}  ~~   \mathsf{err}^ \ottnt{l}   \]
    (by \E{MonPred} on the left and
    \E{MonCPred} on the right). By the assumption in the invariant
    relation, the conditions of the if co-terminate. The two sides
    behave \textit{exactly} the same when the conditions diverge,
    raise blame, or return $ \mathsf{false} $. When they return $ \mathsf{true} $,
    they return related values.

    The more interesting case is when $\ottnt{T}  \ottsym{=}   \ottnt{T_{{\mathrm{1}}}} \mathord{ \rightarrow } \ottnt{T_{{\mathrm{2}}}} $; we must show that:
    \[ \delta_{{\mathrm{1}}}  \ottsym{(}   \mathsf{mon}^{ \ottnt{l} }(  \mathit{x} \mathord{:} \ottnt{C_{{\mathrm{11}}}} \mapsto \ottnt{C_{{\mathrm{12}}}}  ,  \ottnt{v_{{\mathrm{1}}}} )   \ottsym{)}  \sim  \delta_{{\mathrm{2}}}  \ottsym{(}   \mathsf{mon}(  \mathit{x} \mathord{:}  \mathsf{label} ^{ \ottnt{l} }( \ottnt{C_{{\mathrm{21}}}} )  \mapsto  \mathsf{label} ^{ \ottnt{l} }( \ottnt{C_{{\mathrm{22}}}} )   ,  \ottnt{v_{{\mathrm{2}}}} )   \ottsym{)}  \ottsym{:}  \ottnt{T_{{\mathrm{2}}}} \]
    The left-hand side is a value; the right-hand side may or may not
    be a value, depending on whether $\ottnt{v_{{\mathrm{2}}}}$ is proxied or not.
    \begin{itemize}
    \item[($\ottnt{v_{{\mathrm{2}}}} =  \lambda \mathit{x} \mathord{:} \ottnt{T_{{\mathrm{1}}}} .~  \ottnt{e_{{\mathrm{2}}}} $)] Both sides are values. To show
      the logical relation, we let arguments $\ottnt{v_{{\mathrm{12}}}}  \sim  \ottnt{v_{{\mathrm{22}}}}  \ottsym{:}  \ottnt{T_{{\mathrm{1}}}}$ be
      given and must show that
      \[ \begin{array}{l}  \delta_{{\mathrm{1}}}  \ottsym{(}   \mathsf{mon}^{ \ottnt{l} }(  \mathit{x} \mathord{:} \ottnt{C_{{\mathrm{11}}}} \mapsto \ottnt{C_{{\mathrm{12}}}}  ,  \ottnt{v_{{\mathrm{1}}}} )   \ottsym{)}  ~  \ottnt{v_{{\mathrm{12}}}}   \simeq   {} \\  \delta_{{\mathrm{2}}}  \ottsym{(}   \mathsf{mon}(  \mathit{x} \mathord{:}  \mathsf{label} ^{ \ottnt{l} }( \ottnt{C_{{\mathrm{21}}}} )  \mapsto  \mathsf{label} ^{ \ottnt{l} }( \ottnt{C_{{\mathrm{22}}}} )   ,  \ottnt{v_{{\mathrm{2}}}} )   \ottsym{)}  ~  \ottnt{v_{{\mathrm{22}}}}   \ottsym{:}  \ottnt{T_{{\mathrm{2}}}}. \end{array} \]
      Each side unwraps (\E{MonApp} and \E{MonCApp}). Since the LR is
      closed under evaluation, we can use the IH on the domain and
      monitor (reducing on the right to convert to a monitor of
      labeled contracts). Note that there is no codomain monitor
      unless it's in $\ottnt{C_{\ottmv{i}\,{\mathrm{2}}}}$'s closing substitution already.

    \item[($\ottnt{v_{{\mathrm{2}}}} =  \mathsf{mon}(  \mathit{x} \mathord{:} \ottnt{c_{{\mathrm{31}}}} \mapsto \ottnt{c_{{\mathrm{32}}}}  ,  \ottnt{v'_{{\mathrm{2}}}} ) $)] For brevity, let
      $\ottnt{c_{{\mathrm{2}}\,\ottmv{i}}}  \ottsym{=}   \mathsf{label} ^{ \ottnt{l} }( \ottnt{C_{{\mathrm{2}}\,\ottmv{i}}} ) $. The right-hand side steps by
      \E{MonCJoin}:

      \[ \begin{array}{l}
        \delta_{{\mathrm{2}}}  \ottsym{(}   \mathsf{mon}(  \mathit{x} \mathord{:} \ottnt{c_{{\mathrm{21}}}} \mapsto \ottnt{c_{{\mathrm{22}}}}  ,   \mathsf{mon}(  \mathit{x} \mathord{:} \ottnt{c_{{\mathrm{31}}}} \mapsto \ottnt{c_{{\mathrm{32}}}}  ,  \ottnt{v'_{{\mathrm{2}}}} )  )   \ottsym{)} \,  \longrightarrow _{  \mathsf{E}  }  \, {} \\  \quad  \delta_{{\mathrm{2}}}  \ottsym{(}   \mathsf{mon}(  \mathsf{join} (  \mathit{x} \mathord{:} \ottnt{c_{{\mathrm{31}}}} \mapsto \ottnt{c_{{\mathrm{32}}}}  ,  \mathit{x} \mathord{:} \ottnt{c_{{\mathrm{21}}}} \mapsto \ottnt{c_{{\mathrm{22}}}}  )  ,  \ottnt{v'_{{\mathrm{2}}}} )   \ottsym{)} \end{array}
      \]

      which is $ \mathsf{mon}(  \mathit{x} \mathord{:}  \delta_{{\mathrm{2}}} (  \mathsf{join} ( \ottnt{c_{{\mathrm{21}}}} , \ottnt{c_{{\mathrm{31}}}} )  )  \mapsto  \delta_{{\mathrm{2}}} (  \mathsf{join} ( \ottnt{c_{{\mathrm{32}}}} , \ottnt{c_{{\mathrm{22}}}} )  )   ,  \delta_{{\mathrm{2}}}  \ottsym{(}  \ottnt{v'_{{\mathrm{2}}}}  \ottsym{)} ) $.

      Let arguments $\ottnt{v_{{\mathrm{12}}}}  \sim  \ottnt{v_{{\mathrm{22}}}}  \ottsym{:}  \ottnt{T_{{\mathrm{1}}}}$ be given. Now we can unwrap
      each side (by \E{MonApp} and \E{MonCApp}, respectively); we must
      show:

      \[ \begin{array}{l}
         \mathsf{mon}^{ \ottnt{l} }(   \delta_{{\mathrm{1}}} ( \ottnt{C_{{\mathrm{12}}}} )   [  \ottnt{v_{{\mathrm{12}}}}  /  \mathit{x}  ]  ,   \delta_{{\mathrm{1}}}  \ottsym{(}  \ottnt{v_{{\mathrm{1}}}}  \ottsym{)}  ~   \mathsf{mon}^{ \ottnt{l} }(  \delta_{{\mathrm{1}}} ( \ottnt{C_{{\mathrm{11}}}} )  ,  \ottnt{v_{{\mathrm{12}}}} )   )   \simeq  {} \\   \mathsf{mon}(   \delta_{{\mathrm{2}}} (  \mathsf{join} (  \mathsf{wrap} ( \ottnt{c_{{\mathrm{32}}}} , \mathit{x} , \ottnt{c_{{\mathrm{21}}}} )  , \ottnt{c_{{\mathrm{22}}}} )  )   [  \ottnt{v_{{\mathrm{22}}}}  /  \mathit{x}  ]  ,    {} \\  \quad  \delta_{{\mathrm{2}}}  \ottsym{(}  \ottnt{v'_{{\mathrm{2}}}}  \ottsym{)}  ~   \mathsf{mon}(  \delta_{{\mathrm{2}}} (  \mathsf{join} ( \ottnt{c_{{\mathrm{21}}}} , \ottnt{c_{{\mathrm{31}}}} )  )  ,  \ottnt{v_{{\mathrm{22}}}} )    )   \ottsym{:}  \ottnt{T_{{\mathrm{2}}}} \end{array} \]

      We will use monitor congruence to resolve the extra contracts on the
      space-efficient side. Since the LR is closed under evaluation,
      we can step the right-hand side back to separate out the inner
      monitors, letting us recover something that behaves like
      $\ottnt{v_{{\mathrm{1}}}}$.

      That is, to show the goal above, we will show that:

      \[ \begin{array}{l}
         \mathsf{mon}^{ \ottnt{l} }(   \delta_{{\mathrm{1}}} ( \ottnt{C_{{\mathrm{12}}}} )   [  \ottnt{v_{{\mathrm{12}}}}  /  \mathit{x}  ]  ,   \delta_{{\mathrm{1}}}  \ottsym{(}  \ottnt{v_{{\mathrm{1}}}}  \ottsym{)}  ~   \mathsf{mon}^{ \ottnt{l} }(  \delta_{{\mathrm{1}}} ( \ottnt{C_{{\mathrm{11}}}} )  ,  \ottnt{v_{{\mathrm{12}}}} )   )   \simeq  {} \\   \mathsf{mon}(   \delta_{{\mathrm{2}}} ( \ottnt{c_{{\mathrm{22}}}} )   [  \ottnt{v_{{\mathrm{22}}}}  /  \mathit{x}  ]  ,   \delta_{{\mathrm{2}}}  \ottsym{(}   \mathsf{mon}(  \mathit{x} \mathord{:} \ottnt{c_{{\mathrm{31}}}} \mapsto \ottnt{c_{{\mathrm{32}}}}  ,  \ottnt{v'_{{\mathrm{2}}}} )   \ottsym{)}  ~   \mathsf{mon}(  \delta_{{\mathrm{2}}} ( \ottnt{c_{{\mathrm{21}}}} )  ,  \ottnt{v_{{\mathrm{22}}}} )   )   \ottsym{:}  \ottnt{T_{{\mathrm{2}}}}.
      \end{array} \]

      We break the check apart in order to apply the IH. Before we
      explain how to find this relation, why is proving that these two
      terms related sufficient to show that our original terms are
      related? It's sufficient because monitor congruence
      (Lemma~\ref{lem:semonitorcongruence}) allows us to consider each
      of the domain and codomain monitors separately, knowing that
      we'll get identical behavior.
      We'll see that the $ \mathsf{wrap} $ in our original term is
      accounted for by codomain monitor that appears when we go
      through step-by-step.

      By the IH, we know that $ \mathsf{mon}^{ \ottnt{l} }(  \delta_{{\mathrm{1}}} ( \ottnt{C_{{\mathrm{11}}}} )  ,  \ottnt{v_{{\mathrm{12}}}} )   \simeq   \mathsf{mon}(  \delta_{{\mathrm{2}}} ( \ottnt{c_{{\mathrm{21}}}} )  ,  \ottnt{v_{{\mathrm{22}}}} )   \ottsym{:}  \ottnt{T_{{\mathrm{1}}}}$. If these terms terminate or
      diverge, than so will our original right-hand side domain
      monitor of $ \mathsf{join} ( \ottnt{c_{{\mathrm{21}}}} , \ottnt{c_{{\mathrm{31}}}} ) $, by
      Lemma~\ref{lem:semonitorcongruence}. In that case we're done.

      If the terms yield values $\ottnt{v_{{\mathrm{13}}}}  \simeq  \ottnt{v_{{\mathrm{23}}}}  \ottsym{:}  \ottnt{T_{{\mathrm{1}}}}$, we can
      continue. Using monitor congruence to step inside of the
      codomain check (Lemma~\ref{lem:semonitorcongruence}), we have:

      \[ \begin{array}{rl}
                 &  \mathsf{mon}(   \delta_{{\mathrm{2}}} ( \ottnt{c_{{\mathrm{22}}}} )   [  \ottnt{v_{{\mathrm{22}}}}  /  \mathit{x}  ]  ,   \delta_{{\mathrm{2}}}  \ottsym{(}   \mathsf{mon}(  \mathit{x} \mathord{:} \ottnt{c_{{\mathrm{31}}}} \mapsto \ottnt{c_{{\mathrm{32}}}}  ,  \ottnt{v'_{{\mathrm{2}}}} )   \ottsym{)}  ~   \mathsf{mon}(  \delta_{{\mathrm{2}}} ( \ottnt{c_{{\mathrm{21}}}} )  ,  \ottnt{v_{{\mathrm{22}}}} )   )  \\
         \longrightarrow ^{*}_{  \mathsf{E}  } &  \mathsf{mon}(   \delta_{{\mathrm{2}}} ( \ottnt{c_{{\mathrm{22}}}} )   [  \ottnt{v_{{\mathrm{22}}}}  /  \mathit{x}  ]  ,   \delta_{{\mathrm{2}}}  \ottsym{(}   \mathsf{mon}(  \mathit{x} \mathord{:} \ottnt{c_{{\mathrm{31}}}} \mapsto \ottnt{c_{{\mathrm{32}}}}  ,  \ottnt{v'_{{\mathrm{2}}}} )   \ottsym{)}  ~  \ottnt{v_{{\mathrm{23}}}}  ) 
      \end{array} \]

      And we must show:

      \[ \begin{array}{l}
         \mathsf{mon}^{ \ottnt{l} }(   \delta_{{\mathrm{1}}} ( \ottnt{C_{{\mathrm{12}}}} )   [  \ottnt{v_{{\mathrm{12}}}}  /  \mathit{x}  ]  ,   \delta_{{\mathrm{1}}}  \ottsym{(}  \ottnt{v_{{\mathrm{1}}}}  \ottsym{)}  ~  \ottnt{v_{{\mathrm{13}}}}  )   \simeq  {} \\   \mathsf{mon}(   \delta_{{\mathrm{2}}} ( \ottnt{c_{{\mathrm{22}}}} )   [  \ottnt{v_{{\mathrm{22}}}}  /  \mathit{x}  ]  ,   \delta_{{\mathrm{2}}}  \ottsym{(}   \mathsf{mon}(  \mathit{x} \mathord{:} \ottnt{c_{{\mathrm{31}}}} \mapsto \ottnt{c_{{\mathrm{32}}}}  ,  \ottnt{v'_{{\mathrm{2}}}} )   \ottsym{)}  ~  \ottnt{v_{{\mathrm{23}}}}  )   \ottsym{:}  \ottnt{T_{{\mathrm{2}}}}.
      \end{array} \]

      We can step the right hand side to:

      \[  \mathsf{mon}(   \delta_{{\mathrm{2}}} ( \ottnt{c_{{\mathrm{22}}}} )   [  \ottnt{v_{{\mathrm{22}}}}  /  \mathit{x}  ]  ,   \mathsf{mon}(   \delta_{{\mathrm{2}}} ( \ottnt{c_{{\mathrm{32}}}} )   [  \ottnt{v_{{\mathrm{23}}}}  /  \mathit{x}  ]  ,   \delta_{{\mathrm{2}}}  \ottsym{(}  \ottnt{v'_{{\mathrm{2}}}}  \ottsym{)}  ~   \mathsf{mon}( \ottnt{c_{{\mathrm{31}}}} ,  \ottnt{v_{{\mathrm{23}}}} )   )  )  \]

      By monitor congruence, we know $ \mathsf{mon}(  \mathsf{join} ( \ottnt{c_{{\mathrm{21}}}} , \ottnt{c_{{\mathrm{31}}}} )  ,  \ottnt{v_{{\mathrm{22}}}} ) $ and
      $ \mathsf{mon}( \ottnt{c_{{\mathrm{31}}}} ,  \ottnt{v_{{\mathrm{23}}}} ) $ co-terminate: either they both diverge, both
      blame the same label, or reduce to a common value,
      $\ottnt{v_{{\mathrm{24}}}}$. We're done in all but the last case, where we must
      consider:

      \[  \mathsf{mon}(   \delta_{{\mathrm{2}}} ( \ottnt{c_{{\mathrm{22}}}} )   [  \ottnt{v_{{\mathrm{22}}}}  /  \mathit{x}  ]  ,   \mathsf{mon}(   \delta_{{\mathrm{2}}} ( \ottnt{c_{{\mathrm{32}}}} )   [  \ottnt{v_{{\mathrm{23}}}}  /  \mathit{x}  ]  ,   \delta_{{\mathrm{2}}}  \ottsym{(}  \ottnt{v'_{{\mathrm{2}}}}  \ottsym{)}  ~  \ottnt{v_{{\mathrm{24}}}}  )  )  \]

      Recall that $ \mathsf{mon}(  \delta_{{\mathrm{2}}} ( \ottnt{c_{{\mathrm{21}}}} )  ,  \ottnt{v_{{\mathrm{22}}}} )  \,  \longrightarrow ^{*}_{  \mathsf{E}  }  \, \ottnt{v_{{\mathrm{23}}}}$; so
      $ \mathsf{mon}(   \delta_{{\mathrm{2}}} ( \ottnt{c_{{\mathrm{32}}}} )   [  \ottnt{v_{{\mathrm{23}}}}  /  \mathit{x}  ]  ,   \delta_{{\mathrm{2}}}  \ottsym{(}  \ottnt{v'_{{\mathrm{2}}}}  \ottsym{)}  ~  \ottnt{v_{{\mathrm{24}}}}  ) $ is the same as
      $ \mathsf{mon}(   \delta_{{\mathrm{2}}} (  \mathsf{wrap} ( \ottnt{c_{{\mathrm{32}}}} , \mathit{x} , \ottnt{c_{{\mathrm{21}}}} )  )   [  \ottnt{v_{{\mathrm{22}}}}  /  \mathit{x}  ]  ,   \delta_{{\mathrm{2}}}  \ottsym{(}  \ottnt{v'_{{\mathrm{2}}}}  \ottsym{)}  ~  \ottnt{v_{{\mathrm{24}}}}  ) $---that is, the wrapping done in the join captures
      exactly the substitution that would occur if we evaluated each
      layer of function proxying step by step.
      To patch up the derivation, we use substitutivity
      (Definition~\ref{def:implaxioms}, part (\ref{implsubst})) to see
      that if the codomains merged before the substitution, they would
      also do so after the substitution.
      So our modified
      right-hand term is still in lock step with our original, where
      $\ottnt{c_{{\mathrm{32}}}}$ sees a value monitored by $\ottnt{c_{{\mathrm{21}}}}$ but $\ottnt{c_{{\mathrm{22}}}}$
      doesn't.

      We've assumed that $\delta_{{\mathrm{1}}}  \ottsym{(}  \ottnt{v_{{\mathrm{1}}}}  \ottsym{)}  \simeq  \delta_{{\mathrm{2}}}  \ottsym{(}  \ottnt{v_{{\mathrm{2}}}}  \ottsym{)}  \ottsym{:}   \ottnt{T_{{\mathrm{1}}}} \mathord{ \rightarrow } \ottnt{T_{{\mathrm{2}}}} $, so
      we know that $ \delta_{{\mathrm{1}}}  \ottsym{(}  \ottnt{v_{{\mathrm{1}}}}  \ottsym{)}  ~  \ottnt{v_{{\mathrm{13}}}}   \simeq   \mathsf{mon}(   \delta ( \ottnt{c_{{\mathrm{32}}}} )   [  \ottnt{v_{{\mathrm{23}}}}  /  \mathit{x}  ]  ,   \delta  \ottsym{(}  \ottnt{v'_{{\mathrm{2}}}}  \ottsym{)}  ~  \ottnt{v_{{\mathrm{24}}}}  )   \ottsym{:}  \ottnt{T_{{\mathrm{2}}}}$ by
      back-evaluating the right-hand side a few steps. If these terms
      diverge or go to errors together, we are done. Otherwise, they
      evaluate to values $\ottnt{v_{{\mathrm{10}}}}  \simeq  \ottnt{v_{{\mathrm{20}}}}  \ottsym{:}  \ottnt{T_{{\mathrm{2}}}}$. It remains to see
      that:

      \[  \mathsf{mon}^{ \ottnt{l} }(   \delta_{{\mathrm{1}}} ( \ottnt{C_{{\mathrm{12}}}} )   [  \ottnt{v_{{\mathrm{12}}}}  /  \mathit{x}  ]  ,  \ottnt{v_{{\mathrm{10}}}} )   \simeq   \mathsf{mon}(   \delta_{{\mathrm{2}}} ( \ottnt{c_{{\mathrm{22}}}} )   [  \ottnt{v_{{\mathrm{22}}}}  /  \mathit{x}  ]  ,  \ottnt{v_{{\mathrm{20}}}} )   \ottsym{:}  \ottnt{T_{{\mathrm{2}}}} \]

      which we have by the IH. We can return to our original term and
      see that $\ottnt{v_{{\mathrm{20}}}}$ is what will be given to what's left of
      $\ottnt{c_{{\mathrm{22}}}}$ after joining with $ \mathsf{wrap} ( \ottnt{c_{{\mathrm{32}}}} , \mathit{x} , \ottnt{c_{{\mathrm{21}}}} ) $---monitor
      congruence let's us know that these monitors will evaluate the
      same merged and unmerged.
    \end{itemize}
    \fi}
  \end{proof}
\end{lemma}

\begin{lemma}[Unwinding]
  \label{lem:unwinding}
  If $ \emptyset   \vdash    \mu ( \mathit{x} \mathord{:} \ottnt{T} ) . ~  \ottnt{e}   :  \ottnt{T} $, then $ \mu ( \mathit{x} \mathord{:} \ottnt{T} ) . ~  \ottnt{e}  \,  \longrightarrow ^{*}_{ \mathsf{m} }  \, \ottnt{w}$ iff
  there exists an $n$ such that unrolling the fixpoint only $n$ times
  converges to the same value, i.e., $ \ottnt{e}  [   \mu ( \mathit{x} \mathord{:} \ottnt{T} ) . ~     \dots  ~  \ottnt{e}   [   \mu ( \mathit{x} \mathord{:} \ottnt{T} ) . ~  \ottnt{e}  / \mathit{x}  ]   ~  \dots   / \mathit{x}  ]  \,  \longrightarrow ^{*}_{ \mathsf{m} }  \, \ottnt{w}$.
  {\iffull
  \begin{proof}
    From left-to-right, by induction on the evaluation derivation,
    observing that there must be a finite number of unrollings.

    From right-to-left, by induction on $n$, observing that we can
    replace the substitution by its finite unrolling.
  \end{proof}
  \fi}
\end{lemma}

\begin{theorem}[\CPCFc and \CPCFe terms are logically related]
  \label{thm:lr}
  \begin{enumerate}
  \item If $ \mathrm{\Gamma}   \vdash   \ottnt{e}  :  \ottnt{T} $ as a source program then $\mathrm{\Gamma}  \vdash  \ottnt{e}  \simeq  \ottnt{e}  \ottsym{:}  \ottnt{T}$.
  \item If $\mathrm{\Gamma}  \vdash  \ottnt{C}  \ottsym{:}  \ottnt{T}$ as a source program then $\mathrm{\Gamma}  \vdash  \ottnt{C}  \sim  \ottnt{C}  \ottsym{:}  \ottnt{T}$.
  \end{enumerate}
  \begin{proof}
    By mutual induction on the typing relations.
    {\iffull
    \paragraph{Terms \fbox{$ \mathrm{\Gamma}   \vdash   \ottnt{e}  :  \ottnt{T} $}} ~ \\
    Each case proceeds by letting $\mathrm{\Gamma}  \models  \delta$ be given, and then
    showing that both sides co-terminate, i.e., both diverge, raise
    blame, or return related values.

    \begin{itemize}
    \item[(\T{Var})] By the definition of $\mathrm{\Gamma}  \models  \delta$.
    \item[(\T{Const})] By the definition of the LR.
    \item[(\T{Abs})] By the IHs, letting some argument be given to
      show that the bodies behave similarly.
    \item[(\T{Op})] By the IHs.
    \item[(\T{App})] By the IHs.
    \item[(\T{Mu})] We give this case in full, as it is new to the proof.

      Let $\mathrm{\Gamma}  \models  \delta$; we must show that $\delta_{{\mathrm{1}}}  \ottsym{(}   \mu ( \mathit{x} \mathord{:} \ottnt{T} ) . ~  \ottnt{e_{{\mathrm{1}}}}   \ottsym{)}  \simeq  \delta_{{\mathrm{2}}}  \ottsym{(}   \mu ( \mathit{x} \mathord{:} \ottnt{T} ) . ~  \ottnt{e_{{\mathrm{2}}}}   \ottsym{)}  \ottsym{:}  \ottnt{T}$. Each side reduces in a
      single step (\E{Fix}) to $ \delta_{\ottmv{i}}  \ottsym{(}  \ottnt{e_{\ottmv{i}}}  \ottsym{)}  [  \delta_{\ottmv{i}}  \ottsym{(}   \mu ( \mathit{x} \mathord{:} \ottnt{T} ) . ~  \ottnt{e_{\ottmv{i}}}   \ottsym{)} / \mathit{x}  ] $, which can be rearranged into
      $\delta_{\ottmv{i}}  \ottsym{(}   \ottnt{e_{\ottmv{i}}}  [   \mu ( \mathit{x} \mathord{:} \ottnt{T} ) . ~  \ottnt{e_{\ottmv{i}}}  / \mathit{x}  ]   \ottsym{)}$; we can rearrange this to
      something that looks like a dual substitution:
      $ \delta_{\ottmv{i}}  [   \mu ( \mathit{x} \mathord{:} \ottnt{T} ) . ~  \ottnt{e_{{\mathrm{1}}}}  ,  \mu ( \mathit{x} \mathord{:} \ottnt{T} ) . ~  \ottnt{e_{{\mathrm{2}}}}  / \mathit{x}  ]   \ottsym{(}  \ottnt{e_{\ottmv{i}}}  \ottsym{)}$. By unwinding
      (Lemma~\ref{lem:unwinding}), we can see that the fixpoint on
      each side either (a) diverges, or (b) converge to some related
      values in a finite unrolling and can be replaced by a finite
      unrolling. In the former case, we are done because both terms
      diverge; in the latter case, we have $ \mathrm{\Gamma} , \mathit{x} \mathord{:} \ottnt{T}   \models   \delta  [   \mu ( \mathit{x} \mathord{:} \ottnt{T} ) . ~  \ottnt{e_{{\mathrm{1}}}}  ,  \mu ( \mathit{x} \mathord{:} \ottnt{T} ) . ~  \ottnt{e_{{\mathrm{2}}}}  / \mathit{x}  ] $ and can apply the IH.
    \item[(\T{Mon})] By the IHs and Lemma~\ref{lem:lrcontract}.
    \item[(\T{MonC})] Contradictory---doesn't appear in source programs.
    \item[(\T{Blame})] Contradictory---doesn't appear in source programs.
    \end{itemize}

    \paragraph{Contracts \fbox{$\mathrm{\Gamma}  \vdash  \ottnt{C}  \ottsym{:}  \ottnt{T}$}} ~ \\
    \item[(\T{Pred})] By the IH.
    \item[(\T{Fun})] By the IHs.
      \fi}
  \end{proof}
\end{theorem}

\section{Bounds for space efficiency}
\label{sec:bounds}

We have seen that \CPCFe behaves the same as \CPCFc
(Theorem~\ref{thm:lr}), but is \CPCFe actually space efficient?
For programs that don't use dependency, yes! With dependency, the
story is more complicated.

\subsection{The simple case}
\label{sec:simplebounds}

Greenberg showed that for \textit{simple} contracts---without
dependency---we can put a bounds on space~\cite{Greenberg15space}. We
can recover his result in our more general framework.
Observe that a given source program $\ottnt{e}$ starts with a finite
number of predicate contracts in it. As $\ottnt{e}$ runs, no new
predicates appear (because dependent substitutions have no effect),
but predicates may accumulate in stacks. In the worst case, a
predicate stack could contain every predicate contract from the
original program $\ottnt{e}$ exactly once... but no more than that,
because joins remove redundancy!
Function contracts are also bounded: $\ottnt{e}$ starts out with function
contracts of a certain height, and evaluation can only shrink that
height. The leaves of function contracts are labeled with predicate
stacks, so the largest contract we could ever see is of maximum height
with maximal predicate stacks at every leaf.
As the program runs, abutting monitors are joined, giving us a bound
on the total number of monitors in a program (one per non-monitor AST
node).

We can make these ideas formal by first defining what we mean by ``all
the predicates in a program'', and then showing that evaluation
doesn't introduce predicates (Lemma~\ref{lem:predsstep}). We let
$ \mathsf{preds} ( \ottnt{e} ) $ be the set of predicates in a term, where a predicate
is represented as a pair of term and a closing substitution.

\begin{figure}[t]
    \hdr{Predicate extraction}{\qquad \fbox{$ \mathsf{preds} ( \ottnt{e} ) ,  \mathsf{preds} ( \ottnt{C} ) ,  \mathsf{preds} ( \ottnt{c} )  : \mathcal{P}(\ottnt{e} \times (\mathit{Var}\rightharpoonup\ottnt{e}))$}}
    \sidebysidesqueeze[.49][.5][0][t]
    {\[ \begin{array}{r@{~}c@{~}l}
       \mathsf{preds} ( \mathit{x} )  &=&  \emptyset  \\
       \mathsf{preds} ( \ottnt{k} )  &=&  \emptyset  \\
       \mathsf{preds} (  \lambda \mathit{x} \mathord{:} \ottnt{T} .~  \ottnt{e}  )  &=&  \mathsf{preds} ( \ottnt{e} )  \\
       \mathsf{preds} (  \mathsf{mon}^{ \ottnt{l} }( \ottnt{C} ,  \ottnt{e} )  )  &=&   \mathsf{preds} ( \ottnt{C} )   \cup   \mathsf{preds} ( \ottnt{e} )   \\
       \mathsf{preds} (  \mathsf{mon}( \ottnt{c} ,  \ottnt{e} )  )  &=&   \mathsf{preds} ( \ottnt{c} )   \cup   \mathsf{preds} ( \ottnt{e} )   \\
       \mathsf{preds} (  \ottnt{e_{{\mathrm{1}}}}  ~  \ottnt{e_{{\mathrm{2}}}}  )  &=&   \mathsf{preds} ( \ottnt{e_{{\mathrm{1}}}} )   \cup   \mathsf{preds} ( \ottnt{e_{{\mathrm{2}}}} )   \\
       \mathsf{preds} ( \ottnt{e_{{\mathrm{1}}}} \, \ottnt{op} \, \ottnt{e_{{\mathrm{2}}}} )  &=&   \mathsf{preds} ( \ottnt{e_{{\mathrm{1}}}} )   \cup   \mathsf{preds} ( \ottnt{e_{{\mathrm{2}}}} )   \\
       \mathsf{preds} (  \mathsf{if} ~  \ottnt{e_{{\mathrm{1}}}}  ~  \ottnt{e_{{\mathrm{2}}}}  ~~  \ottnt{e_{{\mathrm{3}}}}  )  &=& \\ \multicolumn{3}{r}{   \mathsf{preds} ( \ottnt{e_{{\mathrm{1}}}} )   \cup   \mathsf{preds} ( \ottnt{e_{{\mathrm{2}}}} )    \cup   \mathsf{preds} ( \ottnt{e_{{\mathrm{3}}}} )  } \\
       \mathsf{preds} (  \mathsf{err}^ \ottnt{l}  )  &=&  \emptyset  \\
    \end{array} \]}
    {\[ \begin{array}{r@{~}c@{~}l}
       \mathsf{preds} (  \mathsf{pred}_{ \sigma }( \ottnt{e} )  )  &=& \\
      \multicolumn{3}{r}{  \set{   ( \ottnt{e} , \sigma )   }   \cup   \mathsf{preds} ( \ottnt{e} )   \cup \bigcup_{ [  \mathit{x}  \mapsto  \ottnt{v}  ]  \in \sigma}  \mathsf{preds} ( \ottnt{v} ) } \\
       \mathsf{preds} (  \mathit{x} \mathord{:} \ottnt{C_{{\mathrm{1}}}} \mapsto \ottnt{C_{{\mathrm{2}}}}  )  &=&   \mathsf{preds} ( \ottnt{C_{{\mathrm{1}}}} )   \cup   \mathsf{preds} ( \ottnt{C_{{\mathrm{2}}}} )   \\
      && \\
       \mathsf{preds} ( \mathsf{nil} )  &=&  \emptyset  \\
       \mathsf{preds} (   \mathsf{pred}^{ \ottnt{l} }_{ \sigma }( \ottnt{e} )  ; \ottnt{r}  )  &=&   \set{   ( \ottnt{e} , \sigma )   }   \cup   \mathsf{preds} ( \ottnt{e} )   \cup {} \\
      \multicolumn{3}{r}{\bigcup_{ [  \mathit{x}  \mapsto  \ottnt{e'}  ]  \in \sigma}{ \mathsf{preds} ( \ottnt{e'} ) } \cup  \mathsf{preds} ( \ottnt{r} ) } \\
       \mathsf{preds} (  \mathit{x} \mathord{:} \ottnt{c_{{\mathrm{1}}}} \mapsto \ottnt{c_{{\mathrm{2}}}}  )  &=&   \mathsf{preds} ( \ottnt{c_{{\mathrm{1}}}} )   \cup   \mathsf{preds} ( \ottnt{c_{{\mathrm{2}}}} )  
    \end{array} \]}
    \hdr{Contract size}{\qquad \fbox{$P_{\ottnt{B}} : \mathbb{N}$} \qquad \fbox{$S_{\ottnt{B}} : \mathbb{N}$} \qquad \fbox{$ \mathsf{size} ( \ottnt{C} )  : \mathbb{N}$}}
    \[
      P_{\ottnt{B}} = \abs{\set{\ottnt{e} \, \in \,  \mathsf{preds} ( \ottnt{e} )  \mid \mathrm{\Gamma}  \vdash   \mathsf{pred}_{ \sigma }( \ottnt{e} )   \ottsym{:}  \ottnt{B}}} \qquad
      S_{\ottnt{B}} = L \cdot P_{\ottnt{B}} \cdot \log_2 P_{\ottnt{B}}
    \]
    \[
       \mathsf{size} (  \mathsf{pred}_{ \sigma }( \ottnt{e} )  )  = S_{\ottnt{B}} \text{ when $\emptyset  \vdash   \mathsf{pred}_{ \sigma }( \ottnt{e} )   \ottsym{:}  \ottnt{B}$} \qquad
       \mathsf{size} (  \mathit{x} \mathord{:} \ottnt{C_{{\mathrm{1}}}} \mapsto \ottnt{C_{{\mathrm{2}}}}  )  =  \mathsf{size} ( \ottnt{C_{{\mathrm{1}}}} )  +  \mathsf{size} ( \ottnt{C_{{\mathrm{2}}}} ) 
    \]
  \caption{Predicate extraction and contract size}
  \label{fig:extraction}
\end{figure}

{\iffull
\begin{lemma}
  \label{lem:predsdrop}
  $  \mathsf{preds} (  \mathsf{drop} ( \ottnt{r} ,  \mathsf{pred}( \ottnt{e} )  )  )   \subseteq   \mathsf{preds} ( \ottnt{r} )  $
  \begin{proof}
    By induction on the predicate stack $\ottnt{r}$. When $\ottnt{r}  \ottsym{=}  \mathsf{nil}$,
    the proof is trivial; otherwise, we observe that we only lose
    predicates from $\ottnt{r}$.
  \end{proof}
\end{lemma}

\begin{lemma}
  \label{lem:predslabel}
  $ \mathsf{preds} ( \ottnt{C} )   \ottsym{=}   \mathsf{preds} (  \mathsf{label} ^{ \ottnt{l} }( \ottnt{C} )  ) $
  \begin{proof}
    By induction on $\ottnt{C}$; the predicate case is immediate, while
    the arrow case uses the IHs.
  \end{proof}
\end{lemma}
\fi}

We say program $\ottnt{e}$ \textit{uses simple contracts} when all
predicates in $\ottnt{e}$ are closed and every predicate stack has no
redundancies. As such a program reduces, no new contracts appear (and
contracts may disappear).
{\iffull
Concretely, these requirements are that if $ \mathsf{pred}_{ \sigma }( \ottnt{e'} )  \in
\ottnt{e}$ or $ \mathsf{pred}^{ \ottnt{l} }_{ \sigma }( \ottnt{e'} )  \in \ottnt{e}$, we have $ \operatorname{fv} ( \ottnt{e'} )  =
 \emptyset $ and $\sigma =  \iota $. Restricting programs to simple
contracts is what lets us prove that substitution and joining don't
increase the set of predicates.
The ``no redundancy'' requirement means that if $  \mathsf{pred}^{ \ottnt{l} }_{ \sigma }( \ottnt{e_{{\mathrm{1}}}} )  ; \ottnt{r} 
\in \ottnt{e}$ then $ \forall   \mathsf{pred}^{ \ottnt{l'} }_{ \sigma' }( \ottnt{e_{{\mathrm{2}}}} )  \, \in \, \ottnt{r}  . ~   \mathsf{pred}_{ \sigma }( \ottnt{e_{{\mathrm{1}}}} )  \, \not \supset \,  \mathsf{pred}_{ \sigma' }( \ottnt{e_{{\mathrm{2}}}} )  $ (and that $\ottnt{r}$ itself has no
redundancy, etc.).
The following theorems only hold for programs that use simple contracts.
\fi}

\begin{lemma}
  \label{lem:predssubst}
  $  \mathsf{preds} (  \ottnt{e}  [  \ottnt{e'} / \mathit{x}  ]  )   \subseteq    \mathsf{preds} ( \ottnt{e} )   \cup   \mathsf{preds} ( \ottnt{e'} )   $
  \begin{proof}
    By induction on $\ottnt{e}$\iffull, using the absorptive property of set
    union\fi.
    If $\ottnt{e}$ is a predicate contract, it has no free
    variables (by assumption), so the substitution doesn't hold on to anything.
  \end{proof}
\end{lemma}

\begin{lemma}
  \label{lem:predsjoin}
  If $\emptyset  \vdash  \ottnt{c_{{\mathrm{1}}}}  \ottsym{:}  \ottnt{T}$ and $\emptyset  \vdash  \ottnt{c_{{\mathrm{2}}}}  \ottsym{:}  \ottnt{T}$ then
  $  \mathsf{preds} (  \mathsf{join} ( \ottnt{c_{{\mathrm{1}}}} , \ottnt{c_{{\mathrm{2}}}} )  )   \subseteq    \mathsf{preds} ( \ottnt{c_{{\mathrm{1}}}} )   \cup   \mathsf{preds} ( \ottnt{c_{{\mathrm{2}}}} )   $.
  \begin{proof}
    By induction on $\ottnt{c_{{\mathrm{1}}}}$\iffull\else, ignoring $ \mathsf{wrap} $'s substitution by
    Lemma~\ref{lem:predssubst}\fi.

    \iffull
    If $\ottnt{c_{{\mathrm{1}}}}$ is a predicate stack, then so must be $\ottnt{c_{{\mathrm{2}}}}$. If
    $\ottnt{c_{{\mathrm{1}}}}  \ottsym{=}  \mathsf{nil}$, the proof is immediate. If $\ottnt{c_{{\mathrm{1}}}}$ is a non-empty
    predicate stack, then by Lemma~\ref{lem:predsdrop}.

    In the function contract case, by the IHs. Since we're using
    simple contracts, we know that $ \mathsf{wrap} $ never does
    anything, because predicates have no free variables.
    \fi
  \end{proof}
\end{lemma}

\begin{lemma}[Reduction is non-increasing in simple predicates]
  \label{lem:predsstep}
  If $ \emptyset   \vdash   \ottnt{e}  :  \ottnt{T} $ and $\ottnt{e} \,  \longrightarrow _{ \mathsf{m} }  \, \ottnt{e'}$ then $  \mathsf{preds} ( \ottnt{e'} )   \subseteq   \mathsf{preds} ( \ottnt{e} )  $.
  \begin{proof}
    By induction on the step taken.
    {\iffull
    \begin{itemize}
    \item[(\E{Delta})] Immediate---no predicates.
    \item[(\E{Beta})] By substitution (Lemma~\ref{lem:predssubst}).
    \item[(\E{Fix})] Observe that the fixpoint operator introduces no
      preds other than its body; then by substitution
      (Lemma~\ref{lem:predssubst}).
    \item[(\E{IfTrue})] By definition: $  \mathsf{preds} ( \ottnt{e_{{\mathrm{2}}}} )   \subseteq     \mathsf{preds} (  \mathsf{true}  )   \cup   \mathsf{preds} ( \ottnt{e_{{\mathrm{2}}}} )    \cup   \mathsf{preds} ( \ottnt{e_{{\mathrm{3}}}} )   $.
    \item[(\E{IfFalse})] By definition, as above.
    \item[(\E{MonPred})] By definition: the right-hand side has no new
      predicates and possibly one fewer.
    \item[(\E{MonApp})] The substitution in the codomain is ignored (Lemma~\ref{lem:predssubst}),
      so the right-hand side is just a rearrangement of the left.
    \item[(\E{MonLabel})] By definition and the fact that labeling
      leaves the predicate set alone (Lemma~\ref{lem:predslabel}).
    \item[(\E{MonCNil})] By definition.
    \item[(\E{MonCPred})] As for \E{MonPred} above: the right-hand
      side has no new predicates and possibly one fewer.
    \item[(\E{MonCApp})] As for \E{MonApp}: we can rearrange our
      constituent parts---because substitutions are ignored
      (Lemma~\ref{lem:predssubst}).
    \item[(\E{MonCJoin})] By the monotonicity of $ \mathsf{join} $
      (Lemma~\ref{lem:predsjoin}), knowing that $ \mathsf{wrap} $ does
      nothing.
    \item[(\E{*L}, \E{*R}, \E{If}, \E{Mon}, \E{MonC})] By the IH.
    \item[(\E{*Raise})] Immediate---there are no predicates on the
      right-hand side.
    \end{itemize}
    \fi}
  \end{proof}
\end{lemma}

To compute the concrete bounds, we define $P_{\ottnt{B}}$ as the number of
distinct predicates at the base type $\ottnt{B}$. We can represent a
predicate stack at type $\ottnt{B}$ in $S_{\ottnt{B}}$ bits, where $L$ is the
number of bits needed to represent a blame label.
A given well typed contract $\emptyset  \vdash  \ottnt{C}  \ottsym{:}  \ottnt{T}$ can then be
represented in $ \mathsf{size} ( \ottnt{C} ) $ bits, where each predicate stacks are
represented is $S_{\ottnt{B}}$ bits and function types are represented as
trees of predicate stacks.
Finally, since reduction is non-increasing
(Lemma~\ref{lem:predsstep}), we can bound the amount of space used by
any contract by looking at the source program, $\ottnt{e}$: we can
represent all contracts in our program in at most $s = \max_{\ottnt{C} \in
  \ottnt{e}}  \mathsf{size} ( \ottnt{C} ) $ space---constant for a fixed source program.

Readers familiar with Greenberg's paper (and earlier work, like Herman
et al.~\cite{Herman07space}) will notice that our bounds are more precise, tracking the
number of holes in contracts per type ($ \mathsf{size} ( \ottnt{C} ) $) rather than
simply computing the largest conceivable type
($2^{\mathsf{height}(T)}$).

\subsection{The dependent case}
\label{sec:depbounds}

In the dependent case, we can't \textit{generally} bound the number of contracts by the
size of contracts used in the program. Consider the following term,
where $ n  \in \mathbb{N}$:
\[\begin{array}{l}
  \mathsf{let} ~  \mathsf{downTo}  =  \mu ( \mathit{f} \mathord{:}   \mathsf{Int}  \mathord{ \rightarrow }  \mathsf{Int}   ) . ~  {} \\  \quad   \mathsf{mon}^{ \ottnt{l} }(  \mathit{x} \mathord{:}  \mathsf{pred}(  \lambda \mathit{x} \mathord{:}  \mathsf{Int}  .~  \mathit{x}  \, \ge \, \ottsym{0} )  \mapsto  \mathsf{pred}(  \lambda \mathit{y} \mathord{:}  \mathsf{Int}  .~  \mathit{x}  \, \ge \, \mathit{y} )   ,  {} \\  \qquad   \lambda \mathit{x} \mathord{:}  \mathsf{Int}  .~   \mathsf{if} ~   ( \mathit{x} \,  =  \, \ottsym{0} )   ~  \ottsym{0}  ~~   (  \mathit{f}  ~   ( \mathit{x} \,  -  \, \ottsym{1} )   )    )   ~ \mathsf{in} \\
    \mathsf{downTo}   ~   n  
\end{array} \]
How many different contracts will appear in a run of this program? As
$ \mathsf{downTo} $ runs, we'll see $ n $ different forms of the predicate
$ \mathsf{pred}^{ \ottnt{l} }_{ \sigma_{\ottmv{i}} }(  \lambda \mathit{y} \mathord{:}  \mathsf{Int}  .~  \mathit{x}  \, \ge \, \mathit{y} ) $. We'll have one $\sigma_{\ottmv{n}} =
 [  \mathit{x}  \mapsto   n   ] $ on the first call, $\sigma_{{\ottmv{n}-1}} =  [  \mathit{x}  \mapsto   n  \,  -  \, \ottsym{1}  ] $ on the second call, and so on. But $ n $'s magnitude
doesn't affect our measure of the size of source program's contracts. The number
of distinct contracts that appear will be effectively unbounded.

In the simple case, we get bounds automatically, using the smallest
possible implication relation---syntactic equality. In the dependent
case, it's up to the programmer to identify implications that recover
space efficiency.
We can recover space efficiency for $ \mathsf{downTo} $ by saying
$ \mathsf{pred}_{ \sigma_{{\mathrm{1}}} }(  \lambda \mathit{y} \mathord{:}  \mathsf{Int}  .~  \mathit{x}  \, \ge \, \mathit{y} )  \, \supset \,  \mathsf{pred}_{ \sigma_{{\mathrm{2}}} }(  \lambda \mathit{y} \mathord{:}  \mathsf{Int}  .~  \mathit{x}  \, \ge \, \mathit{y} ) $ iff
$\sigma_{{\mathrm{1}}}  \ottsym{(}  \mathit{x}  \ottsym{)} \le \sigma_{{\mathrm{2}}}  \ottsym{(}  \mathit{x}  \ottsym{)}$. Then the codomain checks from
recursive calls will be able to join:
\[ \begin{array}{rcl}
    \mathsf{downTo}   ~   n   &  \longrightarrow ^{*}_{  \mathsf{E}  }  &  \mathsf{mon}^{ \ottnt{l} }(  \mathsf{pred}_{  [  \mathit{x}  \mapsto   n   ]  }( \dots )  ,  \dots )  \\
                 &  \longrightarrow ^{*}_{  \mathsf{E}  }  &  \mathsf{mon}^{ \ottnt{l} }(  \mathsf{pred}_{  [  \mathit{x}  \mapsto   n   ]  }( \dots )  ,   \mathsf{mon}^{ \ottnt{l} }(  \mathsf{pred}_{  [  \mathit{x}  \mapsto   n  \,  -  \, \ottsym{1}  ]  }( \dots )  ,  \dots )  )  \\
                 &  \longrightarrow ^{*}_{  \mathsf{E}  }  &  \mathsf{mon}^{ \ottnt{l} }(  \mathsf{pred}_{  [  \mathit{x}  \mapsto   n  \,  -  \, \ottsym{1}  ]  }( \dots )  ,  \dots ) 
\end{array} \]
Why are we able to recover space efficiency in this case? Because we
can come up with an easily decidable implication rule for our specific
predicates matching how our function checks narrower and narrower
properties as it recurses.

Recall the mutually recursive $ \mathsf{even} $/$ \mathsf{odd} $ example
(Fig.~\ref{fig:spaceleak}).  {\iffull
\[\begin{array}{lcl}
   \mathsf{let}~  \mathsf{odd}  &=&
    \mathsf{mon}^{  l_{ \mathsf{odd} }  }(  \mathit{x} \mathord{:}  \mathsf{pred}(  \lambda \mathit{x} \mathord{:}  \mathsf{Int}  .~  \mathit{x}  \, \ge \, \ottsym{0} )  \mapsto  \mathsf{pred}(  \lambda \mathit{b} \mathord{:}  \mathsf{Bool}  .~  \mathit{b}  \,  \mathsf{or}  \,  ( \mathit{x} \,  \mathsf{mod}  \, \ottsym{2} \,  =  \, \ottsym{0} )  )   ,  {} \\  &  &  \quad   \lambda \mathit{x} \mathord{:}  \mathsf{Int}  .~   \mathsf{if} ~   ( \mathit{x} \,  =  \, \ottsym{0} )   ~   \mathsf{false}   ~~   (   \mathsf{even}   ~   ( \mathit{x} \,  -  \, \ottsym{1} )   )    )  \\

   \mathsf{and}~  \mathsf{even}  &=&  \lambda \mathit{x} \mathord{:}  \mathsf{Int}  .~   \mathsf{if} ~   ( \mathit{x} \,  =  \, \ottsym{0} )   ~   \mathsf{true}   ~~   (   \mathsf{odd}   ~   ( \mathit{x} \,  -  \, \ottsym{1} )   )   
\end{array}\]
\fi}
We can make this example space-efficient by adding the implication
that: \[  \mathsf{pred}_{ \sigma_{{\mathrm{1}}} }(  \lambda \mathit{b} \mathord{:}  \mathsf{Bool}  .~  \mathit{b}  \,  \mathsf{or}  \,  ( \mathit{x} \,  \mathsf{mod}  \, \ottsym{2} \,  =  \, \ottsym{0} )  )  \, \supset \,  \mathsf{pred}_{ \sigma_{{\mathrm{2}}} }(  \lambda \mathit{b} \mathord{:}  \mathsf{Bool}  .~  \mathit{b}  \,  \mathsf{or}  \,  ( \mathit{x} \,  \mathsf{mod}  \, \ottsym{2} \,  =  \, \ottsym{0} )  )  \]
iff $\sigma_{{\mathrm{1}}}  \ottsym{(}  \mathit{x}  \ottsym{)} \,  +  \, \ottsym{2}  \ottsym{=}  \sigma_{{\mathrm{2}}}  \ottsym{(}  \mathit{x}  \ottsym{)}$.
Suppose we put contracts on both $ \mathsf{even} $ and $ \mathsf{odd} $:
\[\begin{array}{r}
   \mathsf{let}~  \mathsf{odd}  =
    \mathsf{mon}^{  l_{ \mathsf{odd} }  }(  \mathit{x} \mathord{:}  \mathsf{pred}(  \lambda \mathit{x} \mathord{:}  \mathsf{Int}  .~  \mathit{x}  \, \ge \, \ottsym{0} )  \mapsto  \mathsf{pred}(  \lambda \mathit{b} \mathord{:}  \mathsf{Bool}  .~  \mathit{b}  \,  \mathsf{or}  \,  ( \mathit{x} \,  \mathsf{mod}  \, \ottsym{2} \,  =  \, \ottsym{0} )  )   ,  {} \\  \quad   \lambda \mathit{x} \mathord{:}  \mathsf{Int}  .~   \mathsf{if} ~   ( \mathit{x} \,  =  \, \ottsym{0} )   ~   \mathsf{false}   ~~   (   \mathsf{even}   ~   ( \mathit{x} \,  -  \, \ottsym{1} )   )    )  \\

   \multicolumn{1}{l}{\mathsf{and}~  \mathsf{even}  =} \\
   \quad  \mathsf{mon}^{  l_{ \mathsf{even} }  }(  \mathit{x} \mathord{:}  \mathsf{pred}(  \lambda \mathit{x} \mathord{:}  \mathsf{Int}  .~  \mathit{x}  \, \ge \, \ottsym{0} )  \mapsto  \mathsf{pred}(  \lambda \mathit{b} \mathord{:}  \mathsf{Bool}  .~  \mathit{b}  \,  \mathsf{or}  \,  (  ( \mathit{x} \,  +  \, \ottsym{1} )  \,  \mathsf{mod}  \, \ottsym{2} \,  =  \, \ottsym{0} )  )   ,  {} \\  \quad   \lambda \mathit{x} \mathord{:}  \mathsf{Int}  .~   \mathsf{if} ~   ( \mathit{x} \,  =  \, \ottsym{0} )   ~   \mathsf{true}   ~~   (   \mathsf{odd}   ~   ( \mathit{x} \,  -  \, \ottsym{1} )   )    ) 
\end{array}\]
Now our trace of contracts won't be homogeneous; eliding domain contracts:
\[ \begin{array}{rcl}
    \mathsf{odd}   ~  \ottsym{5} 
  & \longrightarrow ^{*}_{  \mathsf{C}  } &  \mathsf{mon}^{  l_{ \mathsf{odd} }  }(  \mathsf{pred}_{  [  \mathit{x}  \mapsto  \ottsym{5}  ]  }( \dots )  ,    \mathsf{even}   ~  \ottsym{4}  )  \\
\iffull
  & \longrightarrow ^{*}_{  \mathsf{C}  } &  \mathsf{mon}^{  l_{ \mathsf{odd} }  }(  \mathsf{pred}_{  [  \mathit{x}  \mapsto  \ottsym{5}  ]  }( \dots )  ,   \mathsf{mon}^{  l_{ \mathsf{even} }  }(  \mathsf{pred}_{  [  \mathit{x}  \mapsto  \ottsym{4}  ]  }( \dots )  ,    \mathsf{odd}   ~  \ottsym{3}  )  )  \\
  & \longrightarrow ^{*}_{  \mathsf{C}  } &  \mathsf{mon}^{  l_{ \mathsf{odd} }  }(  \mathsf{pred}_{  [  \mathit{x}  \mapsto  \ottsym{5}  ]  }( \dots )  ,   \mathsf{mon}^{  l_{ \mathsf{even} }  }(  \mathsf{pred}_{  [  \mathit{x}  \mapsto  \ottsym{4}  ]  }( \dots )  ,  {} \\  &  &  ~~   \mathsf{mon}^{  l_{ \mathsf{odd} }  }(  \mathsf{pred}_{  [  \mathit{x}  \mapsto  \ottsym{3}  ]  }( \dots )  ,    \mathsf{even}   ~  \ottsym{2}  )  )  )  \\
  & \longrightarrow ^{*}_{  \mathsf{C}  } &  \mathsf{mon}^{  l_{ \mathsf{odd} }  }(  \mathsf{pred}_{  [  \mathit{x}  \mapsto  \ottsym{5}  ]  }( \dots )  ,   \mathsf{mon}^{  l_{ \mathsf{even} }  }(  \mathsf{pred}_{  [  \mathit{x}  \mapsto  \ottsym{4}  ]  }( \dots )  ,  {} \\  &  &  ~~   \mathsf{mon}^{  l_{ \mathsf{odd} }  }(  \mathsf{pred}_{  [  \mathit{x}  \mapsto  \ottsym{3}  ]  }( \dots )  ,   \mathsf{mon}^{  l_{ \mathsf{even} }  }(  \mathsf{pred}_{  [  \mathit{x}  \mapsto  \ottsym{2}  ]  }( \dots )  ,    \mathsf{odd}   ~  \ottsym{1}  )  )  )  )  \\
\fi
  & \longrightarrow ^{*}_{  \mathsf{C}  } &  \mathsf{mon}^{  l_{ \mathsf{odd} }  }(  \mathsf{pred}_{  [  \mathit{x}  \mapsto  \ottsym{5}  ]  }( \dots )  ,   \mathsf{mon}^{  l_{ \mathsf{even} }  }(  \mathsf{pred}_{  [  \mathit{x}  \mapsto  \ottsym{4}  ]  }( \dots )  ,  {} \\  &  &  ~~   \mathsf{mon}^{  l_{ \mathsf{odd} }  }(  \mathsf{pred}_{  [  \mathit{x}  \mapsto  \ottsym{3}  ]  }( \dots )  ,   \mathsf{mon}^{  l_{ \mathsf{even} }  }(  \mathsf{pred}_{  [  \mathit{x}  \mapsto  \ottsym{2}  ]  }( \dots )  ,  {} \\  &  &  \quad   \mathsf{mon}^{  l_{ \mathsf{odd} }  }(  \mathsf{pred}_{  [  \mathit{x}  \mapsto  \ottsym{1}  ]  }( \dots )  ,    \mathsf{even}   ~  \ottsym{0}  )  )  )  )  ) 
\end{array} \]
To make these checks space efficient, we'd need several implications;
we write $ \mathsf{odd_p} $ for $ \lambda \mathit{b} \mathord{:}  \mathsf{Bool}  .~  \mathit{b}  \,  \mathsf{or}  \,  ( \mathit{x} \,  \mathsf{mod}  \, \ottsym{2} \,  =  \, \ottsym{0} ) $ and
$ \mathsf{even_p} $ for $ \lambda \mathit{b} \mathord{:}  \mathsf{Bool}  .~  \mathit{b}  \,  \mathsf{or}  \,  (  ( \mathit{x} \,  +  \, \ottsym{1} )  \,  \mathsf{mod}  \, \ottsym{2} \,  =  \, \ottsym{0} ) $.
The following table gives conditions on the implication relation for
the row predicate to imply the column predicate:
\begin{center}
\begin{tabular}{|c|c|c|}
  \hline
  $ \supset $ & $ \mathsf{pred}_{ \sigma_{{\mathrm{2}}} }(  \mathsf{odd_p}  ) $ & $ \mathsf{pred}_{ \sigma_{{\mathrm{2}}} }(  \mathsf{even_p}  ) $ \\
  \hline
  ~~ $ \mathsf{pred}_{ \sigma_{{\mathrm{1}}} }(  \mathsf{odd_p}  ) $ ~~ & ~~ $\sigma_{{\mathrm{1}}}  \ottsym{(}  \mathit{x}  \ottsym{)} \,  +  \, \ottsym{2} \,  =  \, \sigma_{{\mathrm{2}}}  \ottsym{(}  \mathit{x}  \ottsym{)}$ ~~ & ~~ $\sigma_{{\mathrm{1}}}  \ottsym{(}  \mathit{x}  \ottsym{)} \,  +  \, \ottsym{1} \,  =  \, \sigma_{{\mathrm{2}}}  \ottsym{(}  \mathit{x}  \ottsym{)}$ ~~ \\
  \hline
  ~~ $ \mathsf{pred}_{ \sigma_{{\mathrm{1}}} }(  \mathsf{even_p}  ) $ ~~ & $\sigma_{{\mathrm{1}}}  \ottsym{(}  \mathit{x}  \ottsym{)} \,  +  \, \ottsym{1} \,  =  \, \sigma_{{\mathrm{2}}}  \ottsym{(}  \mathit{x}  \ottsym{)}$ & $\sigma_{{\mathrm{1}}}  \ottsym{(}  \mathit{x}  \ottsym{)} \,  +  \, \ottsym{2} \,  =  \, \sigma_{{\mathrm{2}}}  \ottsym{(}  \mathit{x}  \ottsym{)}$ \\
  \hline
\end{tabular}
\end{center}
Having all four of these implications allows us to eliminate any pair
of checks generated by the recursive calls in $ \mathsf{odd} $ and
$ \mathsf{even} $, reducing the codomain checking to constant space---here, just one
check.
We could define a different implication relation, where, say,
$ \mathsf{pred}_{ \sigma_{{\mathrm{1}}} }(  \mathsf{odd_p}  )  \, \supset \,  \mathsf{pred}_{ \sigma_{{\mathrm{2}}} }(  \mathsf{odd_p}  ) $ iff $\sigma_{{\mathrm{1}}}  \ottsym{(}  \mathit{x}  \ottsym{)} \,  \mathsf{mod}  \, \ottsym{2} \,  =  \, \sigma_{{\mathrm{2}}}  \ottsym{(}  \mathit{x}  \ottsym{)} \,  \mathsf{mod}  \, \ottsym{2}$. Such an implication would apply more
generally than those in the table above\iffull---it isn't always obvious how to
define the implication relation\fi.

As usual, there is a trade-off between time and space. It's possible
to write contracts where the necessary implication relation for space efficiency
amounts to checking both contracts.
Consider the following tail-recursive
factorial function:
\[ \begin{array}{lcl}
  \mathsf{let} ~  \mathsf{any}  &=&  \lambda \mathit{z} \mathord{:}  \mathsf{Int}  .~   \mathsf{true}   \\
  \mathsf{let} ~  \mathsf{fact} 
  &=&  \mu ( \mathit{f} \mathord{:}   \mathsf{Int}  \mathord{ \rightarrow }   \mathsf{Int}  \mathord{ \rightarrow }  \mathsf{Int}    ) . ~  {} \\  &  &  ~~   \mathsf{mon}^{ \ottnt{l} }(  \mathit{x} \mathord{:}  \mathsf{pred}(  \mathsf{any}  )  \mapsto  \mathit{acc} \mathord{:}  \mathsf{pred}(  \mathsf{any}  )  \mapsto  \mathsf{pred}(  \lambda \mathit{y} \mathord{:}  \mathsf{Int}  .~  \mathit{x}  \, \ge \, \ottsym{0} )    ,  {} \\  &  &  \quad   \lambda \mathit{x} \mathord{:}  \mathsf{Int}  .~   \lambda \mathit{acc} \mathord{:}  \mathsf{Int}  .~   \mathsf{if} ~   ( \mathit{x} \,  =  \, \ottsym{0} )   ~  \mathit{acc}  ~~   (  \mathit{f}  ~    ( \mathit{x} \,  -  \, \ottsym{1} )   ~   ( \mathit{x} \,  *  \, \mathit{acc} )    )     )  
\end{array} \]
This contract isn't \textit{wrong}, just strange: if you call $ \mathsf{fact} $ with a
negative number, the program diverges and you indeed won't get a value
back out (contracts enforce partial correctness). A call of $  \mathsf{fact}   ~  \ottsym{3} $ yields monitors that check, from
outside to inside, that $\ottsym{3} \, \ge \, \ottsym{0}$ and $\ottsym{2} \, \ge \, \ottsym{0}$ and $\ottsym{1} \, \ge \, \ottsym{0}$
and $\ottsym{0} \, \ge \, \ottsym{0}$.
When should we say that $ \mathsf{pred}_{ \sigma_{{\mathrm{1}}} }(  \lambda \mathit{y} \mathord{:}  \mathsf{Int}  .~  \mathit{x}  \, \ge \, \ottsym{0} )  \, \supset \,  \mathsf{pred}_{ \sigma_{{\mathrm{1}}} }(  \lambda \mathit{y} \mathord{:}  \mathsf{Int}  .~  \mathit{x}  \, \ge \, \ottsym{0} ) $? We could check that $\sigma_{{\mathrm{1}}}  \ottsym{(}  \mathit{x}  \ottsym{)} \, \ge \, \sigma_{{\mathrm{2}}}  \ottsym{(}  \mathit{x}  \ottsym{)}$... but the time cost is just like checking the
original contract.
{\iffull
More starkly, consider the following function with a strange contract
(where $ \mathsf{prime} $ is a predicate identifying prime numbers):
\[ \begin{array}{rcl}
  \mathsf{let} ~  \mathsf{absurd} 
  &=&  \mathsf{mon}^{ \ottnt{l} }(  \mathit{x} \mathord{:}  \mathsf{pred}(  \lambda \mathit{x} \mathord{:}  \mathsf{Int}  .~   \mathsf{true}   )  \mapsto  \mathsf{pred}(   \lambda \mathit{y} \mathord{:}  \mathsf{Int}  .~   \mathsf{prime}    ~  \mathit{x}  )   ,  {} \\  &  &   \lambda \mathit{x} \mathord{:}  \mathsf{Int}  .~   \mathsf{if} ~   ( \mathit{x} \, \subset \, \ottsym{13} )   ~  \ottsym{1}  ~~   (   \mathsf{absurd}   ~  \ottsym{11}  )    ) 
\end{array}\]
How does $  \mathsf{absurd}   ~  \ottsym{47} $ run?
\[ \begin{array}{rcl}
    \mathsf{absurd}   ~  \ottsym{47} 
  &=&  \mathsf{mon}^{ \ottnt{l} }(  \mathsf{pred}_{  [  \mathit{x}  \mapsto  \ottsym{47}  ]  }(   \lambda \mathit{y} \mathord{:}  \mathsf{Int}  .~   \mathsf{prime}    ~  \mathit{x}  )  ,    \mathsf{absurd}   ~  \ottsym{11}  )  \\
  &=&  \mathsf{mon}^{ \ottnt{l} }(  \mathsf{pred}_{  [  \mathit{x}  \mapsto  \ottsym{47}  ]  }(   \lambda \mathit{y} \mathord{:}  \mathsf{Int}  .~   \mathsf{prime}    ~  \mathit{x}  )  ,  {} \\  &  &  ~~   \mathsf{mon}^{ \ottnt{l} }(  \mathsf{pred}_{  [  \mathit{x}  \mapsto  \ottsym{11}  ]  }(   \lambda \mathit{y} \mathord{:}  \mathsf{Int}  .~   \mathsf{prime}    ~  \mathit{x}  )  ,  \ottsym{1} )  ) 
\end{array}\]
What implication relation could resolve these two unrelated checks
into one? We could say that $ \mathsf{pred}_{ \sigma_{{\mathrm{1}}} }(   \lambda \mathit{y} \mathord{:}  \mathsf{Int}  .~   \mathsf{prime}    ~  \mathit{x}  )  \, \supset \,  \mathsf{pred}_{ \sigma_{{\mathrm{2}}} }(   \lambda \mathit{y} \mathord{:}  \mathsf{Int}  .~   \mathsf{prime}    ~  \mathit{x}  ) $ iff $\sigma_{{\mathrm{1}}}  \ottsym{(}  \mathit{x}  \ottsym{)}$ and
$\sigma_{{\mathrm{2}}}  \ottsym{(}  \mathit{x}  \ottsym{)}$ are prime.  There's no time-wise savings, but it would
allow us to collapse tall stacks of monitors.
\fi}

\section{Where should the implication relation come from?}
\label{sec:implication}

The simplest option is to punt: derive the implication relation as the
reflexive transitive closure of a programmer's rules. A programmer might specify how
several different predicates interrelate as follows:
\begin{lstlisting}[frame=single]
y:Int{x1 >= y} implies y:Int{x2 >= y} when x1 <= x2
y:Int{x1 >  y} implies y:Int{x2 >= y} when x1 <= x2 + 1
y:Int{x1 >  y} implies y:Int{x2 >  y} when x1 <= x2
\end{lstlisting}
A default collection of such implications might come with the
language; library programmers should be able to write their own, as
well.
But it is probably unwise to allow programmers to write arbitrary
implications: what if they're wrong?
A good implementation would only accept verified
implications, using a theorem prover or an SMT solver to avoid bogus
implications.

Rather than having programmers write their own implications, we could
try to \textit{automatically} derive the implications. Given a
program, a fixed number of predicates occur, even if an unbounded
number of predicate/closing substitution pairings might occur at
runtime.
Collect all possible predicates from the source program, and consider
each pair of predicates over the same base type,
$ \mathsf{pred}( \ottnt{e_{{\mathrm{1}}}} ) $ and $ \mathsf{pred}( \ottnt{e_{{\mathrm{2}}}} ) $ such that $ \mathrm{\Gamma}   \vdash   \ottnt{e_{\ottmv{i}}}  :   \ottnt{B} \mathord{ \rightarrow }  \mathsf{Bool}   $. We can derive from
the typing derivation the shapes of the respective closing
substitutions, $\sigma_{{\mathrm{1}}}$ and $\sigma_{{\mathrm{2}}}$. What are the conditions
on $\sigma_{{\mathrm{1}}}$ and $\sigma_{{\mathrm{2}}}$ such that $ \mathsf{pred}_{ \sigma_{{\mathrm{1}}} }( \ottnt{e_{{\mathrm{1}}}} )  \, \supset \,  \mathsf{pred}_{ \sigma_{{\mathrm{2}}} }( \ottnt{e_{{\mathrm{2}}}} ) $?
We are looking for a property $P(\sigma_{{\mathrm{1}}},\sigma_{{\mathrm{2}}})$ such that:
\[ \forall  \ottnt{k}  \in \mathcal{K}_{ \ottnt{B} } , ~
     P(\sigma_{{\mathrm{1}}},\sigma_{{\mathrm{2}}}) \wedge  \sigma_{{\mathrm{1}}}  \ottsym{(}  \ottnt{e_{{\mathrm{1}}}}  \ottsym{)}  ~  \ottnt{k}  \,  \longrightarrow ^{*}_{  \mathsf{E}  }  \,  \mathsf{true}  \Rightarrow
      \sigma_{{\mathrm{2}}}  \ottsym{(}  \ottnt{e_{{\mathrm{2}}}}  \ottsym{)}  ~  \ottnt{k}  \,  \longrightarrow ^{*}_{  \mathsf{E}  }  \,  \mathsf{true}  \]
Ideally, $P$ is also efficiently decidable---at least more efficiently
than deciding both predicates.
The problem of finding $P$ can be reduced to that of finding the
weakest precondition for the safety of the following function:
\begin{lstlisting}[frame=single]
fun x:B =>
  let y0 = v`[${}_{10}$]`, ..., yn = v`[${}_{1n}$]` (* `[$\sigma_{{\mathrm{1}}}$]`'s bindings *)
      z0 = v`[${}_{20}$]`, ..., zn = v`[${}_{2m}$]` (* `[$\sigma_{{\mathrm{2}}}$]`'s bindings *) in
  if e1 x then (if e2 x then () else error) else ()
\end{lstlisting}
Since $P$ would be the \textit{weakest} precondition, we would know
that we had found the most general condition for the implication
relation.
Whether or not the most general condition is the \textit{best}
condition depends on context. We should also consider a cost model for
$P$; programmers may want to occasionally trade space for time, not
bothering to join predicates that would be expensive to test.

Finding implication conditions resembles liquid type
inference~\cite{Rondon08liquid,Vazou2013,Jhala14}: programmers get a
small control knob (which expressions can go in $P$) and then an SMT
solver does the rest. The settings are different, though: liquid types
are about verifying programs, while we're executing checks at runtime.

\subsection{Implementation}

Implementation issues abound. How should the free variables in terms
be represented? What kind of refactorings and optimizations can the
compiler do, and how might they interfere with the set of contracts
that appear in a program? When is the right moment in compilation to
fix the implication relation?
More generally, what's the design space of closure representations and
calling conventions for languages with contracts?

\section{Extensions}
\label{sec:extensions}

Generalizing our space-efficient semantics to sums and products does
not seem particularly hard: we'd need contracts with corresponding
shapes, and the join operation would push through such shapes.
Recursive types and datatypes are more
interesting~\cite{Sekiyama15datatypes}. Findler et al.'s lazy contract
checking keeps contracts from changing the asymptotic time complexity
of the program~\cite{Findler08immutable}; we may be able to adapt
their work to avoid changes in asymptotic space complexity, as well.

The predicates here range over base types, but we could also allow
predicates over other types. If we allow predicates over higher types,
how should the adequacy constraint on predicate implication
(Definition~\ref{def:implaxioms}) change?

Impredicative polymorphism in the style of System F would require even
more technical changes.
The introduction of type variables would make our reasoning about names and binders trickier.
In order to support predicates over type variables, we'd need to allow
predicates over higher types---and so our notion of adequacy of
$ \supset $ would change.
In order to support predicates over quantified types, we'd need to
change adequacy again. Adequacy would end up looking like the logical
relation used to show relational parametricity: when would we have
$\forall \alpha. \ottnt{T_{{\mathrm{1}}}}  \supset  \forall \alpha. \ottnt{T_{{\mathrm{2}}}}$? If we
substitute $\ottnt{T'_{{\mathrm{1}}}}$ for $\alpha$ on the left and $\ottnt{T'_{{\mathrm{2}}}}$ for
$\alpha$ on the right (and $\ottnt{T'_{{\mathrm{1}}}}$ and $\ottnt{T'_{{\mathrm{2}}}}$ are somehow
related), then $\ottnt{T_{{\mathrm{1}}}} [ \ottnt{T'_{{\mathrm{1}}}}/\alpha ]  \supset  \ottnt{T_{{\mathrm{2}}}} [
  \ottnt{T'_{{\mathrm{2}}}}/\alpha ]$.
Not only would the technicalities be tricky, implementations would
need to be careful to manage closure representations correctly (e.g.,
what happens if polymorphic code differs for boxed and unboxed
types?).

We don't treat blame as an interesting algebraic structure---it's
enough for our proofs to show that we always produce the same
answer. Changing our calculus to have a more interesting notion of
blame, like \textit{indy} semantics~\cite{Dimoulas11indy} or
involutive blame labels~\cite{Wadler09blame,Wadler15blame}, would be a
matter of pushing a shallow change in the semantics through the
proofs.

Finally, it would make sense to have substitution on predicate stacks
perform joins, saying $  (   \mathsf{pred}^{ \ottnt{l} }_{ \sigma }( \ottnt{e} )  ; \ottnt{r}  )   [  \ottnt{v}  /  \mathit{x}  ]  =
 \mathsf{join} (    \mathsf{pred}^{ \ottnt{l} }_{ \sigma }( \ottnt{e} )   [  \ottnt{v}  /  \mathit{x}  ]  ; \mathsf{nil}  ,  \ottnt{r}  [  \ottnt{v}  /  \mathit{x}  ]  ) $, so that substituting a
value into a predicate stack checks for newly revealed redundancies.
We haven't proved that this change would be sound, which would require
changes to both type and space-efficiency soundness.
{\iffull In particular, we'd need to (a) revise
Lemma~\ref{lem:lrcontract} and (b) simultaneously prove that
substitution and joins preserve types
(Lemmas~\ref{lem:cpcfesubstitution}) and~\ref{lem:cpcfejoin}). \fi}

\section{Related work}
\label{sec:related}

For the technique of space efficiency itself, we refer the reader to
Greenberg~\cite{Greenberg15space} for a full description of related
work.
We have striven to use Greenberg's notation exactly, but we made some
changes in adapting to dependent contracts: the cons operator for
predicate stacks is a semi-colon, to avoid ambiguity; there were
formerly two things named $ \mathsf{join} $, but one has been folded into the
other; and our predicates have closing substitutions to account for
dependency.
We place one more requirement on the implication relation than
Greenberg did: monotonicity under substitution, which we call
\textit{substitutivity}. Substitutions weren't an issue in his
non-dependent system, but we must require that if a join can happen
without having a value for $\mathit{x}$, then the same join happens when we
know $\mathit{x}$'s value.

CPCF was first introduced in several papers by Dimoulas et al. in
2011~\cite{Dimoulas11cpcf,Dimoulas11indy}, and has later been the
subject of studies of blame for dependent function
contracts~\cite{Dimoulas12complete} and static
analysis~\cite{Tobin-Hochstadt12symbolic}. Our exact behavioral
equivalence means we could use results from Tobin-Hochstadt et al.'s
static analysis in terms of \CPCFc to optimize space efficient
programs in \CPCFe.
More interestingly, the predicate implication relation $ \supset $
seems to be doing some of the work that their static analysis does, so
there may be a deeper relationship.

Thiemann introduces a manifest calculus where the compiler optimizes
casts for time efficiency: a theorem prover uses the ``delta'' between
types to synthesize more efficient checks~\cite{Thiemann16delta}. His
deltas and our predicate implication relation are similar. He uses
a separate logical language for predicates and restricts dependency
(codomains can only depend on base values, avoiding abusive
contracts).

Sekiyama et al.~\cite{Sekiyama16fh} also use delayed substitutions in
their polymorphic manifest contract calculus, but for different
technical reasons. While delayed substitutions resemble explicit
substitutions~\cite{AbadiCardelliCurienLevy91JFP} or explicit
bindings~\cite{Grossman00typeabs,Ahmed11blame}, we use delayed
substitutions more selectively and to resolve issues with dependency.

The manifest type system in Greenberg's work is somewhat disappointing
compared to the type system given here. Greenberg works much harder
than we do to prove a stronger type soundness theorem... but that
theorem isn't enough to help materially in proving the soundness of
space efficiency.
Developing the approach to dependency used here was much easier in a
latent calculus, though several bugs along the way would have been
caught early by a stronger type system. Type system complexity trade-offs of this
sort are an old story.

\subsection{Racket's implementation}

If contracts leak space, how is it that they are used so effectively
throughout PLT Racket?
Racket is designed to avoid using contracts in leaky ways.
In Racket, contracts tend to go on module boundaries. Calls inside of
a module then don't trigger contract checks---including recursive
calls, like in the $ \mathsf{even} $/$ \mathsf{odd} $ example.
Racket \textit{will} monitor recursive calls across module boundaries,
and these checks can indeed lead to space leaks.
In our terms, Racket tends to implement contract checks on
recursive functions as follows:
\[ \begin{array}{r@{}l}
   \mathsf{downTo}  =  \mathsf{mon}^{ \ottnt{l} }( &   \mathit{x} \mathord{:}  \mathsf{pred}(  \lambda \mathit{x} \mathord{:}  \mathsf{Int}  .~  \mathit{x}  \, \ge \, \ottsym{0} )  \mapsto  \mathsf{pred}(  \lambda \mathit{y} \mathord{:}  \mathsf{Int}  .~  \mathit{x}  \, \ge \, \mathit{y} )   ,  {} \\  &   \mu ( \mathit{f} \mathord{:}   \mathsf{Int}  \mathord{ \rightarrow }  \mathsf{Int}   ) . ~   \lambda \mathit{x} \mathord{:}  \mathsf{Int}  .~   \mathsf{if} ~   ( \mathit{x} \,  =  \, \ottsym{0} )   ~  \ottsym{0}  ~~   (  \mathit{f}  ~   ( \mathit{x} \,  -  \, \ottsym{1} )   )     ) 
\end{array}
\]
Note that calling $  \mathsf{downTo}   ~   n  $ will merely check that the final
result is less than $ n $---none of the intermediate values. Our
version of $ \mathsf{downTo} $ above puts the contract \textit{inside} the
recursive knot, forcing checks every time (Sec.~\ref{sec:depbounds}).

Racket also offers a less thorough form of space efficiency\iffull via the
\texttt{tail-marks-match?} predicate.\footnote{From
  \texttt{racket/collects/racket/contract/private/arrow.rkt}.}\else. \fi
We can construct a program where Racket will avoid redundant checks, but
but wrapping the underlying function with the same contract twice
leads to a space leak (Figure~\ref{fig:racketse})\iffull. 
We use low-level contract mechanisms here, to highlight the
compromises Racket makes for space efficiency, ``strik[ing] a balance
between common ways that tail recursion is broken and checking that
would not be too expensive in the case that there wouldn't have been
any pile-up of redundant wrappers''.\else.\fi\footnote{Robby Findler, personal
  correspondence, 2016-05-19.}

\begin{figure*}[t]
\begin{lstlisting}[language=lisp]
(define (count-em-integer? x)
  (printf "checking ~s\n" x)
  (integer? x))
(letrec
  ([f (contract (-> any/c count-em-integer?)
        (lambda (x) (if (zero? x) x (f (- x 1))))
        'pos 'neg)])
  (f 3))
\end{lstlisting}
\iffull
(a) Space-efficient program
\begin{lstlisting}[language=lisp]
(letrec
  ([f (contract (-> any/c count-em-integer?)
        (contract (-> any/c count-em-integer?)
          (lambda (x) (if (zero? x) x (f (- x 1))))
           'pos 'neg)
        'pos 'neg)])
  (f 3))
\end{lstlisting}
(a) Space-leaking program
\fi
\caption{Space-efficiency in Racket}
\label{fig:racketse}
\end{figure*}
Finally, contracts are first-class in Racket\iffull: their monitors take two expressions,
one for the contract and one to be monitored\fi. Computing new contracts
at runtime breaks our framing of space-efficiency: if we can't
predetermine which contracts arise at runtime, we can't fix an
implication relation in advance.
We hope that \CPCFe is close enough to Racket's internal model to
provide insight into how to achieve space efficiency for at least some
contracts in Racket.

\section{Conclusion}
\label{sec:conclusion}

We have translated Greenberg's original result~\cite{Greenberg15space}
from a manifest calculus~\cite{Greenberg10contracts} to a latent
one~\cite{Dimoulas11cpcf,Dimoulas11indy}. In so doing, we have:
offered a simpler explanation of the original result; isolated the
parts of the type system required for space bounds\iffull, which were
intermingled with complexities from conflating contracts and types\fi;
and, extended the original result, by covering more features
(dependency and nontermination) and more precisely bounding non-dependent programs.

\vspace*{-1em}
\subsubsection*{Acknowledgments.}
The existence of this paper is due to comments from Sam
Tobin-Hochstadt and David Van Horn that I chose to interpret as
encouragement.
Robby Findler provided the Racket example and helped correct and
clarify a draft; Sam Tobin-Hochstadt also offered corrections and
suggestions.
The reviews offered helpful comments, too.

\vspace*{-1em}

\bibliographystyle{splncs03}
\bibliography{mgree}

\end{document}

%% file: cpcf.tex
\newcommand{\ottdrule}[4][]{{\displaystyle\frac{\begin{array}{l}#2\end{array}}{#3}\quad\ottdrulename{#4}}}
\newcommand{\ottusedrule}[1]{\[#1\]}
\newcommand{\ottpremise}[1]{ #1 \\}
\newenvironment{ottdefnblock}[3][]{ \framebox{\mbox{#2}} \quad #3 \\[0pt]}{}

\newcommand{\ottnt}[1]{\mathit{#1}}
\newcommand{\ottmv}[1]{\mathit{#1}}

\newcommand{\ottsym}[1]{#1}

\newcommand{\ottdrulename}[1]{\textsc{#1}}

\newcommand{\ottdruleTVar}[1]{\ottdrule[#1]{%
\ottpremise{ \mathit{x}  \mathord{:}  \ottnt{T}  \in  \mathrm{\Gamma} }%
}{
 \mathrm{\Gamma}   \vdash   \mathit{x}  :  \ottnt{T} }{%
{\ottdrulename{TVar}}{}%
}}

\newcommand{\ottdruleTConst}[1]{\ottdrule[#1]{%
}{
 \mathrm{\Gamma}   \vdash   \ottnt{k}  :   \mathsf{ty} ( \ottnt{k} )  }{%
{\ottdrulename{TConst}}{}%
}}

\newcommand{\ottdruleTAbs}[1]{\ottdrule[#1]{%
\ottpremise{  \mathrm{\Gamma} , \mathit{x} \mathord{:} \ottnt{T_{{\mathrm{1}}}}    \vdash   \ottnt{e_{{\mathrm{12}}}}  :  \ottnt{T_{{\mathrm{2}}}} }%
}{
 \mathrm{\Gamma}   \vdash    \lambda \mathit{x} \mathord{:} \ottnt{T_{{\mathrm{1}}}} .~  \ottnt{e_{{\mathrm{12}}}}   :   \ottnt{T_{{\mathrm{1}}}} \mathord{ \rightarrow } \ottnt{T_{{\mathrm{2}}}}  }{%
{\ottdrulename{TAbs}}{}%
}}

\newcommand{\ottdruleTRec}[1]{\ottdrule[#1]{%
\ottpremise{  \mathrm{\Gamma} , \mathit{x} \mathord{:} \ottnt{T}    \vdash   \ottnt{e}  :  \ottnt{T} }%
}{
 \mathrm{\Gamma}   \vdash    \mu ( \mathit{x} \mathord{:} \ottnt{T} ) . ~  \ottnt{e}   :  \ottnt{T} }{%
{\ottdrulename{TRec}}{}%
}}

\newcommand{\ottdruleTOp}[1]{\ottdrule[#1]{%
\ottpremise{ \mathsf{ty} (\mathord{ \ottnt{op} })   \ottsym{=}    \ottnt{T_{{\mathrm{1}}}} \mathord{ \rightarrow }   \ottnt{T_{{\mathrm{2}}}} \mathord{ \rightarrow } \ottnt{T}    }%
\ottpremise{  \mathrm{\Gamma}   \vdash   \ottnt{e_{{\mathrm{1}}}}  :  \ottnt{T_{{\mathrm{1}}}}   \quad   \mathrm{\Gamma}   \vdash   \ottnt{e_{{\mathrm{2}}}}  :  \ottnt{T_{{\mathrm{2}}}}  }%
}{
 \mathrm{\Gamma}   \vdash   \ottnt{e_{{\mathrm{1}}}} \, \ottnt{op} \, \ottnt{e_{{\mathrm{2}}}}  :  \ottnt{T} }{%
{\ottdrulename{TOp}}{}%
}}

\newcommand{\ottdruleTApp}[1]{\ottdrule[#1]{%
\ottpremise{  \mathrm{\Gamma}   \vdash   \ottnt{e_{{\mathrm{1}}}}  :   \ottnt{T_{{\mathrm{1}}}} \mathord{ \rightarrow } \ottnt{T_{{\mathrm{2}}}}    \quad   \mathrm{\Gamma}   \vdash   \ottnt{e_{{\mathrm{2}}}}  :  \ottnt{T_{{\mathrm{1}}}}  }%
}{
 \mathrm{\Gamma}   \vdash    \ottnt{e_{{\mathrm{1}}}}  ~  \ottnt{e_{{\mathrm{2}}}}   :  \ottnt{T_{{\mathrm{2}}}} }{%
{\ottdrulename{TApp}}{}%
}}

\newcommand{\ottdruleTIf}[1]{\ottdrule[#1]{%
\ottpremise{  \mathrm{\Gamma}   \vdash   \ottnt{e_{{\mathrm{1}}}}  :   \mathsf{Bool}    \quad     \mathrm{\Gamma}   \vdash   \ottnt{e_{{\mathrm{2}}}}  :  \ottnt{T}   \quad   \mathrm{\Gamma}   \vdash   \ottnt{e_{{\mathrm{3}}}}  :  \ottnt{T}    }%
}{
 \mathrm{\Gamma}   \vdash    \mathsf{if} ~  \ottnt{e_{{\mathrm{1}}}}  ~  \ottnt{e_{{\mathrm{2}}}}  ~~  \ottnt{e_{{\mathrm{3}}}}   :  \ottnt{T} }{%
{\ottdrulename{TIf}}{}%
}}

\newcommand{\ottdruleTMon}[1]{\ottdrule[#1]{%
\ottpremise{  \mathrm{\Gamma}   \vdash   \ottnt{e}  :  \ottnt{T}   \quad  \mathrm{\Gamma}  \vdash  \ottnt{C}  \ottsym{:}  \ottnt{T} }%
}{
 \mathrm{\Gamma}   \vdash    \mathsf{mon}^{ \ottnt{l} }( \ottnt{C} ,  \ottnt{e} )   :  \ottnt{T} }{%
{\ottdrulename{TMon}}{}%
}}

\newcommand{\ottdruleTMonC}[1]{\ottdrule[#1]{%
\ottpremise{  \mathrm{\Gamma}   \vdash   \ottnt{e}  :  \ottnt{T}   \quad  \mathrm{\Gamma}  \vdash  \ottnt{c}  \ottsym{:}  \ottnt{T} }%
}{
 \mathrm{\Gamma}   \vdash    \mathsf{mon}( \ottnt{c} ,  \ottnt{e} )   :  \ottnt{T} }{%
{\ottdrulename{TMonC}}{}%
}}

\newcommand{\ottdruleTBlame}[1]{\ottdrule[#1]{%
}{
 \mathrm{\Gamma}   \vdash    \mathsf{err}^ \ottnt{l}   :  \ottnt{T} }{%
{\ottdrulename{TBlame}}{}%
}}

\newcommand{\ottdruleTPred}[1]{\ottdrule[#1]{%
\ottpremise{  \mathrm{\Gamma}  \ottsym{,}  \mathrm{\Gamma}'   \vdash   \ottnt{e}  :   \ottnt{B} \mathord{ \rightarrow }  \mathsf{Bool}     \quad  \mathrm{\Gamma}'  \vdash  \sigma }%
}{
\mathrm{\Gamma}  \vdash   \mathsf{pred}_{ \sigma }( \ottnt{e} )   \ottsym{:}  \ottnt{B}}{%
{\ottdrulename{TPred}}{}%
}}

\newcommand{\ottdruleTFun}[1]{\ottdrule[#1]{%
\ottpremise{ \mathrm{\Gamma}  \vdash  \ottnt{C_{{\mathrm{1}}}}  \ottsym{:}  \ottnt{T_{{\mathrm{1}}}}  \quad   \mathrm{\Gamma} , \mathit{x} \mathord{:} \ottnt{T_{{\mathrm{1}}}}   \vdash  \ottnt{C_{{\mathrm{2}}}}  \ottsym{:}  \ottnt{T_{{\mathrm{2}}}} }%
}{
\mathrm{\Gamma}  \vdash   \mathit{x} \mathord{:} \ottnt{C_{{\mathrm{1}}}} \mapsto \ottnt{C_{{\mathrm{2}}}}   \ottsym{:}   \ottnt{T_{{\mathrm{1}}}} \mathord{ \rightarrow } \ottnt{T_{{\mathrm{2}}}} }{%
{\ottdrulename{TFun}}{}%
}}

\newcommand{\ottdruleTCNil}[1]{\ottdrule[#1]{%
}{
\mathrm{\Gamma}  \vdash  \mathsf{nil}  \ottsym{:}  \ottnt{B}}{%
{\ottdrulename{TCNil}}{}%
}}

\newcommand{\ottdruleTCPred}[1]{\ottdrule[#1]{%
\ottpremise{ \mathrm{\Gamma}  \vdash   \mathsf{pred}_{ \sigma }( \ottnt{e} )   \ottsym{:}  \ottnt{B}  \quad  \mathrm{\Gamma}  \vdash  \ottnt{r}  \ottsym{:}  \ottnt{B} }%
}{
\mathrm{\Gamma}  \vdash    \mathsf{pred}^{ \ottnt{l} }_{ \sigma }( \ottnt{e} )  ; \ottnt{r}   \ottsym{:}  \ottnt{B}}{%
{\ottdrulename{TCPred}}{}%
}}

\newcommand{\ottdruleTCFun}[1]{\ottdrule[#1]{%
\ottpremise{ \mathrm{\Gamma}  \vdash  \ottnt{c_{{\mathrm{1}}}}  \ottsym{:}  \ottnt{T_{{\mathrm{1}}}}  \quad   \mathrm{\Gamma} , \mathit{x} \mathord{:} \ottnt{T_{{\mathrm{1}}}}   \vdash  \ottnt{c_{{\mathrm{2}}}}  \ottsym{:}  \ottnt{T_{{\mathrm{2}}}} }%
}{
\mathrm{\Gamma}  \vdash   \mathit{x} \mathord{:} \ottnt{c_{{\mathrm{1}}}} \mapsto \ottnt{c_{{\mathrm{2}}}}   \ottsym{:}   \ottnt{T_{{\mathrm{1}}}} \mathord{ \rightarrow } \ottnt{T_{{\mathrm{2}}}} }{%
{\ottdrulename{TCFun}}{}%
}}

\newcommand{\ottdruleTId}[1]{\ottdrule[#1]{%
}{
\emptyset  \vdash   \iota }{%
{\ottdrulename{TId}}{}%
}}

\newcommand{\ottdruleTMap}[1]{\ottdrule[#1]{%
\ottpremise{ \mathrm{\Gamma}  \vdash  \sigma  \quad    \mathit{x} \mathord{:} \ottnt{T}    \vdash   \ottnt{v}  :  \ottnt{T}  }%
}{
 \mathrm{\Gamma} , \mathit{x} \mathord{:} \ottnt{T}   \vdash   \sigma  [  \mathit{x}  \mapsto  \ottnt{v}  ] }{%
{\ottdrulename{TMap}}{}%
}}

\newcommand{\ottdruleEDelta}[1]{\ottdrule[#1]{%
}{
\ottnt{k_{{\mathrm{1}}}} \, \ottnt{op} \, \ottnt{k_{{\mathrm{2}}}} \,  \longrightarrow _{ \mathsf{m} }  \, \denot{ op } \, \ottsym{(}  \ottnt{k_{{\mathrm{1}}}}  \ottsym{,}  \ottnt{k_{{\mathrm{2}}}}  \ottsym{)}}{%
{\ottdrulename{EDelta}}{}%
}}

\newcommand{\ottdruleEBeta}[1]{\ottdrule[#1]{%
}{
  (  \lambda \mathit{x} \mathord{:} \ottnt{T} .~  \ottnt{e}  )   ~  \ottnt{v}  \,  \longrightarrow _{ \mathsf{m} }  \,  \ottnt{e}  [  \ottnt{v} / \mathit{x}  ] }{%
{\ottdrulename{EBeta}}{}%
}}

\newcommand{\ottdruleEFix}[1]{\ottdrule[#1]{%
}{
 \mu ( \mathit{x} \mathord{:} \ottnt{T} ) . ~  \ottnt{e}  \,  \longrightarrow _{ \mathsf{m} }  \,  \ottnt{e}  [   \mu ( \mathit{x} \mathord{:} \ottnt{T} ) . ~  \ottnt{e}  / \mathit{x}  ] }{%
{\ottdrulename{EFix}}{}%
}}

\newcommand{\ottdruleEIfTrue}[1]{\ottdrule[#1]{%
}{
 \mathsf{if} ~   \mathsf{true}   ~  \ottnt{e_{{\mathrm{2}}}}  ~~  \ottnt{e_{{\mathrm{3}}}}  \,  \longrightarrow _{ \mathsf{m} }  \, \ottnt{e_{{\mathrm{2}}}}}{%
{\ottdrulename{EIfTrue}}{}%
}}

\newcommand{\ottdruleEIfFalse}[1]{\ottdrule[#1]{%
}{
 \mathsf{if} ~   \mathsf{false}   ~  \ottnt{e_{{\mathrm{2}}}}  ~~  \ottnt{e_{{\mathrm{3}}}}  \,  \longrightarrow _{ \mathsf{m} }  \, \ottnt{e_{{\mathrm{3}}}}}{%
{\ottdrulename{EIfFalse}}{}%
}}

\newcommand{\ottdruleEMonPred}[1]{\ottdrule[#1]{%
}{
 \mathsf{mon}^{ \ottnt{l} }(  \mathsf{pred}_{ \sigma }( \ottnt{e_{{\mathrm{1}}}} )  ,  \ottnt{v_{{\mathrm{2}}}} )  \,  \longrightarrow _{  \mathsf{C}  }  \,  \mathsf{if} ~   (  \sigma  \ottsym{(}  \ottnt{e_{{\mathrm{1}}}}  \ottsym{)}  ~  \ottnt{v_{{\mathrm{2}}}}  )   ~  \ottnt{v_{{\mathrm{2}}}}  ~~   \mathsf{err}^ \ottnt{l}  }{%
{\ottdrulename{EMonPred}}{}%
}}

\newcommand{\ottdruleEMonApp}[1]{\ottdrule[#1]{%
}{
  \mathsf{mon}^{ \ottnt{l} }(  \mathit{x} \mathord{:} \ottnt{C_{{\mathrm{1}}}} \mapsto \ottnt{C_{{\mathrm{2}}}}  ,  \ottnt{v_{{\mathrm{1}}}} )   ~  \ottnt{v_{{\mathrm{2}}}}  \,  \longrightarrow _{  \mathsf{C}  }  \,  \mathsf{mon}^{ \ottnt{l} }(  \ottnt{C_{{\mathrm{2}}}}  [  \ottnt{v_{{\mathrm{2}}}}  /  \mathit{x}  ]  ,   \ottnt{v_{{\mathrm{1}}}}  ~   \mathsf{mon}^{ \ottnt{l} }( \ottnt{C_{{\mathrm{1}}}} ,  \ottnt{v_{{\mathrm{2}}}} )   ) }{%
{\ottdrulename{EMonApp}}{}%
}}

\newcommand{\ottdruleEMon}[1]{\ottdrule[#1]{%
\ottpremise{\ottnt{e} \,  \longrightarrow _{  \mathsf{C}  }  \, \ottnt{e'}}%
}{
 \mathsf{mon}^{ \ottnt{l} }( \ottnt{C} ,  \ottnt{e} )  \,  \longrightarrow _{  \mathsf{C}  }  \,  \mathsf{mon}^{ \ottnt{l} }( \ottnt{C} ,  \ottnt{e'} ) }{%
{\ottdrulename{EMon}}{}%
}}

\newcommand{\ottdruleEMonRaise}[1]{\ottdrule[#1]{%
}{
 \mathsf{mon}^{ \ottnt{l} }( \ottnt{C} ,   \mathsf{err}^ \ottnt{l'}  )  \,  \longrightarrow _{  \mathsf{C}  }  \,  \mathsf{err}^ \ottnt{l'} }{%
{\ottdrulename{EMonRaise}}{}%
}}

\newcommand{\ottdruleEMonLabel}[1]{\ottdrule[#1]{%
}{
 \mathsf{mon}^{ \ottnt{l} }( \ottnt{C} ,  \ottnt{e} )  \,  \longrightarrow _{  \mathsf{E}  }  \,  \mathsf{mon}(  \mathsf{label} ^{ \ottnt{l} }( \ottnt{C} )  ,  \ottnt{e} ) }{%
{\ottdrulename{EMonLabel}}{}%
}}

\newcommand{\ottdruleEMonCNil}[1]{\ottdrule[#1]{%
}{
 \mathsf{mon}( \mathsf{nil} ,  \ottnt{v_{{\mathrm{1}}}} )  \,  \longrightarrow _{  \mathsf{E}  }  \, \ottnt{v_{{\mathrm{1}}}}}{%
{\ottdrulename{EMonCNil}}{}%
}}

\newcommand{\ottdruleEMonCPred}[1]{\ottdrule[#1]{%
}{
 \mathsf{mon}(   \mathsf{pred}^{ \ottnt{l} }_{ \sigma }( \ottnt{e} )  ; \ottnt{r}  ,  \ottnt{v_{{\mathrm{1}}}} )  \,  \longrightarrow _{  \mathsf{E}  }  \,  \mathsf{if} ~   (  \sigma  \ottsym{(}  \ottnt{e}  \ottsym{)}  ~  \ottnt{v_{{\mathrm{1}}}}  )   ~   \mathsf{mon}( \ottnt{r} ,  \ottnt{v_{{\mathrm{1}}}} )   ~~   \mathsf{err}^ \ottnt{l}  }{%
{\ottdrulename{EMonCPred}}{}%
}}

\newcommand{\ottdruleEMonCApp}[1]{\ottdrule[#1]{%
}{
  \mathsf{mon}(  \mathit{x} \mathord{:} \ottnt{c_{{\mathrm{1}}}} \mapsto \ottnt{c_{{\mathrm{2}}}}  ,  \ottnt{v_{{\mathrm{1}}}} )   ~  \ottnt{v_{{\mathrm{2}}}}  \,  \longrightarrow _{  \mathsf{E}  }  \,  \mathsf{mon}(  \ottnt{c_{{\mathrm{2}}}}  [  \ottnt{v_{{\mathrm{2}}}}  /  \mathit{x}  ]  ,   \ottnt{v_{{\mathrm{1}}}}  ~   \mathsf{mon}( \ottnt{c_{{\mathrm{1}}}} ,  \ottnt{v_{{\mathrm{2}}}} )   ) }{%
{\ottdrulename{EMonCApp}}{}%
}}

\newcommand{\ottdruleEMonC}[1]{\ottdrule[#1]{%
\ottpremise{ \ottnt{e}  \neq   \mathsf{mon}( \ottnt{c'} ,  \ottnt{e''} )   \quad  \ottnt{e} \,  \longrightarrow _{  \mathsf{E}  }  \, \ottnt{e'} }%
}{
 \mathsf{mon}( \ottnt{c} ,  \ottnt{e} )  \,  \longrightarrow _{  \mathsf{E}  }  \,  \mathsf{mon}( \ottnt{c} ,  \ottnt{e'} ) }{%
{\ottdrulename{EMonC}}{}%
}}

\newcommand{\ottdruleEMonCJoin}[1]{\ottdrule[#1]{%
}{
 \mathsf{mon}( \ottnt{c_{{\mathrm{2}}}} ,   \mathsf{mon}( \ottnt{c_{{\mathrm{1}}}} ,  \ottnt{e} )  )  \,  \longrightarrow _{  \mathsf{E}  }  \,  \mathsf{mon}(  \mathsf{join} ( \ottnt{c_{{\mathrm{1}}}} , \ottnt{c_{{\mathrm{2}}}} )  ,  \ottnt{e} ) }{%
{\ottdrulename{EMonCJoin}}{}%
}}

\newcommand{\ottdruleEMonCRaise}[1]{\ottdrule[#1]{%
}{
 \mathsf{mon}( \ottnt{c} ,   \mathsf{err}^ \ottnt{l}  )  \,  \longrightarrow _{  \mathsf{E}  }  \,  \mathsf{err}^ \ottnt{l} }{%
{\ottdrulename{EMonCRaise}}{}%
}}

\newcommand{\ottdruleEAppL}[1]{\ottdrule[#1]{%
\ottpremise{\ottnt{e_{{\mathrm{1}}}} \,  \longrightarrow _{ \mathsf{m} }  \, \ottnt{e'_{{\mathrm{1}}}}}%
}{
 \ottnt{e_{{\mathrm{1}}}}  ~  \ottnt{e_{{\mathrm{2}}}}  \,  \longrightarrow _{ \mathsf{m} }  \,  \ottnt{e'_{{\mathrm{1}}}}  ~  \ottnt{e_{{\mathrm{2}}}} }{%
{\ottdrulename{EAppL}}{}%
}}

\newcommand{\ottdruleEAppR}[1]{\ottdrule[#1]{%
\ottpremise{\ottnt{e_{{\mathrm{2}}}} \,  \longrightarrow _{ \mathsf{m} }  \, \ottnt{e'_{{\mathrm{2}}}}}%
}{
 \ottnt{v_{{\mathrm{1}}}}  ~  \ottnt{e_{{\mathrm{2}}}}  \,  \longrightarrow _{ \mathsf{m} }  \,  \ottnt{v_{{\mathrm{1}}}}  ~  \ottnt{e'_{{\mathrm{2}}}} }{%
{\ottdrulename{EAppR}}{}%
}}

\newcommand{\ottdruleEOpL}[1]{\ottdrule[#1]{%
\ottpremise{\ottnt{e_{{\mathrm{1}}}} \,  \longrightarrow _{ \mathsf{m} }  \, \ottnt{e'_{{\mathrm{1}}}}}%
}{
\ottnt{e_{{\mathrm{1}}}} \, \ottnt{op} \, \ottnt{e_{{\mathrm{2}}}} \,  \longrightarrow _{ \mathsf{m} }  \, \ottnt{e'_{{\mathrm{1}}}} \, \ottnt{op} \, \ottnt{e_{{\mathrm{2}}}}}{%
{\ottdrulename{EOpL}}{}%
}}

\newcommand{\ottdruleEOpR}[1]{\ottdrule[#1]{%
\ottpremise{\ottnt{e_{{\mathrm{2}}}} \,  \longrightarrow _{ \mathsf{m} }  \, \ottnt{e'_{{\mathrm{2}}}}}%
}{
\ottnt{v_{{\mathrm{1}}}} \, \ottnt{op} \, \ottnt{e_{{\mathrm{2}}}} \,  \longrightarrow _{ \mathsf{m} }  \, \ottnt{v_{{\mathrm{1}}}} \, \ottnt{op} \, \ottnt{e'_{{\mathrm{2}}}}}{%
{\ottdrulename{EOpR}}{}%
}}

\newcommand{\ottdruleEIf}[1]{\ottdrule[#1]{%
\ottpremise{\ottnt{e_{{\mathrm{1}}}} \,  \longrightarrow _{ \mathsf{m} }  \, \ottnt{e'_{{\mathrm{1}}}}}%
}{
 \mathsf{if} ~  \ottnt{e_{{\mathrm{1}}}}  ~  \ottnt{e_{{\mathrm{2}}}}  ~~  \ottnt{e_{{\mathrm{3}}}}  \,  \longrightarrow _{ \mathsf{m} }  \,  \mathsf{if} ~  \ottnt{e'_{{\mathrm{1}}}}  ~  \ottnt{e_{{\mathrm{2}}}}  ~~  \ottnt{e_{{\mathrm{3}}}} }{%
{\ottdrulename{EIf}}{}%
}}

\newcommand{\ottdruleEAppLRaise}[1]{\ottdrule[#1]{%
}{
  \mathsf{err}^ \ottnt{l}   ~  \ottnt{e_{{\mathrm{2}}}}  \,  \longrightarrow _{ \mathsf{m} }  \,  \mathsf{err}^ \ottnt{l} }{%
{\ottdrulename{EAppLRaise}}{}%
}}

\newcommand{\ottdruleEAppRRaise}[1]{\ottdrule[#1]{%
}{
 \ottnt{v_{{\mathrm{1}}}}  ~   \mathsf{err}^ \ottnt{l}   \,  \longrightarrow _{ \mathsf{m} }  \,  \mathsf{err}^ \ottnt{l} }{%
{\ottdrulename{EAppRRaise}}{}%
}}

\newcommand{\ottdruleEOpLRaise}[1]{\ottdrule[#1]{%
}{
 \mathsf{err}^ \ottnt{l}  \, \ottnt{op} \, \ottnt{e_{{\mathrm{2}}}} \,  \longrightarrow _{ \mathsf{m} }  \,  \mathsf{err}^ \ottnt{l} }{%
{\ottdrulename{EOpLRaise}}{}%
}}

\newcommand{\ottdruleEOpRRaise}[1]{\ottdrule[#1]{%
}{
\ottnt{v_{{\mathrm{1}}}} \, \ottnt{op} \,  \mathsf{err}^ \ottnt{l}  \,  \longrightarrow _{ \mathsf{m} }  \,  \mathsf{err}^ \ottnt{l} }{%
{\ottdrulename{EOpRRaise}}{}%
}}

\newcommand{\ottdruleEIfRaise}[1]{\ottdrule[#1]{%
}{
 \mathsf{if} ~   \mathsf{err}^ \ottnt{l}   ~  \ottnt{e_{{\mathrm{2}}}}  ~~  \ottnt{e_{{\mathrm{3}}}}  \,  \longrightarrow _{ \mathsf{m} }  \,  \mathsf{err}^ \ottnt{l} }{%
{\ottdrulename{EIfRaise}}{}%
}}








%% file: tfp.bbl
\begin{thebibliography}{10}
\providecommand{\url}[1]{\texttt{#1}}
\providecommand{\urlprefix}{URL }

\bibitem{AbadiCardelliCurienLevy91JFP}
Abadi, M., Cardelli, L., Curien, P.L., L{\'e}vy, J.J.: Explicit substitutions.
  Journal of {F}unctional {P}rogramming ({JFP})  1(4),  375--416 (1991)

\bibitem{Ahmed11blame}
Ahmed, A., Findler, R.B., Siek, J., Wadler, P.: Blame for all. In: Principles
  of {P}rogramming {L}anguages ({POPL}) (2011)

\bibitem{Dimoulas11cpcf}
Dimoulas, C., Felleisen, M.: On contract satisfaction in a higher-order world.
  TOPLAS  33(5),  16:1--16:29 (Nov 2011)

\bibitem{Dimoulas11indy}
Dimoulas, C., Findler, R.B., Flanagan, C., Felleisen, M.: Correct blame for
  contracts: no more scapegoating. In: Principles of {P}rogramming {L}anguages
  ({POPL}) (2011)

\bibitem{Dimoulas12complete}
Dimoulas, C., Tobin-Hochstadt, S., Felleisen, M.: Complete monitors for
  behavioral contracts. In: Seidl, H. (ed.) Programming Languages and Systems,
  LNCS, vol. 7211, pp. 214--233. Springer Berlin Heidelberg (2012)

\bibitem{Findler02contracts}
Findler, R.B., Felleisen, M.: Contracts for higher-order functions. In:
  International {C}onference on {F}unctional {P}rogramming ({ICFP}) (2002)

\bibitem{Findler08immutable}
Findler, R.B., Guo, S.Y., Rogers, A.: Lazy contract checking for immutable data
  structures. In: Chitil, O., Horv\'{a}th, Z., Zs\'{o}k, V. (eds.)
  Implementation and Application of Functional Languages, pp. 111--128.
  Springer-Verlag, Berlin, Heidelberg (2008)

\bibitem{Racket}
Flatt, M., PLT: Reference: Racket. Tech. Rep. PLT-TR-2010-1, PLT Design Inc.
  (2010), \url{http://racket-lang.org/tr1/}

\bibitem{Garcia13threesomes}
Garcia, R.: Calculating threesomes, with blame. In: International {C}onference
  on {F}unctional {P}rogramming ({ICFP}) (2013)

\bibitem{Greenberg13thesis}
Greenberg, M.: Manifest Contracts. Ph.D. thesis, University of Pennsylvania
  (November 2013)

\bibitem{Greenberg15space}
Greenberg, M.: Space-efficient manifest contracts. In: Principles of
  {P}rogramming {L}anguages ({POPL}) (2015)

\bibitem{Greenberg10contracts}
Greenberg, M., Pierce, B.C., Weirich, S.: Contracts made manifest. In:
  Principles of {P}rogramming {L}anguages ({POPL}) (2010)

\bibitem{Grossman00typeabs}
Grossman, D., Morrisett, G., Zdancewic, S.: Syntactic type abstraction. TOPLAS
  22(6),  1037--1080 (Nov 2000)

\bibitem{Herman07space}
Herman, D., Tomb, A., Flanagan, C.: Space-efficient gradual typing. In: Trends
  in {F}unctional {P}rogramming ({TFP}). pp. 404--419 (Apr 2007)

\bibitem{Herman10space}
Herman, D., Tomb, A., Flanagan, C.: Space-efficient gradual typing. Higher
  Order Symbol. Comput.  23(2),  167--189 (Jun 2010)

\bibitem{Jhala14}
Jhala, R.: Refinement types for haskell. In: Programming {L}anguages {M}eets
  {P}rogram {V}erification ({PLPV}). pp. 27--27. ACM, New York, NY, USA (2014)

\bibitem{Meyer92Eiffel}
Meyer, B.: Eiffel: the language. Prentice-Hall, Inc. (1992)

\bibitem{Plotkin77pcf}
Plotkin, G.: {LCF} considered as a programming language. Theoretical Computer
  Science  5(3),  223 -- 255 (1977)

\bibitem{RacketContracts}
{R}acket contract system (2013)

\bibitem{Rondon08liquid}
Rondon, P.M., Kawaguchi, M., Jhala, R.: Liquid types. In: Programming
  {L}anguage {D}esign and {I}mplementation ({PLDI}) (2008)

\bibitem{Sekiyama16fh}
Sekiyama, T., Igarashi, A., Greenberg, M.: Polymorphic manifest contracts,
  revised and resolved. TOPLAS  (2016), accepted in September; to appear

\bibitem{Sekiyama15datatypes}
Sekiyama, T., Nishida, Y., Igarashi, A.: Manifest contracts for datatypes. In:
  Principles of {P}rogramming {L}anguages ({POPL}). pp. 195--207. ACM, New
  York, NY, USA (2015)

\bibitem{Siek15coercions}
Siek, J., Thiemann, P., Wadler, P.: Blame, coercion, and threesomes: Together
  again for the first time. In: Programming {L}anguage {D}esign and
  {I}mplementation ({PLDI}) (2015)

\bibitem{Siek06gradual}
Siek, J.G., Taha, W.: Gradual typing for functional languages. In: Scheme and
  {F}unctional {P}rogramming {W}orkshop (September 2006)

\bibitem{Siek10threesomes}
Siek, J.G., Wadler, P.: Threesomes, with and without blame. In: Principles of
  {P}rogramming {L}anguages ({POPL}). pp. 365--376. ACM, New York, NY, USA
  (2010)

\bibitem{Thiemann16delta}
Thiemann, P.: A delta for hybrid type checking. In: Wadler Festschrift. pp.
  411--432. LNCS 9600, Springer Switzerland

\bibitem{Tobin-Hochstadt12symbolic}
Tobin-Hochstadt, S., Van~Horn, D.: Higher-order symbolic execution via
  contracts. In: OOPSLA. pp. 537--554. ACM, New York, NY, USA (2012)

\bibitem{Vazou2013}
Vazou, N., Rondon, P.M., Jhala, R.: Abstract refinement types. In: Felleisen,
  M., Gardner, P. (eds.) European {S}ymposium on {P}rogramming ({ESOP}). pp.
  209--228. Springer Berlin Heidelberg, Berlin, Heidelberg (2013)

\bibitem{Wadler15blame}
Wadler, P.: {A Complement to Blame}. In: Ball, T., Bodik, R., Krishnamurthi,
  S., Lerner, B.S., Morrisett, G. (eds.) SNAPL. LIPIcs, vol.~32, pp. 309--320.
  Schloss Dagstuhl--Leibniz-Zentrum fuer Informatik (2015)

\bibitem{Wadler09blame}
Wadler, P., Findler, R.B.: Well-typed programs can't be blamed. In: European
  {S}ymposium on {P}rogramming ({ESOP}) (2009)

\end{thebibliography}
